\documentclass[english,notitlepage,nofootinbib,superscriptaddress]{revtex4-1}
\usepackage[T1]{fontenc}
\usepackage[latin9]{inputenc}
\usepackage[usenames,dvipsnames]{color}  %%%%%%% usenames,dvipsnames required to get BrickRed 
\definecolor{darkred}{rgb}{0.6,0,0}
\definecolor{brown}{rgb}{0.59, 0.29, 0.0}
\usepackage{amsmath}
\usepackage{graphicx}
\usepackage[colorlinks=true,citecolor=darkred,urlcolor=darkred, pdfborder={0 0 0}]{hyperref}
\usepackage{ulem}
 
\newcommand {\ignore}[1]{}
\def\321{$\mathrm{SU(3) \otimes SU(2) \otimes U(1)}$ }

\def\lsim{\raise0.3ex\hbox{$\;<$\kern-0.75em\raise-1.1ex\hbox{$\sim\;$}}}
\def\lfv{lepton flavour violation }
\def\clfv{charged lepton flavour violation }

\def\Clfvg{Charged lepton flavour violating }
\def\clfvg{charged lepton flavour violating }

\def\SM{$\mathrm{SU(3)_c \otimes SU(2)_L \otimes U(1)_Y}$ }
\newcommand{\sm}{{Standard Model }}
\def \znbb {$0\nu\beta\beta$ }
\def\vev#1{\left\langle #1\right\rangle}
\def\gsim{\raise0.3ex\hbox{$\;>$\kern-0.75em\raise-1.1ex\hbox{$\sim\;$}}}
\def\lsim{\raise0.3ex\hbox{$\;<$\kern-0.75em\raise-1.1ex\hbox{$\sim\;$}}}

\def\3211{$\mathrm{SU(3) \otimes SU(2)_L \otimes U(1)_R \otimes U(1)_{B-L}}$ }
\def\321{$\mathrm{SU(3) \otimes SU(2) \otimes U(1)}$ }
\def\422{$\mathrm{SU(4) \otimes SU(2) \otimes SU(2)_R}$ }
\newcommand {\black} {\color{black}}

\definecolor{linkcolor}{rgb}{0,0,0.5}

\newcommand{\vvvv}{ \rule[-2ex]{0pt}{6ex} }
\makeatletter

\providecommand{\tabularnewline}{\\}

\usepackage{slashed}

\makeatother

\usepackage{babel}
\begin{document}

    \title{\color{BrickRed} Towards deconstructing the simplest seesaw mechanism}

\author{Sanjoy Mandal}\email{smandal@kias.re.kr}
\affiliation{Korea Institute for Advanced Study, Seoul 02455, Korea}

\author{O. G. Miranda}\email{omr@fis.cinvestav.mx}
\affiliation{Departamento de F\'{\i}sica, Centro de   Investigaci{\'o}n y de Estudios Avanzados del IPN\\ Apartado Postal   14-740 07000 Ciudad de M\'exico, Mexico}

\author{G. Sanchez Garcia}\email{gsanchez@fis.cinvestav.mx}
\affiliation{Departamento de F\'{\i}sica, Centro de   Investigaci{\'o}n y de Estudios Avanzados del IPN\\ Apartado Postal   14-740 07000 Ciudad de M\'exico, Mexico}

\author{J. W. F. Valle}\email{valle@ific.uv.es}
\affiliation{AHEP Group, Institut de F\'{i}sica Corpuscular --   C.S.I.C./Universitat de Val\`{e}ncia, Parc Cientific de Paterna.\\   C/Catedratico Jos\'e Beltr\'an, 2 E-46980 Paterna (Val\`{e}ncia) - SPAIN}

\author{Xun-Jie Xu}\email{xuxj@ihep.ac.cn}
\affiliation{Institute of High Energy Physics, Chinese Academy of Sciences, Beijing 100049, China}.

\date{\today}
\begin{abstract}

  The triplet or type-II seesaw mechanism is the simplest way to endow neutrinos with mass in the Standard Model (SM).
  Here we review its associated theory and phenomenology, including restrictions from $S$, $T$, $U$ parameters, neutrino experiments, \clfv as well as collider searches.
  We also examine restrictions coming from requiring consistency of electroweak symmetry breaking, i.e. perturbative unitarity and stability of the vacuum. 
  Finally, we discuss novel effects associated to the scalar mediator of neutrino mass generation namely,
  (i) rare processes, e.g. $l_\alpha \to  l_\beta \gamma$ decays, at the intensity frontier, and also
  (ii) four-lepton signatures in colliders at the high-energy frontier.
  These can be used to probe neutrino properties in an important way, providing
  a test of the absolute neutrino mass and mass-ordering, as well as of the atmospheric octant.
    They may also provide the first evidence for charged lepton flavour violation in nature.
  In contrast, neutrino non-standard interaction strengths are found to lie below current detectability.

\end{abstract}
\maketitle

\section{Introduction}
\label{sec:introduction}

The discovery of neutrino oscillations~\cite{Kajita:2016cak,McDonald:2016ixn} has opened a new era in particle physics, implying the need for small neutrino mass
and, as a result, providing the first strong evidence for new physics beyond the Standard Model (SM)~\cite{Valle:2015pba}.
So far, we have not been able to underpin the ultimate mechanism responsible for neutrino mass generation.
An attractive possibility is the seesaw mechanism, which leads to small active neutrino masses through the exchange of heavy lepton or scalar mediators.
Most generally the seesaw is formulated within the SM picture, i.e., using the \SM gauge group~\cite{Schechter:1980gr,Schechter:1981cv}. 

Here we re-examine the simplest scalar-mediated \SM seesaw mechanism, i.e., the type-II seesaw with explicit breaking of lepton number.
We discuss both its theoretical consistency and the resulting restrictions, including the current experimental constraints,
and those from electroweak precision data.
We show that future high-energy colliders can play a key role in probing neutrino properties such as the atmospheric octant,
the neutrino mass-ordering, and the absolute-neutrino-mass-scale, as recently sketched in~\cite{Miranda:2022xbi}.

Moreover, such high-energy collider experiments may also provide the
discovery site for processes with charged lepton flavor non-conservation in nature~\cite{Bernabeu:1987gr,Rius:1989gk,AguilarSaavedra:2012fu,Das:2012ii,Deppisch:2013cya}.
In contrast, current \clfv limits severely restrict the strength of neutrino non-standard interactions (NSI),
posing a real challenge for the next generation of neutrino experiments. \\[-.4cm]

  After a brief review of the status of the type-II seesaw mechanism we show how it provides a neat realization of the idea that
  neutrino properties can be well-studied also within high energy experiments.
The paper extends the presentation given in~\cite{Miranda:2022xbi}, providing the full theoretical discussion and relevant technical details.
The organization is as follows. In Sec.~\ref{sec:type-ii-seesaw}, we briefly describe the type-II seesaw model with explicit violation of lepton number.
In Sec.~\ref{sec:constraints}, we analyze the constraints,
including consistency of the scalar sector in~\ref{sec:stab-unit-pert},
electroweak precision data in~\ref{STU-par},
constraints from neutrino physics in~\ref{sec:neutr-mass-constr},
from \clfv bounds in~\ref{sec:charg-lept-flav},
and from collider experiments in~\ref{sec:collider-constraints}. 
We show how most of these restrictions can be significant.\\[-.4cm]
 
In Sec.~\ref{sec:collider-signature}, we describe how the type-II seesaw model can lead to promising signatures at hadron colliders such as LHC and planned FCC, 
as well as future $e^+ e^-$ colliders, such as the ILC, CLIC, and the CEPC project in China.
Altogether, as discussed in Sec.~\ref{sec:compl-betw-clfv}, we find that current and future experiments can ultimately fully deconstruct the type-II seesaw mechanism.
We discuss the interplay between collider probes and \clfv searches, illustrating the complementarity of high-energy and high-intensity frontiers.
As a result one will be able to probe important current unknowns from neutrino experiments, such as the atmospheric octant, the mass ordering, and the absolute neutrino mass.

In Sec.~\ref{sec:nsi-from-type}, we examine the possibility of having sizeable non-standard neutrino interactions (NSI) given the constraints from searches for \lfv\ described previously.
Summary and outlook are given in Sec.~\ref{sec:outlook}. Complementary material is presented in Appendices \ref{app:RGEs}, \ref{app:STU} and \ref{app:decay-width}.

\section{Type-II seesaw mechanism}
\label{sec:type-ii-seesaw}

Small neutrino masses required by neutrino oscillation experiments can be generated by the non-renormalizable dimension-5 Weinberg operator:
\begin{equation}
  {\cal L} \propto \frac{1}{\Lambda} {L}\Phi\Phi L.
  \label{eq:LLHH}
\end{equation}
Out of the three possible UV-completions of the Weinberg operator, here we focus on the simplest, the type-II seesaw
mechanism~\cite{Schechter:1980gr,Schechter:1981cv}~\footnote{The type-II is the simplest seesaw mechanism, hence originally named type-I in the above references.}.
It introduces one heavy $SU(2)_{L}$ triplet scalar $\Delta=(\Delta^{++},\Delta^{+},\Delta^{0})^{T}$, with hypercharge $Y_{\Delta}= 2$. 
It is convenient to describe $\Delta$ in its matrix form as: 
\begin{equation}
\Delta=\left(\begin{array}{cc}
\Delta^{+}/\sqrt{2} & \Delta^{++}\\
\Delta^{0} & -\Delta^{+}/\sqrt{2}
\end{array}\right),\label{eq:t2-7}
\end{equation}
which is a traceless bidoublet under $SU(2)_{L}$, i.e., $\Delta\to U\Delta U^{\dagger}$ for $U\in SU(2)_{L}$. 
This triplet couples only to the SM lepton doublets, and to the scalar Higgs doublet $\Phi = (\Phi^{+}~~ \Phi^{0})^{T}$.
The relevant Yukawa Lagrangian reads: 
\begin{equation}
{\cal L}_{{\rm type-II}}=\left[iY_{\Delta\alpha\beta}L_{\alpha}^{T}C^{-1}
    %\widetilde{L}_{\alpha}^{\dagger}
    \tau_2\Delta L_{\beta}+\text{h.c.}\right]+\left(D_{\mu}\Phi\right)^{\dagger}\left(D^{\mu}\Phi\right) +\left(D_{\mu}\Delta\right)^{\dagger}\left(D^{\mu}\Delta\right)-V(\Phi,\Delta),\label{app:x-3}
\end{equation}
where $Y_{\Delta\alpha \beta}$ is a complex symmetric matrix, $L_{\alpha}$ are the lepton doublets, $C$ is the charge conjugation operator, and $D_{\mu}$ denotes the covariant derivative of the
corresponding scalar field.
We can assume without loss of generality that the charged lepton Yukawa coupling is in diagonal form.
Note that after symmetry breaking, the first term in Eq.~(\ref{app:x-3}) will generate the Majorana mass term for light neutrinos as
\begin{align}
{\cal L}_{\text{Majorana}}=\overline{\nu_{\alpha L}^c} m_{\alpha\beta}^{\nu}\nu_{\beta L} + \text{h.c.}
\end{align} 
The scalar potential $V(\Phi,\Delta)$ is given as, 
\begin{equation}
\begin{aligned}
 V(\Phi,\Delta) = & -m_{\Phi}^{2}\Phi^{\dagger}\Phi + \frac{\lambda}{4}(\Phi^{\dagger}\Phi)^{2} +  \tilde{M}_{\Delta}^{2}{\rm Tr}\left[\Delta^{\dagger}\Delta\right]+\lambda_{2}\left[{\rm Tr}\Delta^{\dagger}\Delta\right]^{2}+\lambda_{3}{\rm Tr}\left[\Delta^{\dagger}\Delta\right]^{2}\\
  &+ \left[\mu \Phi^{T}i\sigma_{2}\Delta^{\dagger}\Phi+\text{h.c.}\right]+\lambda_{1}(\Phi^{\dagger}\Phi){\rm Tr}\left[\Delta^{\dagger}\Delta\right]+\lambda_{4}\Phi^{\dagger}\Delta\Delta^{\dagger}\Phi.\label{eq:potential}
\end{aligned}
\end{equation}

The Higgs triplet UV-completion of the Weinberg operator is characterized by a small induced vacuum expectation value (VEV) for the triplet.
Indeed, within most of the parameter space, the minimization of the potential can flavour nonzero VEVs $v_{\Phi}$ and $v_\Delta$ for the doublet and the triplet,
respectively~\footnote{For more details, see Ref.~\cite{Xu:2016klg}.}.
Minimization of the total potential $V(\Phi,\Delta) $ leads to the relations: 
\begin{align}
 \tilde{M}_\Delta^2&=\displaystyle M_\Delta^2-\frac{1}{2}\left[2v_\Delta^2(\lambda_2+\lambda_3)+v_\Phi^2(\lambda_1+\lambda_4)\right],\,\,\text{ with } M_\Delta^2\equiv \frac{v_\Phi^2\mu}{\sqrt{2}v_\Delta}.  \label{Tadpole1} \\
m_\Phi^2&=\displaystyle \frac{1}{2}\left[\frac{v_\Phi^2\lambda}{2}+v_\Delta^2(\lambda_1+\lambda_4)-2\sqrt{2}\mu v_\Delta\right].
\label{Tadpole2}
\end{align}
In the limit $M_{\Delta} \gg v_{\Phi} $, which is consistent with all the constraints (see below), we can solve  Eq. (\ref{Tadpole1}) for $v_{\Delta}$.
Keeping terms of $\mathcal{O}(v_{\Phi}/M_{\Delta})$ we get the small induced triplet vacuum expectation value : 
\begin{equation}
\boxed{
v_{\Delta} \approx \frac{\mu v_{\Phi}^{2}}{\sqrt{2}\tilde{M}_{\Delta}^{2}}}~. \label{V-triplet-approx}
\end{equation}
Eq.~(\ref{V-triplet-approx}) explicitly shows that the smallness of the triplet VEV can be induced either by a small $\mu$, or by a large value for $\tilde{M}_\Delta$.
A small $\mu$ is in agreement with t'Hooft's naturalness argument, since it is the only parameter that sources lepton number violation in
Eq.~\eqref{eq:potential}~\footnote{Of course, within a more complete theory one can give a dynamical meaning to the parameter $\mu$~\cite{Schechter:1981cv}.}.
As a result, in the limit $\mu \to 0$, lepton number symmetry is recovered. 
After spontaneous symmetry breaking, we can write the doublet and triplet neutral fields as:
\begin{equation}
\Delta = \frac{1}{\sqrt{2}}\begin{pmatrix}
 \Delta^{+}& \sqrt{2}\Delta^{++}\\ 
 v_{\Delta} + h_{\Delta} + i\eta_{\Delta}& - ~\Delta^{+}
\end{pmatrix},~~~~~~~
\Phi = \frac{1}{\sqrt{2}}\begin{pmatrix}
\sqrt{2}{\Phi}^{+}\\ 
v + h_{\Phi} + i\eta_{\Phi}
\end{pmatrix}.
\end{equation}
Our scalar sector contains 10 degrees of freedom. After EW breaking, seven remain as physical fields with definite masses.
  These scalar fields are the charged Higgs bosons $H^{\pm\pm}$, $H^{\pm}$, and the neutral Higgs bosons $h$, $H^0$ and $A^0$.
  The doubly-charged Higgs $H^{\pm\pm}$ is simply the $\Delta^{\pm\pm}$ present in $\Delta$.
  The two singly-charged scalar fields $\Phi^{\pm}$ and $\Delta^{\pm}$ from $\Phi$ and $\Delta$ mix, giving the singly-charged Higgs boson $H^{\pm}$ and the unphysical charged Goldstone
  $G^{\pm}$~\footnote{Note that, the mixing within the singly-charged Higgs sector is characterized by $\tan\beta_{\pm}=\sqrt{2}\frac{v_\Delta}{v_\Phi}$, hence can be neglected for small $v_\Delta$.}.
  Similarly, $\eta_{\Delta}$ and $\eta_{\Phi}$ will mix and give rise to the CP-odd scalar $A^0$ and the neutral Goldstone boson $G^0$ which becomes the longitudinal mode of $Z$.
  Finally, two CP-even fields $h_\Delta$ and $h_\Phi$ mix and give rise to the SM Higgs boson $h$ and a heavy Higgs boson $H^0$.
  For a detailed discussion of charged and neutral scalar mass matrices and mixings, see~\cite{Arhrib:2011uy,Bonilla:2015eha,Chun:2019hce}.

The physical masses of doubly- and singly-charged Higgs bosons $H^{\pm \pm}$ and $H^{\pm}$  can be written as:
    \begin{eqnarray}
m_{H^{++}}^2=M_\Delta^2-v_\Delta^2\lambda_3-\frac{\lambda_4}{2}v_\Phi^2,\label{eq:mhpp}~~~
m_{H^+}^2= \left(M_\Delta^2-\frac{\lambda_4}{4}v_\Phi^2\right)\left(1+\frac{2v_\Delta^2}{v_\Phi^2}\right),\label{eq:mhp}
\end{eqnarray} 
whereas the CP-odd Higgs field $A^0$ has the following mass:
\begin{eqnarray}
m_{A^{0}}^2 &= &M_\Delta^2\left(1+\frac{4v_\Delta^2}{v_\Phi^2}\right). \label{mA}
\end{eqnarray}  
The common $M_{\Delta}^{2}$ is expected, since the scalars come from the same triplet. Finally, the CP-even Higgs bosons $h$ and $H^0$ have the following physical masses:
\begin{eqnarray}
m_{h}^2&=&\frac{1}{2}[A+C-\sqrt{(A-C)^2+4B^2}], \label{eq:mh0}\\
m_{H^0}^2&=&\frac{1}{2}[A+C+\sqrt{(A-C)^2+4B^2}], \label{eq:mH0}
\end{eqnarray}
where,
\begin{eqnarray}
  A &=& \frac{\lambda}{2}{v_\Phi^2}, \; \;
  B =v_\Phi ( -\sqrt{2}\mu+(\lambda_1+\lambda_4)v_\Delta) , \; \; 
  C = M_\Delta^2+2(\lambda_2+\lambda_3)v_\Delta^2 .
\label{eq:ABC}
\end{eqnarray}

For $v_\Delta\ll v_\Phi$, the masses of the physical Higgs bosons can be approximated as follows:
\begin{align}
\boxed{
m_{H^{\pm\pm}}^2\simeq M_\Delta^2-\frac{\lambda_4}{2}v_\Phi^2,\,\,\,\, m_{H^\pm}^2\simeq M_\Delta^2-\frac{\lambda_4}{4}v_\Phi^2,\,\,\,\, m_h^2\simeq 2\lambda v_\Phi^2\,\,\,\,\text{and}\,\,\,\,m_{H^0}^2\approx m_{A^0}^2\simeq M_{\Delta}^2 }~,
\end{align}
so their mass-squared differences are given by:
\begin{align}
m_{H^{\pm\pm}}^2-m_{H^\pm}^2\approx m_{H^\pm}^2-m_{H^0/A^0}^2\approx -\frac{\lambda_4}{4}v_\Phi^2.
\end{align}
We further define the two mass-splittings as follows:
\begin{align}
\delta m_1=m_{H^\pm}-m_{H^0},\,\,\,\,\,\,\,\, \delta m_2= m_{H^{\pm\pm}}-m_{H^\pm}.
\label{eq:dm}
\end{align}
With the assumptions $v_\Delta\ll v_\Phi$ and $M_{\Delta}^2\gg |\lambda_4| v_\Phi^2$, the two mass-splittings $\delta m_{1,2}$ can be approximated as 
\begin{align}
\delta m_{1,2}\approx -\frac{\lambda_4}{8}\frac{v_\Phi^2}{M_\Delta}.
\end{align}
Hence, the tree-level mass-splitting is dictated by the dimensionless quartic coupling $\lambda_4$, so $\delta m_1\approx \delta m_2$ at tree-level.
Note also that a relatively large value of this quartic coupling at the electroweak scale can become non-perturbative at high energies even below the Planck scale.
Hence $\lambda_4$ should be small, and as a result, the mass-splitting $\delta m_{1,2}$ should also be small.

Note also that, three distinct mass spectra are expected depending on the value (sign) of $\lambda_4$, as:
$(1).\,\, \lambda_4=0:\,\,\delta m\approx 0\,( m_{H^{\pm\pm}}\simeq m_{H^\pm}\simeq m_{H^0/A}) $,
$(2).\,\, \lambda_4<0:\,\,\delta m>0\,( m_{H^{\pm\pm}} > m_{H^\pm} > m_{H^0/A})$ and
$(3).\,\, \lambda_4>0:\,\,\delta m<0\,( m_{H^{\pm\pm}} < m_{H^\pm} < m_{H^0/A})$. 
These will be important when discussing the collider constraints in the triplet seesaw, Sec.~\ref{sec:collider-constraints}.
 
\begin{figure}[h]
\centering
\includegraphics[width=0.45\textwidth,height=4cm]{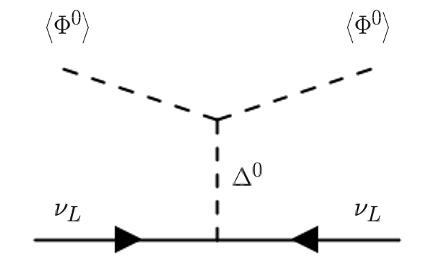}
\caption{Feynman diagram generating Majorana masses in type-II seesaw mechanism. } 
\label{fig:neutrino-mass}
\end{figure}
On the other hand, the Yukawa interaction in the first term in Eq.~(\ref{app:x-3}) is responsible for generating light Majorana neutrino masses.
After spontaneous symmetry breaking, we have:  
\begin{equation}
\boxed{  m_{\alpha\beta}^{\nu}\equiv (Y_{\Delta})_{\alpha\beta}\frac{v_{\Delta}}{\sqrt{2}}}~.\label{eq:s-18}
\end{equation}
This mass matrix $ m_{\alpha\beta}^{\nu}$ in the flavor basis must be diagonalized in order to obtain the physical neutrino masses.
The smallness of the latter will follow from the small VEV, Eq. (\ref{V-triplet-approx}).

\section{Constraints} 
\label{sec:constraints}

We now examine the constraints that follow from electroweak precision tests, theoretical consistency, as well as from various experiments.
Theoretical consistency requirements include stability of the electroweak vacuum as well as perturbative unitarity.

\subsection{Stability, Unitarity and Perturbativity} 
\label{sec:stab-unit-pert}

The stability of the vacuum requires the potential given in Eq.~\eqref{eq:potential} to be bounded from below when the scalar fields become large in any field space direction.
At large field values, the potential Eq.~\eqref{eq:potential} is generically dominated by the part containing the quartic terms, i.e. 
\begin{equation}
V^{(4)}(\Phi, \Delta) =\frac{\lambda}{4}(\Phi^\dagger{\Phi})^2
+\lambda_1(\Phi^\dagger{\Phi})\text{Tr}(\Delta^{\dagger}{\Delta})+\lambda_2(\text{Tr}\Delta^{\dagger}{\Delta})^2
+\lambda_3\text{Tr}(\Delta^{\dagger}{\Delta})^2
+\lambda_4{\Phi^\dagger\Delta\Delta^{\dagger}\Phi} .
\label{eq:Vquartic}
\end{equation}
The quartic piece of the potential suffices to obtain the conditions for tree-level stability. 
Taking into account all field directions, the necessary and sufficient conditions for the potential to be bounded from below~(BFB) have been studied in Ref.~\cite{Bonilla:2015eha,Arhrib:2011uy}.
They can be written as follows, 
\begin{eqnarray}
&& \lambda > 0 \;\;{\rm \&}\;\; \lambda_2+\lambda_3 > 0  \;\;{\rm \&}  \;\;\lambda_2+\frac{\lambda_3}{2} > 0 
\label{eq:BFBgen1} \\
&& {\rm \&} \;\;\lambda_1+ \sqrt{\lambda(\lambda_2+\lambda_3)} > 0 \;\;{\rm \&}\;\;
\lambda_1+ \sqrt{\lambda(\lambda_2+\frac{\lambda_3}{2})} > 0  \label{eq:BFBgen2}\\
&& {\rm \&} \;\; \lambda_1+\lambda_4+\sqrt{\lambda(\lambda_2+\lambda_3)} > 0 \;\; {\rm \&} \;\; 
\lambda_1+\lambda_4+\sqrt{\lambda(\lambda_2+ \frac{\lambda_3}{2})} > 0.  \label{eq:BFBgen3}
\end{eqnarray}
The above conditions ensure a BFB potential for all possible directions in field space and provide the most general necessary and sufficient BFB conditions at tree-level. 
As quartic couplings change with energy due to renormalisation group evolution, the BFB conditions must be satisfied at all energy scales for a consistent vacuum.\\[-.4cm]

In addition, quartic couplings can also be constrained by demanding tree-level unitarity, which should be preserved in a variety of scattering processes
such as scalar-scalar scattering, gauge-boson-gauge-boson scattering, and scalar-gauge-boson scattering. 
These have been studied in great detail for the type-II seesaw model in Ref.~\cite{Arhrib:2011uy}. 
Demanding tree-level unitarity to be preserved for different elastic scattering processes, one obtains the following constraints: 
\begin{align}
&|\lambda_1 + \lambda_4| \leq \kappa \pi,\,\,  |\lambda_1| \leq \kappa \pi,\,\, |2 \lambda_1 + 3 \lambda_4| \leq 2 \kappa \pi,\,\, |\lambda| \leq  2 {\kappa} \pi,\,\, |\lambda_2| \leq  \frac{\kappa}{2} \pi,\,\, |\lambda_2 + \lambda_3| \leq  \frac{\kappa}{2} \pi, \\
&|\lambda + 4 \lambda_2 + 8 \lambda_3 \pm \sqrt{(\lambda - 4 \lambda_2 - 8 \lambda_3)^2
+ 16 \lambda_4^2} \;| \leq  4 \kappa \pi,  \\
&| 3 \lambda + 16 \lambda_2 + 12 \lambda_3 \pm \sqrt{(3 \lambda - 16 \lambda_2 - 12 \lambda_3)^2
+ 24 (2 \lambda_1 +\lambda_4)^2} \;| \leq  4 \kappa \pi,  \\
&|2 \lambda_1 - \lambda_4| \leq 2 \kappa \pi,\,\, |2 \lambda_2 - \lambda_3| \leq  \kappa \pi. \label{eq:unit10} 
\end{align}
where $\kappa=8$. Moreover, we also impose the perturbativity condition, that is, all quartic couplings should be less than $4\pi$ up to the Planck scale.

Let us now examine the stability of the electroweak vacuum in more detail.  
To see how the couplings evolve with energy we use the full two-loop renormalisation group equations (RGEs) governing the evolution of the Higgs quartic coupling. \\[-.4cm]
\newpage
\underline{\bf Effective Theory}\\[-.4cm]

 Below the scale $\mu_r=M_\Delta$ one needs to integrate out the triplet field, so that the resulting theory is the Standard Model plus an effective dimension-five Weinberg operator, 
\begin{align}
\mathcal{L}_\nu^{d=5}=\kappa_0 L\Phi\Phi L,\,\,\text{with}\,\,\kappa_0=\frac{Y_\Delta\mu}{M_\Delta^2}.
\end{align}
Hence, below the scale $\mu_r = M_\Delta$, only the Standard Model couplings as well as the coefficient $\kappa_0$ will evolve.   

\begin{figure}[h]
\centering
\includegraphics[width=0.45\textwidth]{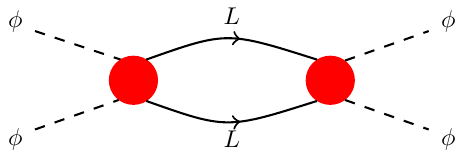}
\caption{Effect of Weinberg's effective operator on the Higgs quartic interaction in the effective theory~\cite{Bonilla:2015kna}.} 
\label{fig:quartic-from-neutrino-mass}
\end{figure}

\begin{figure}[h]
\centering
\includegraphics[width=0.3\textwidth]{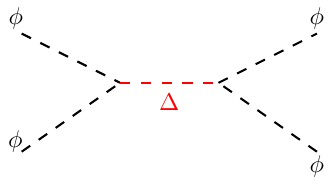}~~~
\includegraphics[width=0.2\textwidth]{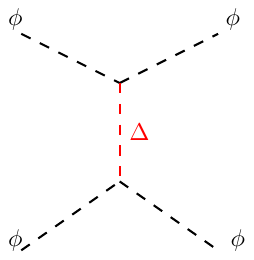}~~~
\includegraphics[width=0.2\textwidth]{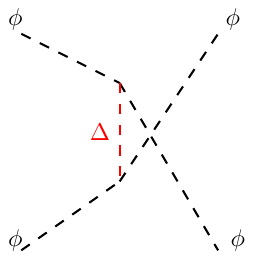}
\caption{Contributions to the quartic coupling $\lambda$ in effective theory obtained by integrating out the triplet.} 
\label{fig:contribution-lambda}
\end{figure}

As seen in Fig.~\ref{fig:neutrino-mass}, the same operator which generates the neutrino mass below the scale $\mu_r=M_\Delta$,
also provides a correction to the running of the Higgs quartic self-coupling $\lambda$ below that scale, see~Fig.~\ref{fig:quartic-from-neutrino-mass}. 
This generic contribution is of order $v_\Delta^2\kappa_0^2$ and hence negligible~\cite{Ng:2015eia}, as $\kappa_0$ is very small to account for the small neutrino masses. 
However, in the type-II seesaw, 
there are tree-level contributions to the effective potential arising from Fig.~\ref{fig:contribution-lambda}. 
These are obtained by integrating out the triplet and introducing threshold corrections to the Standard Model Higgs quartic coupling $\lambda$ at the scale $\mu_r=M_\Delta$. 


The effective potential below the scale $\mu_r=M_\Delta$ is then given as: 
%%%%%%%%%%%%%%%%%%%%%%%%%%%%%%%%%%%%%%%%%%%%%%%%%%%%%%%%%%%%%%%%%%%%%%
\begin{align}
V=-m_{\Phi}^2 \Phi^{\dagger}\Phi+\frac{\lambda'}{4}(\Phi^{\dagger}\Phi)^2,
\end{align}
where $\lambda'$ is given as:
\begin{align}
\lambda'=\lambda-\frac{4\mu^2}{M_\Delta^2}.
\label{eq:matching}
\end{align}
Eq.~(\ref{eq:matching}) shows how the matching condition at the scale $\mu_r=M_\Delta$ induces a positive shift in the Higgs quartic coupling, $\delta\lambda=\frac{4\mu^2}{M_\Delta^2}$.
This corresponds to a larger Higgs quartic coupling above the scale $\mu_r=M_\Delta$ and improves the chances of keeping $\lambda$ positive all the way at higher scales.  

\begin{figure}[h]
\centering
\includegraphics[width=0.45\textwidth]{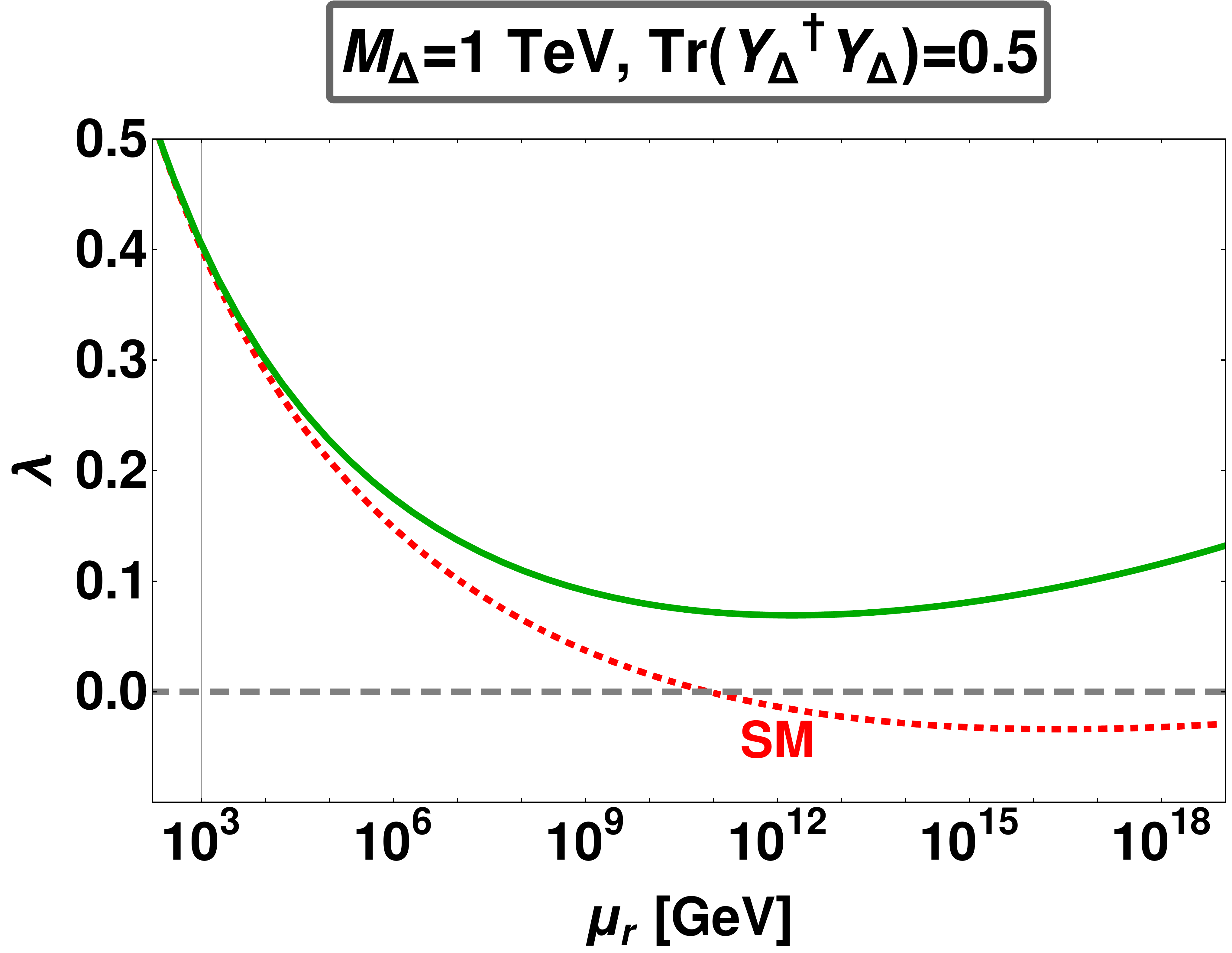}~~~
\includegraphics[width=0.45\textwidth]{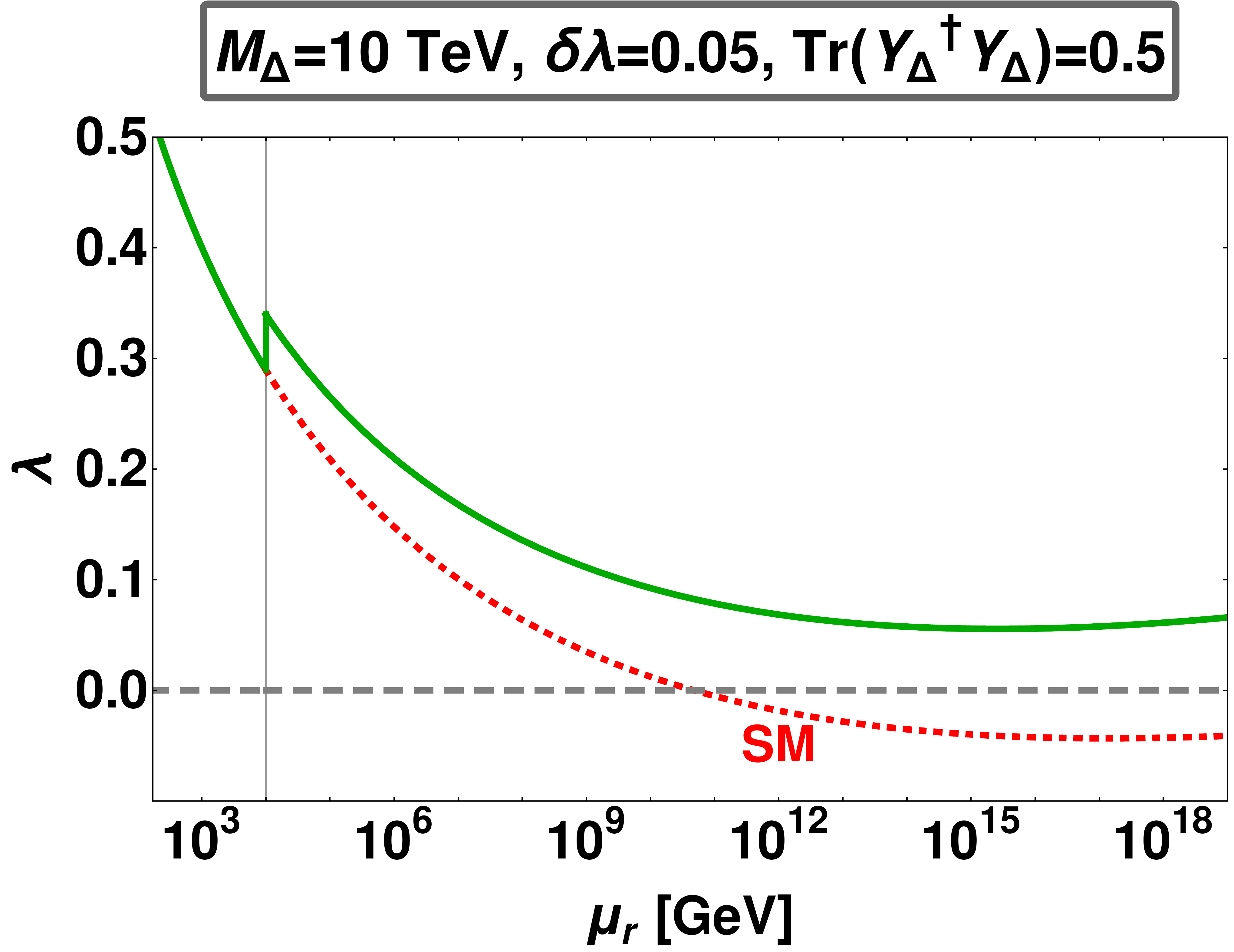}
\caption{The evolution of the SM quartic coupling $\lambda_{\text{SM}}$ is shown by the red dashed curves.
  This nearly coincides with the evolution of the scalar quartic coupling $\lambda$ below $\mu_r=M_\Delta$.
  The solid green curve illustrates the positivity of the quartic coupling throughout its evolution in both cases.} 
\label{fig:RGE}
\end{figure}
\underline{\bf Full Theory}\\[-.4cm]

Building upon the discussion of the previous section, we now turn to the region above the scale $\mu_r=M_\Delta$. 
As discussed earlier, below the $\mu_r=M_\Delta$ scale, the theory is an effective Standard Model supplemented by the dimension-five Weinberg operator. 
However, above the $\mu_r=M_\Delta$ scale, the theory is
UV-complete. Therefore, all the new couplings, such as the Yukawa coupling $Y_\Delta$ and the quartic couplings $\lambda_{1,2,3,4}$
will take part in the system of RGEs, and will consequently affect the evolution of the Standard Model couplings, especially $\lambda$. 
In Appendix~\ref{app:RGEs}, we present the one-loop RGEs for the full theory, and in Fig.~\ref{fig:RGE} we give illustrative examples of the evolution of $\lambda$. 

Notice that below the scale $M_\Delta$, the RGEs are the ones with ${\lambda_{1,2,3,4}}$ removed and $\lambda$ replaced by $\lambda'$. 
Above $M_\Delta$ one needs to include $\beta_{\lambda_{1,2,3,4}}$ and find $\lambda$ using the full RGEs with the matching condition as in Eq.~(\ref{eq:matching}) at $\mu_r=M_\Delta$. 
As far as the new Yukawa coupling $Y_\Delta$ is concerned, it can be obtained by substituting $Y_\Delta= \theta(\mu_r-M_\Delta) Y_\Delta$ on the right side of the RGEs of the full theory. \\[-.4cm] 

In Fig.~\ref{fig:RGE}, we show the running of the quartic coupling $\lambda$ for two benchmarks.
In the left panel we fix $M_\Delta=1$~TeV, $\lambda_{1,2,3,4}=0.1$ and $\text{Tr}\left(Y_\Delta^{\dagger}Y_\Delta\right)=0.5$.
In the right panel we fix $M_\Delta=10$~TeV, $\lambda_{1,2,3,4}=0.05$, and also~\footnote{Note that this large Yukawa coupling is chosen just for illustration, and may not be consistent with neutrino masses. However, if stability holds even for such large Yukawa values, then it will be certainly consistent with smaller Yukawas adequate for neutrino masses.} $\text{Tr}\left(Y_\Delta^{\dagger}Y_\Delta\right)=0.5$. We choose the parameter $\mu$ in such a way that the positive shift $\delta\lambda=0.05$.
One sees from the left panel that, just with the effect of running, the coupling $\lambda$ can be kept positive  
all the way up to Planck scale even for relatively large Yukawa coupling $\text{Tr}(Y_\Delta^{\dagger}Y_\Delta)$ as long as we choose adequately large values for the quartic couplings $\lambda_{1,2,3,4}$.  
In the right panel, we show that if the positive shift $\delta\lambda$ coming from the matching condition~Eq.~\eqref{eq:matching} is relatively large, this can also help achieving vacuum stability. 
The evolution of other quartic couplings and stability conditions given in Eq.~\eqref{eq:BFBgen1}-\eqref{eq:BFBgen3} are not shown to avoid cluttering the plot.
In conclusion, one sees that even for relatively large Yukawa coupling $Y_\Delta$, one can have a wide range of parameters with stable vacuum all the way up to the Planck scale.

\subsection{Electroweak precision $S$, $T$, $U$ parameters}
\label{STU-par}
%%%%%%%%%%%%%%%%%%%%%%%%%%%%%%%%%%%%%%%%%%%%%%%%%%%%%%%%%%%%%%%%%%%

The presence of the Higgs triplet in the type-II seesaw will modify the prediction for different radiative corrections, especially for the so-called $\rho$ parameter~\cite{Schechter:1980gr}.
This will imply restrictions on the vacuum expectation value, $v_\Delta$, that will translate into theoretical restrictions for the mass of the triplet scalars. 

The triplet VEV $v_{\Delta}$ leads to a tree-level contribution to the $\rho$ parameter which is defined by $\rho\equiv m_{W}^{2}/(m_{Z}^{2}c_{W}^{2})$.
The \sm predicts $\rho=1$ at tree level. At loop level, it is redefined so as to absorb loop corrections so that the SM value remains $1$. 
The global fit of electroweak precision data currently gives~\cite{ParticleDataGroup:2020ssz}, 
\begin{equation}
\rho_{\text{global fit}}=1.00038\pm0.00020.\label{eq:m-1}
\end{equation}
In the type-II seesaw, $m_{W}^{2}$ and $m_{Z}^{2}$ are modified to $m_{W}^{2}=g^{2}(v_{\Phi}^{2}+2v_{\Delta}^{2})/4$ and $m_{Z}^{2}=g^{2}(v_{\Phi}^{2}+4v_{\Delta}^{2})/(4c_{W}^{2})$.
This leads to the theoretical relation
\begin{equation}
\rho_{\text{th.}}=\frac{v_{\Phi}^{2}+2v_{\Delta}^{2}}{v_{\Phi}^{2}+4v_{\Delta}^{2}}\approx1-2\frac{v_{\Delta}^{2}}{v^{2}},\label{eq:m-2}
\end{equation}
where the electroweak VEV $v\equiv\sqrt{v_{\Phi}^{2}+2v_{\Delta}^{2}}\approx246$ GeV. Comparing Eq.~\eqref{eq:m-2} with Eq.~\eqref{eq:m-1}, we obtain  
\begin{equation}
v_{\Delta}\leq2.6\ {\rm GeV}\ {\rm at}\ 3\sigma\text{ C.L.}\label{eq:m-18}
\end{equation}

Since the triplet couples to the $W$ and $Z$ bosons, it affects the oblique parameters (often referred to as the $S$, $T$, $U$ parameters)~\cite{Peskin:1991sw}. 
The general form of these are given in Appendix~\ref{app:STU}. Assuming that the triplet is heavy and the mass splittings amongst its components are small, the result reads as:
\begin{align}
S & \approx-\frac{(2-4s_{W}^{2}+5s_{W}^{4})m_{Z}^{2}}{30\pi m_{H^{\pm}}^{2}}-\frac{2(\delta m_{1}+\delta m_{2})}{3\pi m_{H^{\pm}}}\thinspace,\label{eq:m-S}\\
T & \approx\frac{\delta m_{1}^{2}+\delta m_{2}^{2}}{6\pi s_{W}^{2}m_{W}^{2}}\thinspace,\label{eq:m-T}\\
U & \approx\frac{\delta m_{1}-\delta m_{2}}{3\pi m_{H^{\pm}}}\thinspace,\label{eq:m-U}
\end{align}
where $\delta m_1$ and $\delta m_2$ are given in Eq.~\eqref{eq:dm}.
The $U$ parameter is highly suppressed, since in type-II seesaw we have $\delta m_{1}\approx\delta m_{2}$.  
The $T$ parameter is mainly sensitive to the mass splitting and it is amplified by the $s_{W}^{2}$ in the denominator.
The $S$ parameter, in principle, could place a constraint on the value of $m_{H^{\pm}}$. 

When $U$ is fixed at zero, the current global fit of electroweak precision data (EWPD) gives~\cite{ParticleDataGroup:2020ssz}:
\begin{align}
S & =0.00\pm0.07\thinspace,\label{eq:m-19}\\
T & =0.05\pm0.06\thinspace.\label{eq:m-20}
\end{align}
Comparing Eq.~\eqref{eq:m-20} with Eq.~\eqref{eq:m-T}, we obtain 
\begin{equation}
|m_{H^{\pm}}-m_{H^{0}}|\approx|m_{H^{\pm\pm}}-m_{H^{\pm}}|<45.5\ {\rm GeV}\ \ {\rm at}\ 90\%\text{ C.L.}\label{eq:m-21}
\end{equation}
As for the $S$ parameter, the first term in Eq.~\eqref{eq:m-S} is negligible ($\lesssim0.01$ for $m_{H^{\pm}}\gtrsim100$ GeV) and
the second term gives $m_{H^{\pm}}>169\ {\rm GeV}$ (90 \%C.L.) for $|m_{H^{\pm\pm}}-m_{H^{\pm}}|=45.5\ {\rm GeV}$. 
This is less stringent than the bound from direct searches at the LHC to be discussed below.

\subsection{Constraints from neutrino physics}% and \lfv} 
\label{sec:neutr-mass-constr}

We now examine the observational restrictions on the type-II seesaw model parameters
that follow from neutrino oscillation experiments, as well as the absolute neutrino mass scale probes.  \\[-.4cm]

\underline{\bf Neutrino oscillations}\\[-.4cm]

The discovery of neutrino oscillations made in solar and atmospheric neutrino studies was quickly confirmed by reactor and accelerator-based experiments which provided a better determination
of the oscillation  parameters.
Current experimental data converge towards a consistent global picture called the ``three-neutrino paradigm'' which provides tight restrictions on the allowed neutrino mass and mixing
parameters~\cite{deSalas:2020pgw,10.5281/zenodo.4726908}
(note, however, that the octant of the atmospheric angle, the mass-ordering, and CP determination still require further input from the next generation of oscillation searches).
These constraints apply to any model of neutrino mass generation, such as the type-II seesaw.

If the type-II seesaw mechanism is responsible for the small neutrino masses, then the Yukawa couplings from Eq.~(\ref{eq:s-18}),
must reproduce the corresponding flavor neutrino mass matrix. This matrix is related to the  mass basis matrix through
\begin{equation}
  \label{eq:m-nu}
 m_{\alpha \beta} = [U \textup{diag}\left\{ m_1, m_2, m_3\right\} U^{T}]_{\alpha\beta},  
\end{equation}
with $U$ denoting the neutrino mixing matrix, and $m_{i}$ the eigenvalues of the massive neutrino states. Note that the matrix $U$ depends on three mixing angles
($\theta_{12}, \theta_{13}, \theta_{23}$), and three CP phases. The Dirac phase affects oscillations, while the two Majorana phases are crucial in the description of neutrinoless double beta decay
as discussed below. Note that we express the mixing matrix in a symmetrical way, where each phase is associated to each of the mixing angles~\cite{Schechter:1980gr}.
The best fit values for the mixing angles as well as for the squared mass differences are determined using a global analysis of the neutrino oscillation data~\cite{deSalas:2020pgw}. \\[-.4cm]

Neutrinoless double-beta decay (\znbb for short) is the prime lepton number violating process~\cite{Vergados:2012xy,Dolinski:2019nrj,Jones:2021cga} that is sensitive not only to the
 absolute scale of  neutrino mass but also to the Majorana phases inaccessible to oscillation experiments.
The Majorana-type mass term of the type-II seesaw model implies that the exchange of light neutrinos yields a \znbb decay amplitude given as  
\begin{equation}
  \label{eq:mbb}
\vev{ m_{\beta\beta}} = \left|\sum_{j=1}^3 U_{ej}^2 m_j\right| =\left|c^{2}_{12}c^{2}_{13} m_1 + s^{2}_{12}c^{2}_{13} m_2 e^{2i\phi_{12} }+ s^{2}_{13} m_3 e^{2i\phi_{13}}\right|~.
\end{equation}
Here the neutrino mixing matrix is expressed in terms of the symmetrical parametrization~\cite{Schechter:1980gr}.
The latter provides a most transparent description of \znbb~\cite{Rodejohann:2011vc}, in contrast with the convention adopted by the PDG. 

\begin{figure}
\begin{center}
\includegraphics[width=0.49\textwidth]{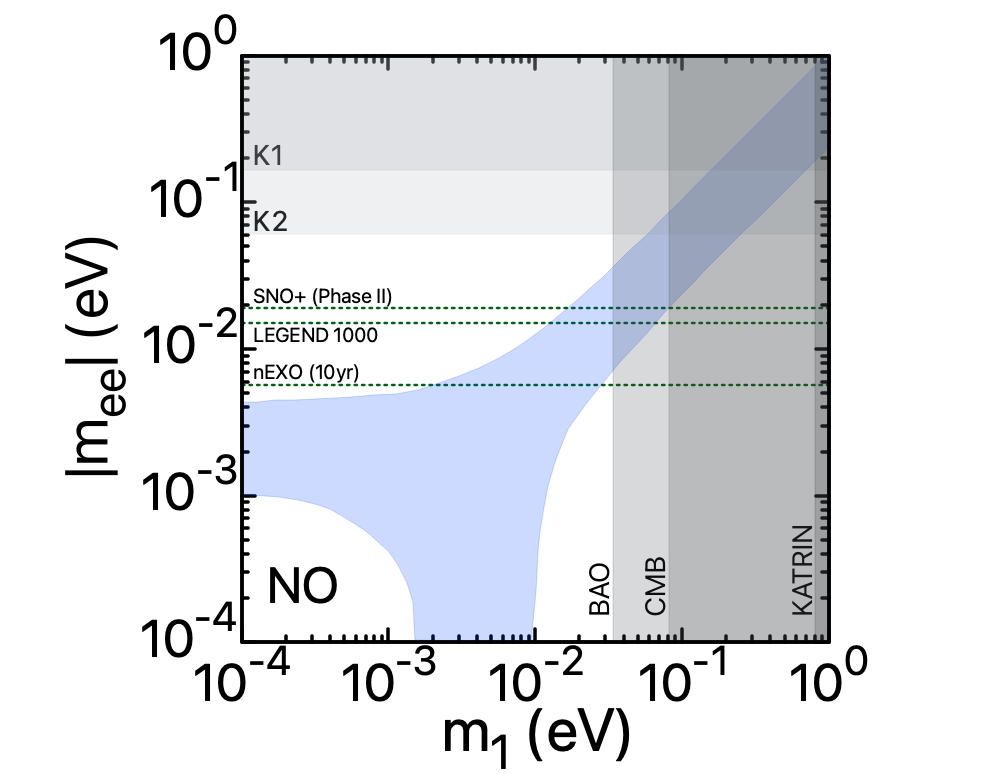}
\includegraphics[width=0.49\textwidth]{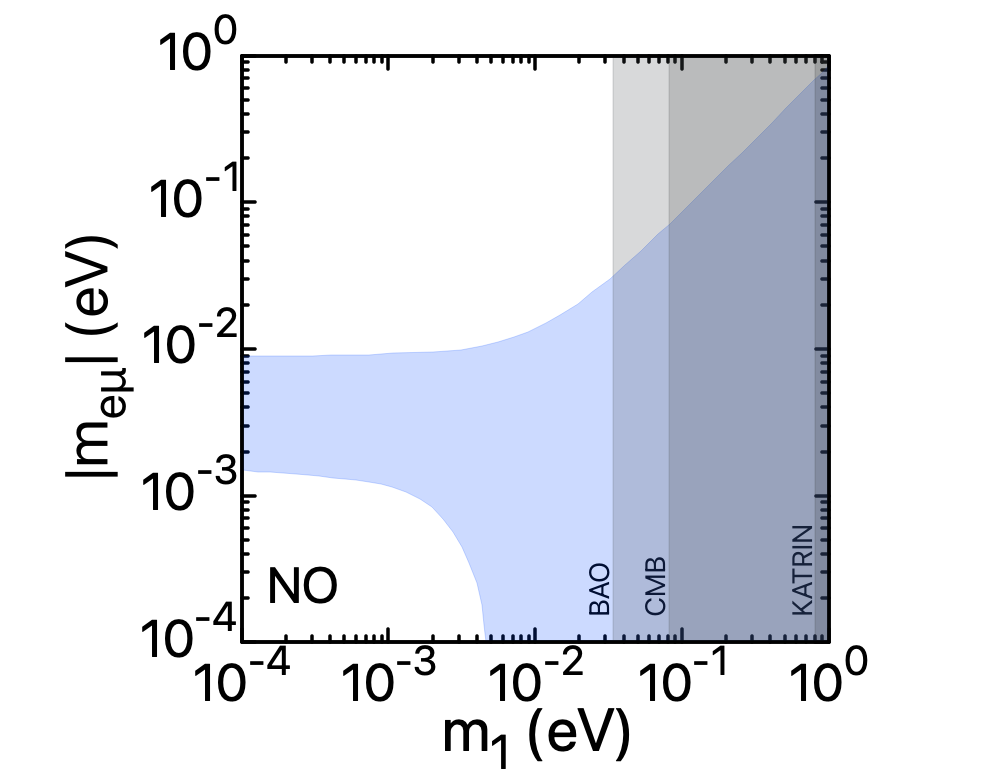}
\includegraphics[width=0.49\textwidth]{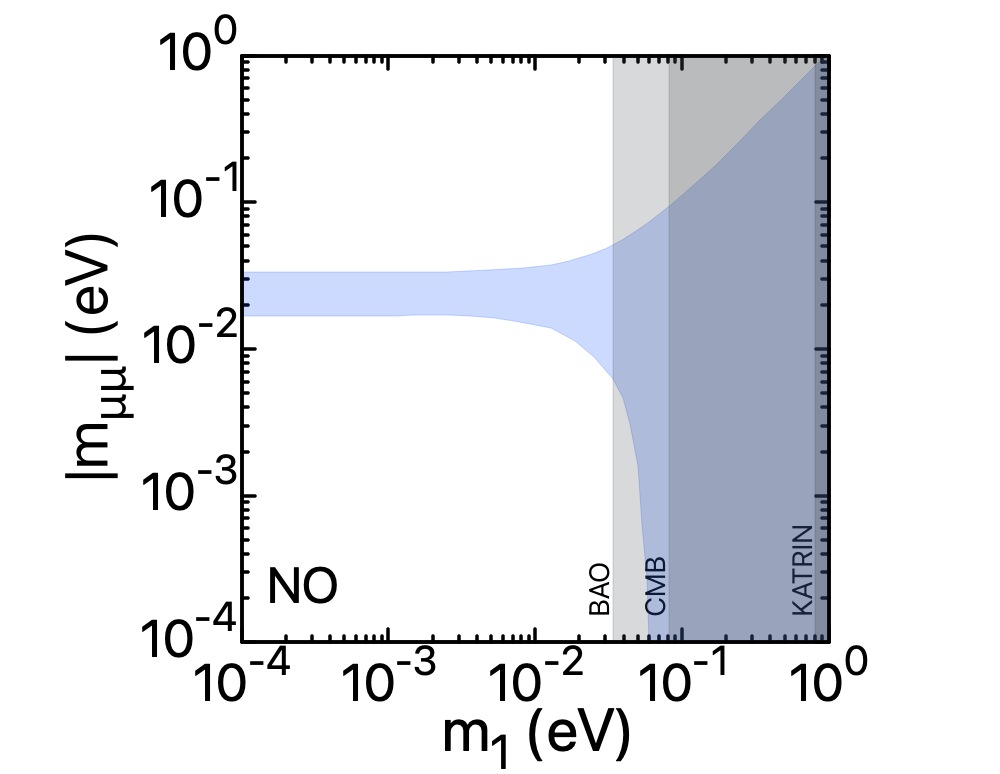}
\includegraphics[width=0.49\textwidth]{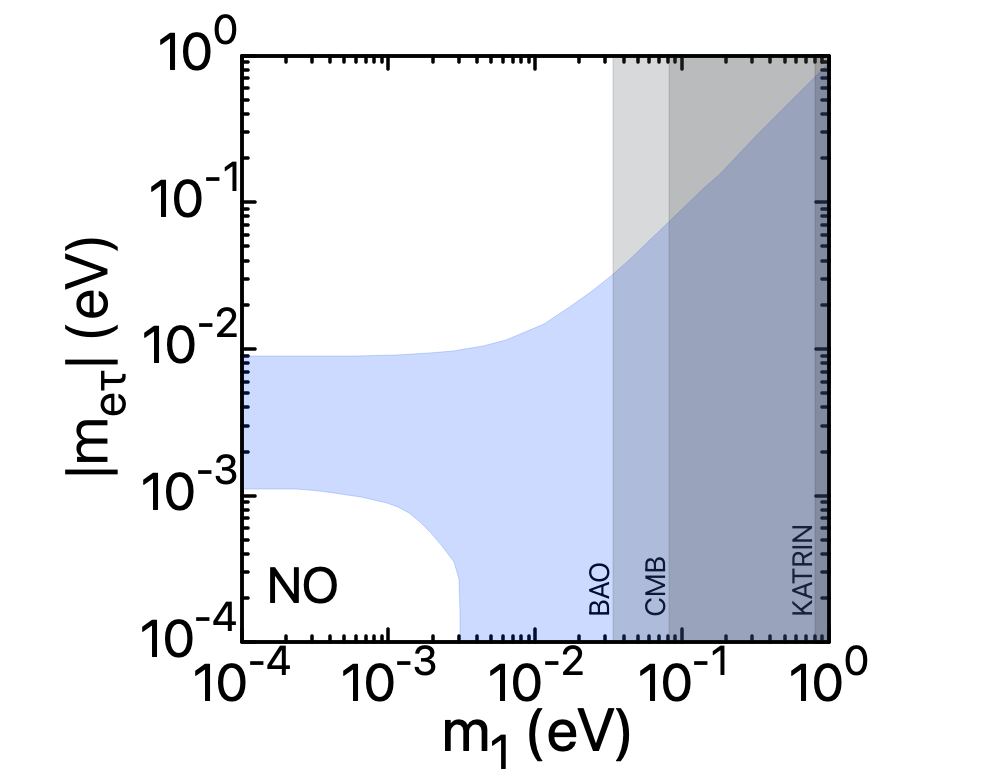}
\includegraphics[width=0.49\textwidth]{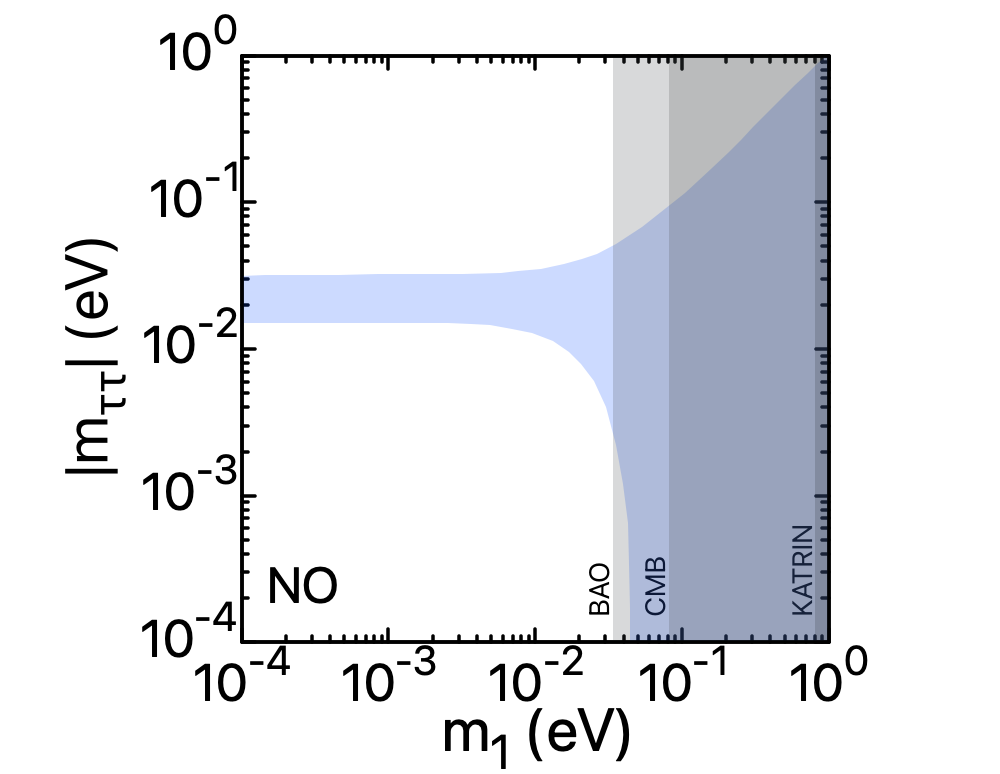}
\includegraphics[width=0.49\textwidth]{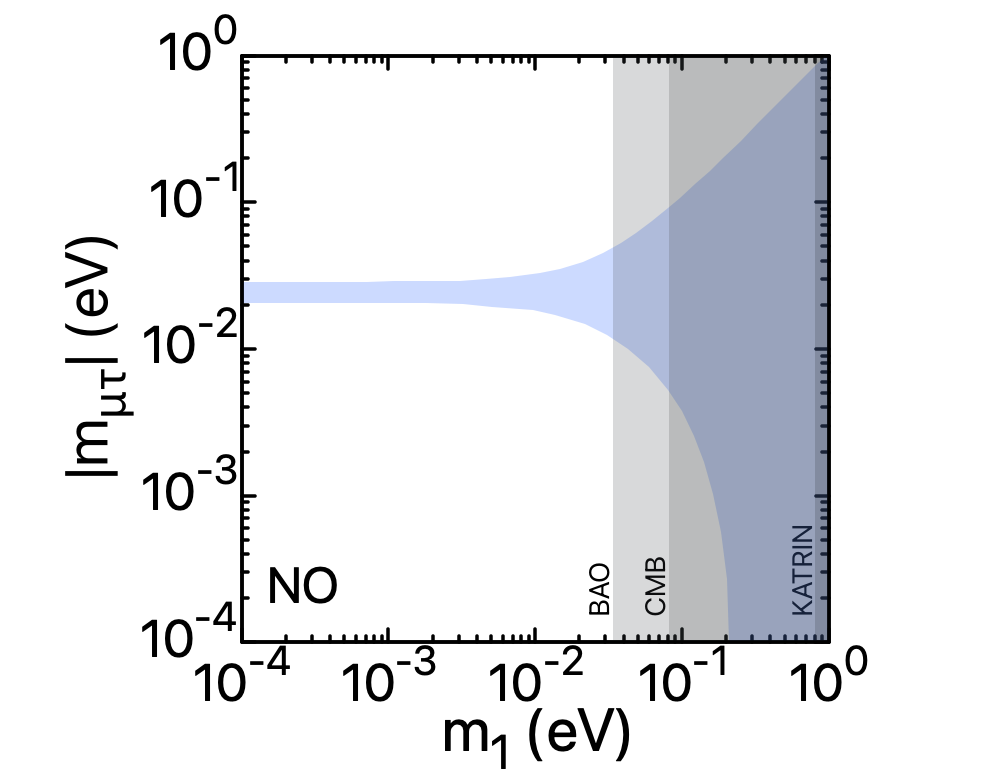}
\end{center}
\caption{ The allowed neutrino mass matrix element regions in the flavor basis, as a function of $m_{1}$, are shown in blue, for normal neutrino mass ordering, {\bf NO}.
  They are arranged as follows: $|m_{ee}|$ (top-left), $|m_{e\mu}|$ (top-right), $|m_{\mu\mu}|$ (center-left),  $|m_{e\tau}|$ (center-right),  $|m_{\tau\tau}|$ (bottom-left),
  and $|m_{\mu\tau}|$ (bottom-right). Mixing angles, squared mass differences, and $\delta$ are taken within their 3$\sigma$ ranges~\cite{deSalas:2020pgw}.
Vertical shaded bands are the KATRIN \cite{Aker:2021gma}, CMB \cite{Aghanim:2018eyx}, and CMB+BAO \cite{eBOSS:2020yzd} limits for $m_{1}$.
The top-left panel coincides with the effective \znbb mass parameter $|m_{\beta\beta}|$ and the shaded horizontal bands marked K1 and K2 are excluded by
KAMLAND-Zen~\cite{KamLAND-Zen:2016pfg}, see text.
}
\label{fig:dbd-bounds}
\end{figure}

Notice that the amplitude $\vev{ m_{\beta\beta}}$ coincides precisely with the $|m_{ee}|$ entry in Eq. (\ref{eq:m-nu}).
Given the current oscillation results one can determine the allowed ranges for the expected \znbb amplitude.
As depicted in the upper-left panel in Fig.~\ref{fig:dbd-bounds} (the other panels will be explained below)
the result depends on whether the ordering of the neutrino mass spectrum is normal {\bf (NO)} or inverted {\bf (IO)}.
Thanks to the effect of Majorana phases in Eq.~(\ref{eq:mbb}) the amplitude can vanish due to possible destructive interference among the three neutrinos.
This is a key feature of the normal neutrino mass ordering, currently preferred by the global oscillation fit~\cite{deSalas:2020pgw},
One sees how oscillations leave an important imprint upon neutrinoless double beta decay studies. 
Limits on $\vev{ m_{\beta\beta}}$ from \znbb searches are also indicated, taking into account the uncertainty in the nuclear matrix elements relevant for the computation of the decay rates.
This leads to the K1 and K2 lines, indicating the current 95\%CL limits from the KamLAND-Zen~400 experiment, for extreme \znbb nuclear matrix element choices~\cite{KamLAND-Zen:2016pfg}.
  Other bounds come from GERDA~\cite{Agostini:2020xta}, CUORE~\cite{Adams:2019jhp} and EXO-200~\cite{Anton:2019wmi}.
There is a reasonable chance that, perhaps, \znbb decay could be seen in the next round of experiments such as SNO+Phase-II~\cite{SNO:2015wyx},
LEGEND-1000~\cite{LEGEND:2017cdu} and nEXO-10 yr~\cite{nEXO:2017nam}.
According to the black-box theorem~\cite{Schechter:1981bd}, this would be a major discovery, as it would imply that at least one of the neutrinos is a Majorana particle. \\[-.4cm] 

Probing the absolute scale of neutrino masses is also the goal of single beta decay experiments such as KATRIN which leads to an upper limit $m_{\beta}<0.8$~eV (90\% CL)~\cite{Aker:2021gma,KATRIN:2019yun}.
 Absolute neutrino masses can also be measured through cosmological observations, like those of the cosmic microwave background anisotropy spectrum or of the clustering of cosmic structures.
 The vertical shaded band in Fig.~\ref{fig:dbd-bounds} indicates the current sensitivity of cosmological data~\cite{Aghanim:2018eyx} on $m_\text{lightest}$. \\[-.4cm]

  We now summarize the above neutrino mass constraints. 
  To do so we perform a scan over the parameter space to calculate the allowed modulus for each of the elements of the mass matrix in the flavor basis as a function of $m_1$.
  For this analysis we consider the range $m_1 < 0.8$~eV, in agreement with KATRIN's upper limit for $m_{\beta}$~\cite{Aker:2021gma}.
  We generate different combinations of random values for the $U$ parameters within their 3$\sigma$ limits, taking into account the most recent neutrino oscillation analysis~\cite{deSalas:2020pgw}.
  We varied the Majorana phases randomly in the range between 0 and 2$\pi$.
  The resulting values for $|m_{ee}|$ obtained after this scan, are depicted in the upper-left panel in Fig.~\ref{fig:dbd-bounds} as a shaded blue region.
    The interval found in this scan, for each of the neutrino mass entries, in units of eV, is given by:
\begin{equation}
\left| Y_{\Delta\alpha\beta}\right| \frac{v_{\Delta}}{\sqrt{2}} = |m_{\alpha\beta}| \in \begin{pmatrix}
(6.77\times10^{-6},0.80) & (6.79\times10^{-6},0.66) &(4.07\times10^{-6},0.68) \\ 
 (6.79\times10^{-6},0.66)& (1.05\times10^{-4},0.80) & (5.53\times10^{-5},0.80)\\ 
(4.07\times10^{-6},0.68) & (5.53\times10^{-5},0.79) & (1.64\times10^{-5},0.80) 
\end{pmatrix}  .
\label{eq:massPattern}
\end{equation}
%    }

Note that, as a result of the limits of the sampling region, all absolute values of the neutrino neutrino mass entries are bounded from above.
Moreover, the $ee$-entry is bounded from below. These are just artifacts of the sampling procedure.

\subsection{Charged lepton flavor violaton}
\label{sec:charg-lept-flav}

Processes with charged lepton flavor violation (cLFV)  are those that do not conserve lepton family number in transitions between $e$, $\mu$ and $\tau$ families. 
Their existence is expected, at some level, from the observation of neutrino oscillations. 
Within the SM, leptonic flavor is conserved at any order of perturbation theory.  
Flavor can be violated in the absence of neutrino mass within specific type-I seesaw models~\cite{Bernabeu:1987gr,Branco:1989bn,Rius:1989gk}. 

\begin{figure}[h] \centering
\includegraphics[width=0.4\textwidth]{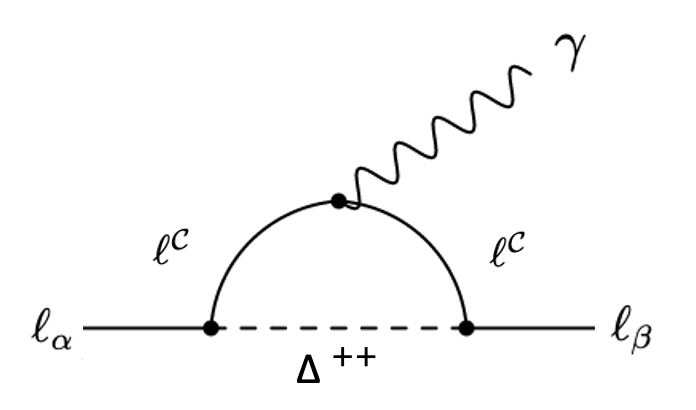}
\includegraphics[width=0.4\textwidth]{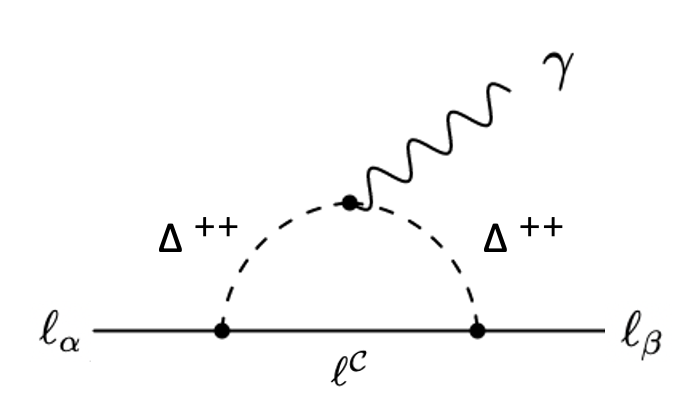}
\caption{\label{fig:Tree-Level1} Feynman diagrams for the doubly-charged contribution to the cLFV process $\ell_{\alpha} \to  \gamma \ell_{\beta}$. }
 \end{figure}

\begin{figure}[h] \centering
\includegraphics[width=0.4\textwidth]{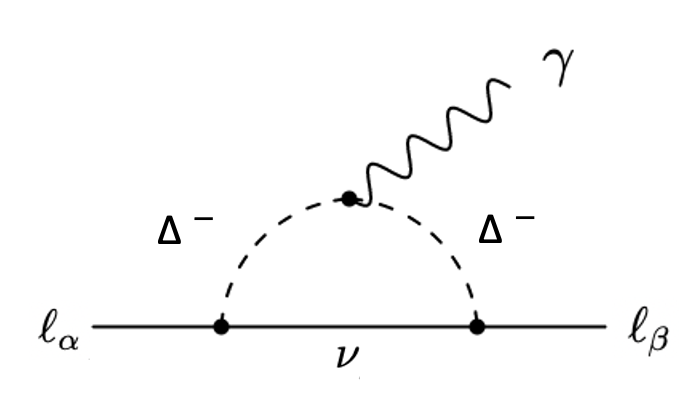}
\includegraphics[width=0.4\textwidth]{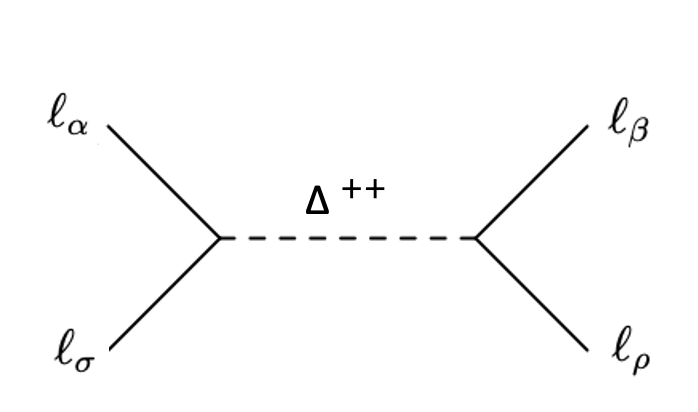}
\caption{\label{fig:Tree-Level2}  Left: Feynman diagram for the singly-charged contribution to the cLFV process $\ell_{\alpha} \to  \gamma \ell_{\beta}$.
  Right: Feynman diagram for the doubly-charged contribution to the cLFV process $\ell_{\alpha} \to  \ell_{\beta}\ell_{\rho}\ell_{\sigma}$. }
 \end{figure}

Within the type-II seesaw mechanism, the same triplet Yukawa coupling responsible for neutrino masses also induce cLFV effects.
In particular, processes $l_\alpha \to  l_\beta \gamma$ are predicted at one-loop level, with contributions from both singly- and doubly-charged scalars,
while the processes $l_\alpha \to  \bar{l_\beta}l_\beta l_\beta$ proceed at the tree level, as seen in Figs.~\ref{fig:Tree-Level1} and \ref{fig:Tree-Level2}. 

 It is straightforward to determine the corresponding expressions for the branching ratios. 
 For example, for  $\mu \to  e\gamma$ and $\mu \to  eee$ the corresponding branching ratios are given as~\cite{Cai:2017jrq,Forero:2011pc}:
\begin{equation}
BR(\mu \to  e\gamma) \approx \frac{\alpha}{192\pi}\frac{\left | (Y_{\Delta}^{\dagger}Y_{\Delta})_{e\mu} \right |^{2}}{G_{F}^{2}}\left ( \frac{1}{m_{H^{+}}^{2}} + \frac{8}{m_{H^{++}}^2} \right )^{2},
\label{eq:BRmutoegamma}
\end{equation}

\begin{equation}
BR(\mu \to  eee) = \frac{1}{G_{F}^{2}}\frac{\left | (Y_{\Delta}^{\dagger})_{ee}(Y_{\Delta})_{\mu e} \right |^{2}}{m_{H^{++}}^4}.
\label{eq:BRmutoeee}
\end{equation}

In a way similar to section~\ref{sec:neutr-mass-constr} we can determine the allowed ranges for the other entries of $(Y_{\Delta})_{\alpha\beta}$.
One can obtain simple expressions that constrain the model parameters such as the Yukawa couplings and charged scalar masses using the negative
searches for \clfvg transitions, such as $\mu \to  e\gamma$ decay, $\mu \to \bar{e}ee$ decay and similar $\tau$ decays which were searched, e.g., in BaBar. 
We summarize the current experimental status of radiative \clfv in the second column of Table~\ref{table:lfv}. 
These are the experimental restrictions on the decay branching ratios~\cite{ParticleDataGroup:2020ssz}.
To obtain the third column we assumed that the difference between $m_{H^{++}}$ and $m_{H^{+}}$ is relatively small, see Sec.\ref{STU-par}.
The branching ratios are inversely proportional to $M^2_{\Delta}$ at first order. 
The resulting constraints from each channel are listed in the third column of Table~\ref{tab:lfv}.   
Notice that from processes that have three leptons in the final state we can constrain the product of two Yukawa couplings,
while the radiative processes allow us to constrain the sum of three of such products.
This will be important for the study of Non-Standard neutrino interaction parameters, as we will see at the end. 
Moreover, one must also include the restrictions from negative searches at collider experiments, such the LHC.
\begin{table*}
\begin{ruledtabular}
\begin{tabular}{llr}
Processes & Experimental limits on BR  & Bounds on $G_{F}^{-1}M_{\Delta}^{-2}Y_{\Delta}^{2}$\phantom{xxxxx}\tabularnewline
\hline 
\vvvv $\mu\to  e\gamma$ & ${\rm BR}<4.2\times10^{-13}$ & $G_{F}^{-1}M_{\Delta}^{-2}|Y_{\Delta}^{\dagger}Y_{\Delta}|_{e\mu}<2.1\times10^{-5} = L_{1}$\tabularnewline
\vvvv $\mu\to 3e$ & ${\rm BR}<1.0\times10^{-12}$ & $G_{F}^{-1}M_{\Delta}^{-2}|Y_{\Delta}^{\dagger}|_{\mu e}|Y_{\Delta}|_{ee}<2.0\times10^{-6}= L_{2}$\tabularnewline
\vvvv $\tau\to 3e$ & ${\rm BR}<2.7\times10^{-8}$ & $G_{F}^{-1}M_{\Delta}^{-2}|Y_{\Delta}^{\dagger}|_{\tau e}|Y_{\Delta}|_{ee}<7.9\times10^{-4}= L_{3}$\tabularnewline
\vvvv $\tau\to  e^{+}e^{-}\mu^{-}$ & ${\rm BR}<1.8\times10^{-8}$ & $G_{F}^{-1}M_{\Delta}^{-2}|Y_{\Delta}^{\dagger}|_{\tau e}|Y_{\Delta}|_{e\mu}<4.5\times10^{-4}= L_{4}$\tabularnewline
\vvvv $\tau\to  e\gamma$ & ${\rm BR}<3.3\times10^{-8}$ & $G_{F}^{-1}M_{\Delta}^{-2}|Y_{\Delta}^{\dagger}Y_{\Delta}|_{e\tau}<1.4\times10^{-2}= L_{5}$\tabularnewline
\vvvv $\tau\to \mu\gamma$ & ${\rm BR}<4.4\times10^{-8}$ & $G_{F}^{-1}M_{\Delta}^{-2}|Y_{\Delta}^{\dagger}Y_{\Delta}|_{\mu\tau}<1.6\times10^{-2}= L_{6}$\tabularnewline
\vvvv $\tau\to \mu^{+}\mu^{-}e^{-}$ & ${\rm BR}<2.7\times10^{-8}$ & $G_{F}^{-1}M_{\Delta}^{-2}|Y_{\Delta}^{\dagger}|_{\tau\mu}|Y_{\Delta}|_{\mu e}<5.6\times10^{-4}= L_{7}$\tabularnewline
\vvvv $\tau\to  e^{+}\mu^{-}\mu^{-}$ & ${\rm BR}<1.7\times10^{-8}$ & $G_{F}^{-1}M_{\Delta}^{-2}|Y_{\Delta}^{\dagger}|_{\tau e}|Y_{\Delta}|_{\mu\mu}<6.3\times10^{-4}= L_{8}$\tabularnewline
\vvvv $\tau\to \mu^{+}e^{-}e^{-}$ & ${\rm BR}<1.5\times10^{-8}$ & $G_{F}^{-1}M_{\Delta}^{-2}|Y_{\Delta}^{\dagger}|_{\tau\mu}|Y_{\Delta}|_{ee}<5.9\times10^{-4}= L_{9}$\tabularnewline
\vvvv $\tau\to 3\mu$ & ${\rm BR}<2.1\times10^{-8}$ & $G_{F}^{-1}M_{\Delta}^{-2}|Y_{\Delta}^{\dagger}|_{\tau\mu}|Y_{\Delta}|_{\mu\mu}<6.9\times10^{-4}= L_{10}$\tabularnewline
\end{tabular}
\end{ruledtabular}
\caption{\label{tab:lfv} Constraints from cLFV processes on the Yukawa coupling matrix $Y_{\Delta}$. Experimental limits on the branching ratios (BR) are
taken from Ref.~\cite{ParticleDataGroup:2020ssz}, and the bounds on $G_{F}^{-1}M_{\Delta}^{-2}Y_{\Delta}^{2}$ are adapted from Ref.~\cite{Dev:2017ouk}.}
\label{table:lfv}
\end{table*}

\subsection{Collider constraints} 
\label{sec:collider-constraints}

The Higgs triplet interacts with gauge bosons, the lepton doublets, as well as the scalar Higgs doublet.
Its decay modes have been explored at the LEP and LHC experiments to some extent.
  Unfortunately, most of the experimental searches for doubly-charged Higgs bosons make specific model assumptions concerning their decay modes,
  whose features do not fully cover the type-II seesaw expectations.
  On the other hand, searches for the singly-charged and neutral scalars do not apply to the type-II seesaw, since the relevant couplings are neutrino-mass-suppressed
  ($v_\Delta$-suppression, see Eq.~(\ref{eq:s-18})) in the type-II seesaw. 
  This applies, for example, to their couplings to quarks, which arise only due to mixing, as the only triplet scalar Yukawa coupling involves leptons, Eq.~(\ref{app:x-3}).
  These inadequacies undermine the robustness of the constraints obtained.
  In this subsection we present a dedicated analysis of the decay modes of the type-II seesaw scalars which
  should serve as basis for deriving truly robust constraints on their masses.\\[-.4cm]

\underline{\bf The doubly charged scalar}\\[-.4cm]  

We start with the case of small mass splitting $\delta m\approx 0$.
  Depending on the magnitude of the triplet VEV $v_{\Delta}$, the doubly-charged Higgs $H^{\pm\pm}$ may mainly decay to same-sign dileptons or gauge bosons. 
Clearly, for small $v_{\Delta}$, the $H^{\pm\pm}$ predominantly decays into same-sign leptonic states $H^{\pm\pm}\to\ell^\pm\ell^\pm$, whereas for larger $v_\Delta$,
the gauge boson decay mode $H^{\pm\pm}\to W^\pm W^\pm$ becomes dominant~\cite{Chun:2003ej,Melfo:2011nx,Chun:2019hce}.  The relevant decay widths are given as:
\begin{equation}
\Gamma (H^{\pm \pm} \to l^{\pm}_i l^{\pm} _j)=\Gamma_{l_i l _j}=\frac{m_H^{\pm \pm} } {(1+\delta_{ij}) 8 \pi}   \left |\frac{m_{ij}^{\nu}}{v_{\Delta}} \right |^2, \, \, m^{\nu}=Y_{\Delta} v_{\Delta}/\sqrt{2},
\label{eq:Hpptoll}
\end{equation}
\begin{equation}
\Gamma (H^{\pm \pm} \to W^{\pm} W^{\pm})=\Gamma_{W^{\pm }W^{\pm }}=\frac{g^2 v^2_{\Delta}}{8 \pi m_{H^{\pm \pm}}} \sqrt{1- \frac{4}{r^2_W}} \left[ \left (2+(r_W/2-1)^2 \right ) \right ].
\end{equation}
where $r_W=\frac{m_{H^{\pm \pm}}}{M_W}$. Here $m^{\nu}$ denotes the neutrino mass matrix, $i,j$ are the generation indices, $\Gamma_{l_i l_j}$ and $\Gamma_{W^{\pm} W^{\pm}}$
are the partial decay widths for the $H^{\pm \pm} \to l^{\pm}_i l^{\pm}_j$, and $H^{\pm \pm} \to W^{\pm} W^{\pm}$  channels, respectively.

For the case of positive mass splitting $(\delta m >0)$, one must consider additonal decay channels
\begin{align}
\Gamma(H^{\pm\pm} \to H^\pm W^{\pm *})=
\frac{9g^4m_{H^{\pm \pm}}\cos^2\beta_\pm}{128\pi^3} G\left(\frac{m_{H^\pm}^2}{m_{H^{\pm \pm}}^2},\frac{m_W^2}{m_{H^{\pm \pm}}^2}\right),
\end{align} 
where $\tan\beta_{\pm}=\frac{\sqrt{2}v_\Delta}{v_\Phi}$ and $G(x,y)$ is defined in the Appendix.~\ref{app:decay-width}. 
\begin{figure}[h]
\centering
\includegraphics[width=0.45\textwidth]{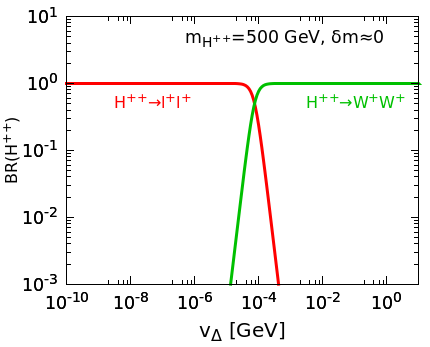}
\includegraphics[width=0.45\textwidth]{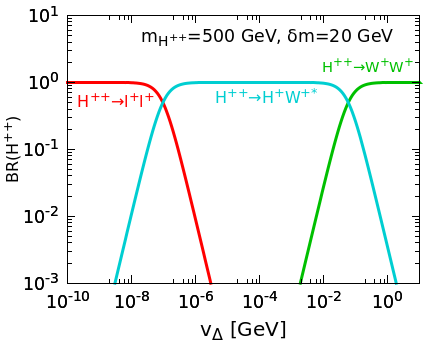}
\caption{
  $H^{\pm\pm}$ branching ratios for $m_{H^{\pm\pm}}=500$~GeV. The left and right panels are for $\delta m\approx 0$ and $\delta m=20$ GeV, respectively.
  Red, green and cyan lines correspond to $H^{\pm\pm}\to l^{\pm}l^{\pm}$, $H^{\pm\pm}\to W^{\pm}W^{\pm}$ and $H^{\pm\pm}\to H^{\pm}W^{\pm *}$, respectively.} 
\label{fig:BRHpp}
\end{figure}

In Fig.~\ref{fig:BRHpp} we display the $H^{\pm\pm}$ branching ratio to the inclusive dilepton channel $l^{\pm}l^{\pm}$,
diboson channel $W^{\pm}W^{\pm}$, and the cascade channel $H^{+}W^{+*}$ for $m_{H^{\pm\pm}}=500$~GeV, with $\delta m\approx 0$~(left panel) or $\delta m=20$~GeV~(right panel). 
One clearly sees that the collider search strategy for the charged Higgs boson crucially depends on the value of the triplet VEV $v_\Delta$ and mass splitting $\delta m$.
The same qualitative behaviour of the $H^{\pm\pm}$ branching ratios holds for different $H^{\pm\pm}$ masses.
%%%
 The direct limit on the $H^{\pm\pm}$ has been derived from collider searches of multi-lepton final states. 
 Below we discuss the existing constraints on $m_{H^{\pm\pm}}$ from LEP and LHC searches for the case $\delta m\approx 0$. 
\begin{itemize}
\item Constraint from LEP-II: LEP has searched for $H^{++}H^{--}$ pair production through s-channel $\gamma/Z$ exchange,
  with subsequent decay of $H^{\pm\pm}$ into charged lepton pairs, which is the dominant decay mode for $v_\Delta\leq 0.1$~MeV.
  This constrains the mass $m_{H^{\pm\pm}}>97.3$~GeV~\cite{Abdallah:2002qj} at $95\%$ C.L. 
\item Constraints from doubly-charged Higgs boson production in three and four lepton final states at 13~TeV LHC:
  These searches analysed Drell-Yan production $pp\to H^{++}H^{--}$ and subsequent decay in the $H^{\pm\pm}\to\ell^{\pm}\ell^{\pm}$ channel.
  The CMS collaboration looked for various leptonic final states such as $ee,\mu\mu,\tau\tau,e\mu,e\tau$ and $\mu\tau$. 
They also studied the associated production of $H^{\pm\pm}H^{\mp}$ through s-channel $W^{\pm}$ exchange, followed by $H^{\pm\pm}$ decay to same-sign di-lepton and $H^{\pm}\to\ell^{\pm}\nu$.
  This combined channel of Drell-Yan production and associated production gives the constarint $m_{H^{\pm\pm}}>820$~GeV~\cite{CMS-PAS-HIG-16-036} at $95\%$ C.L for $e,\mu$ flavor.
  ATLAS searches only include the Drell-Yan production and the bound is $m_{H^{\pm\pm}}>870$~GeV at $95\%$ C.L~\cite{Aaboud:2017qph}. 
  Note that these limits hold only for small triplet VEV, $v_\Delta<0.1$~MeV. For larger triplet $v_\Delta>0.1$ MeV, the gauge boson decay mode is dominant. 
  Hence for this region, a search via pair-production $H^{++}H^{--}$, with subsequent decay into same-sign gauge bosons and further to leptonic final states is required.    
  The ATLAS collaboration has looked for this channel and constrained the doubly-charged Higgs mass $m_{H^{\pm\pm}}$ to lie above 220 GeV~\cite{Aaboud:2018qcu}. 
  Hence, we can conclude that low masses, $m_{H^{\pm\pm}}>220$~GeV, are still allowed, as long as the triplet VEV is ``large'', $v_\Delta>10^{-4}$~GeV.
  For lower triplet VEVs  $v_\Delta < 10^{-4}$~GeV, the doubly-charged Higgs mass constraint is much more stringent, $m_{H^{\pm\pm}} > 870$ GeV.
\end{itemize}
The above-mentioned constraints do not apply to the full parameter space, but rather to a specific range.
  For example, the search in Refs.~\cite{CMS-PAS-HIG-16-036,Aaboud:2017qph} only holds for $\delta m=0$ and $v_\Delta < 10^{-4}$ GeV,
  while the the search in Ref.~\cite{Aaboud:2018qcu} is only valid for $\delta m=0$ and $v_\Delta > 10^{-4}$ GeV.
  
  Notice that in a realistic type-II seesaw scenario the branching fractions of the triplet-like scalars into different lepton flavours are determined by the neutrino
  oscillation parameters.
  However, most of the aforementioned limits are derived in the context of simplified scenarios that do not take into account the footprints of the low-energy neutrino parameters.
  
  Moreover, the triplet scalars in this model are in general non-degenerate~($|\delta m|\neq 0$).  
  For moderate values of triplet VEV $v_\Delta$ and relatively large mass splitting $|\delta m|\sim 10$ GeV, cascade decays quickly dominate over the leptonic and diboson decay modes,
  see the right panel of Fig.~\ref{fig:BRHpp}.  
  Thus, for nonzero mass splitting, the cascade decays are entitled to play a significant role in the phenomenology, resulting in collider signatures
  that may differ significantly from those of the degenerate scenario.
  In order to probe the full parameter space, one should take into account all the complexity of tripet scalar decays, see Sec.~\ref{sec:collider-signature} below.\\[-.4cm]

\underline{\bf The singly charged scalar}\\[-.4cm]

Let us now turn our attention to the singly-charged scalar, $H^{\pm}$, starting with all the relevant partial decay widths,
for the case $\delta m=0$. They are given as~\cite{Ashanujjaman:2021txz,FileviezPerez:2008jbu,Aoki:2011pz,Rizzo:1980gz,Keung:1984hn,Djouadi:1997rp}:  
\begin{align}
&\Gamma(H^\pm \to q\bar{q}')=\frac{3m_{H^\pm}^3\sin^2\beta_\pm}{8\pi v_\Phi^2} 
\left[\left(\frac{m_q^2}{m_{H^\pm}^2}+\frac{m_{q'}^2}{m_{H^\pm}^2}\right)\left(1-\frac{m_q^2}{m_{H^\pm}^2}-\frac{m_{q'}^2}{m_{H^\pm}^2}\right)-4\frac{m_q^2}{m_{H^\pm}^2}\frac{m_{q'}^2}{m_{H^\pm}^2}\right] \lambda^{\frac{1}{2}}\left(\frac{m_q^2}{m_{H^\pm}^2},\frac{m_{q'}^2}{m_{H^\pm}^2}\right),
\\
&\Gamma(H^\pm \to \ell_i^\pm\nu_j)=\frac{m_{H^\pm}}{8\pi v_\Phi^2} \left( \delta_{ij}m_i^2\sin^2\beta_\pm +|Y^{ij}_\Delta|^2v_\Phi^2\cos^2\beta_\pm \right) \left(1-\frac{m_i^2}{m_{H^\pm}^2}\right)^2,
\\
&\Gamma(H^\pm \to W^\pm Z) =\frac{g^4v_\Delta^2\cos^2\beta_\pm}{32\pi \cos^2\theta_w m_{H^\pm}}\left[\lambda\left(\frac{m_W^2}{m_{H^\pm}^2},\frac{m_Z^2}{m_{H^\pm}^2}\right)\right]^{1/2}\left[2+\frac{m_{H^\pm}^4}{4m_W^2m_Z^2}\left(1-\frac{m_W^2}{m_{H^\pm}^2}-\frac{m_Z^2}{m_{H^\pm}^2}\right)^2\right],
\end{align}
where the relevant functions we have listed in Appendix.~\ref{app:decay-width}. 

For the case of non-zero mass splitting $|\delta m|\neq 0$, one also has the following decay channels for $H^{\pm}$: 
\begin{align}
&\Gamma(H^\pm \to H^0/A W^{\pm *}) =\frac{9g^4m_{H^\pm}}{512\pi^3}\xi_{H^\pm W^\mp H^0/A}^2G\left(\frac{m_{H^0/A}^2}{m_{H^\pm}^2},\frac{m_W^2}{m_{H^\pm}^2}\right);\text{  for  }\delta m>0,
\\
&\Gamma(H^\pm \to H^{\pm \pm} W^{\mp *})=
\frac{9g^4m_{H^\pm}\cos^2\beta_\pm}{128\pi^3} G\left(\frac{m_{H^{\pm \pm}}^2}{m_{H^\pm}^2},\frac{m_W^2}{m_{H^\pm}^2}\right); \text{  for  }\delta m < 0.
\end{align} 
\begin{figure}[t]
\begin{center}
\includegraphics[width=0.49\textwidth]{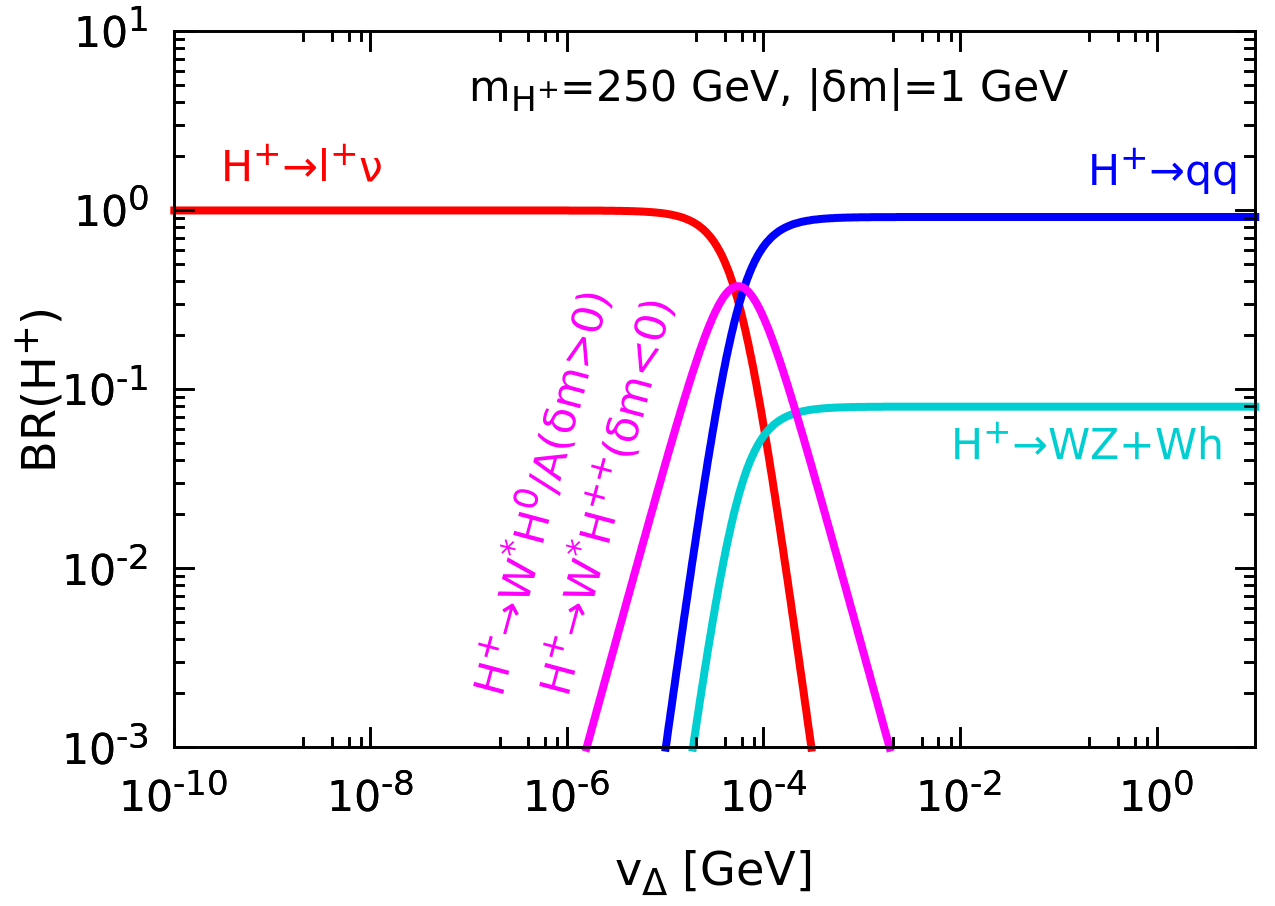}
\includegraphics[width=0.49\textwidth]{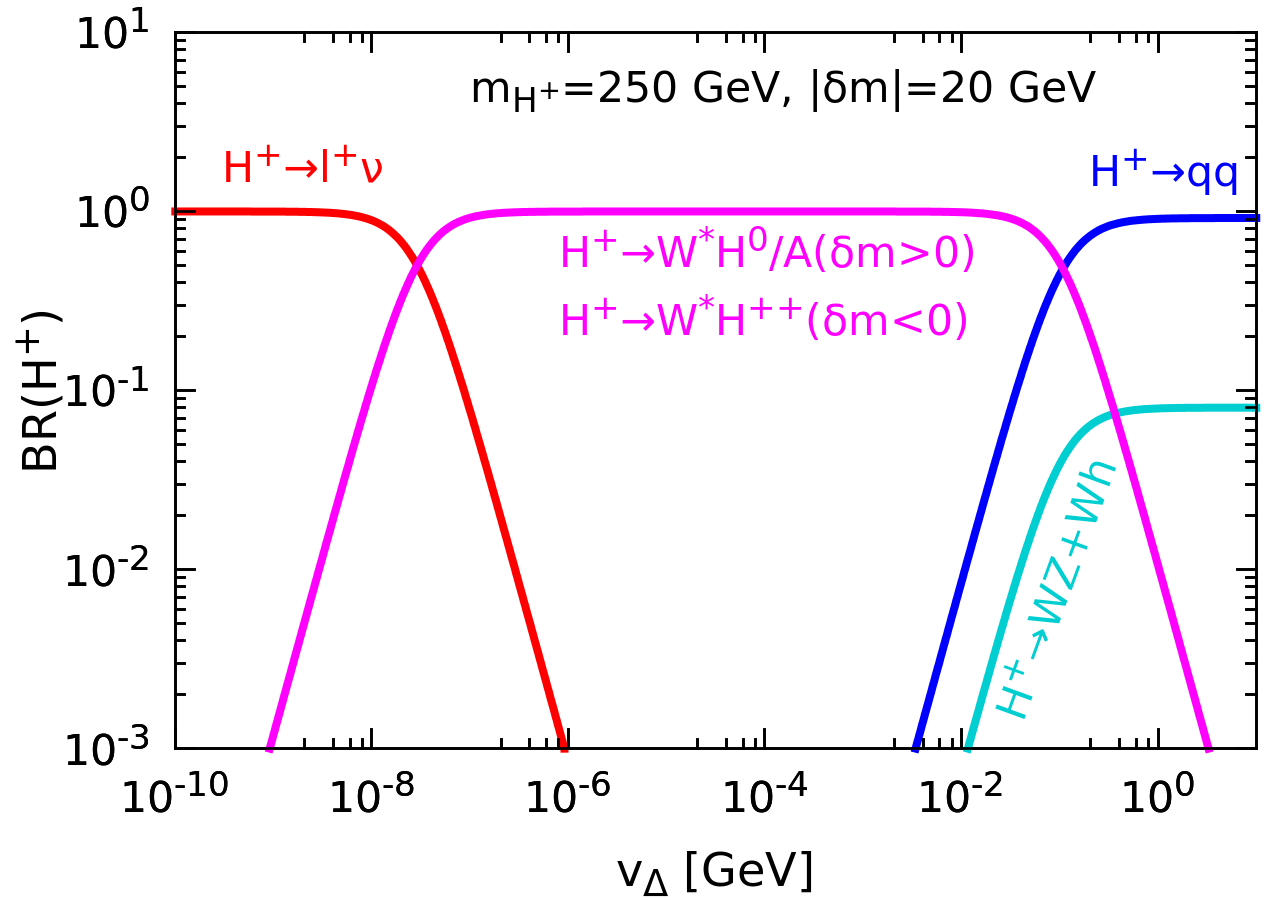}
\includegraphics[width=0.49\textwidth]{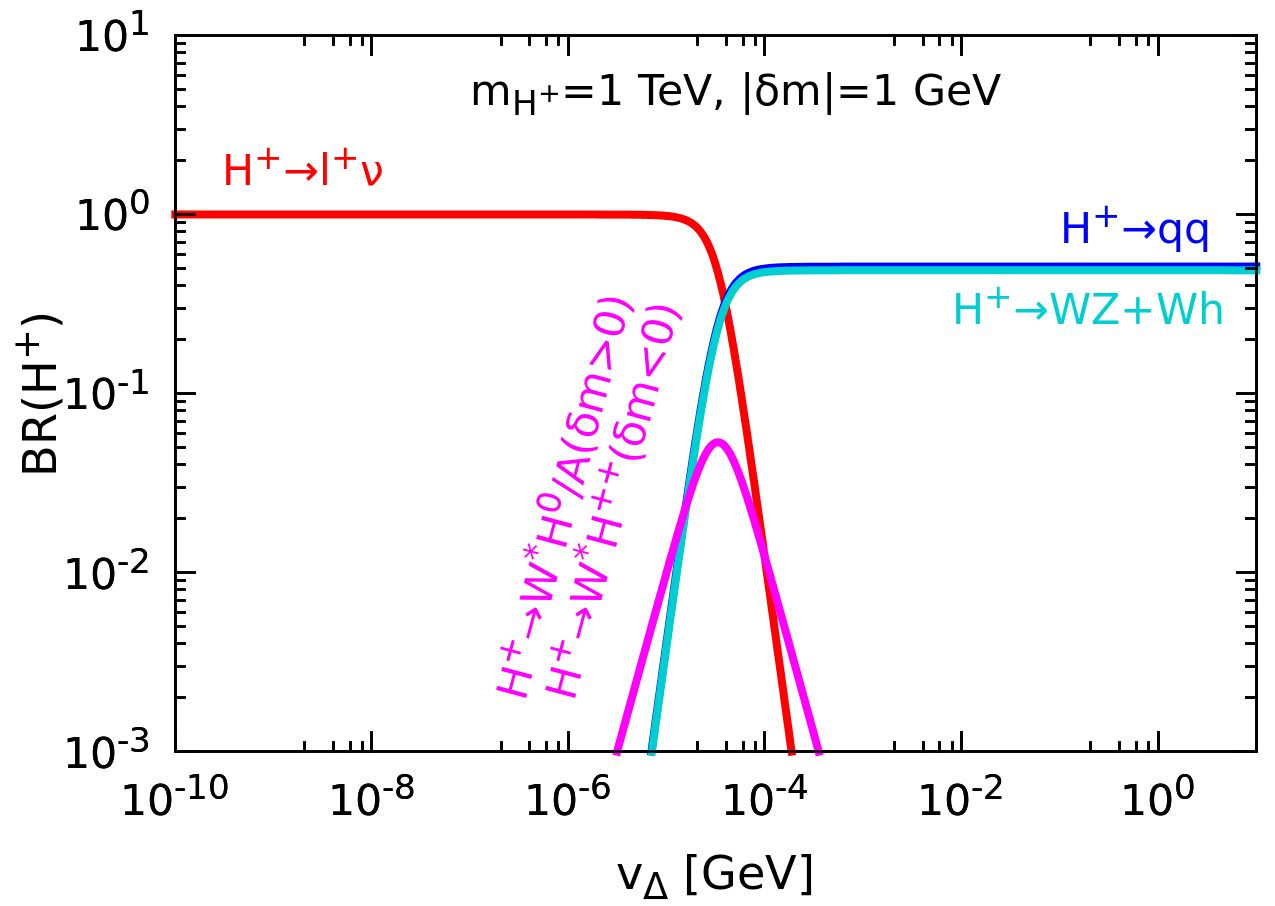}
\includegraphics[width=0.49\textwidth]{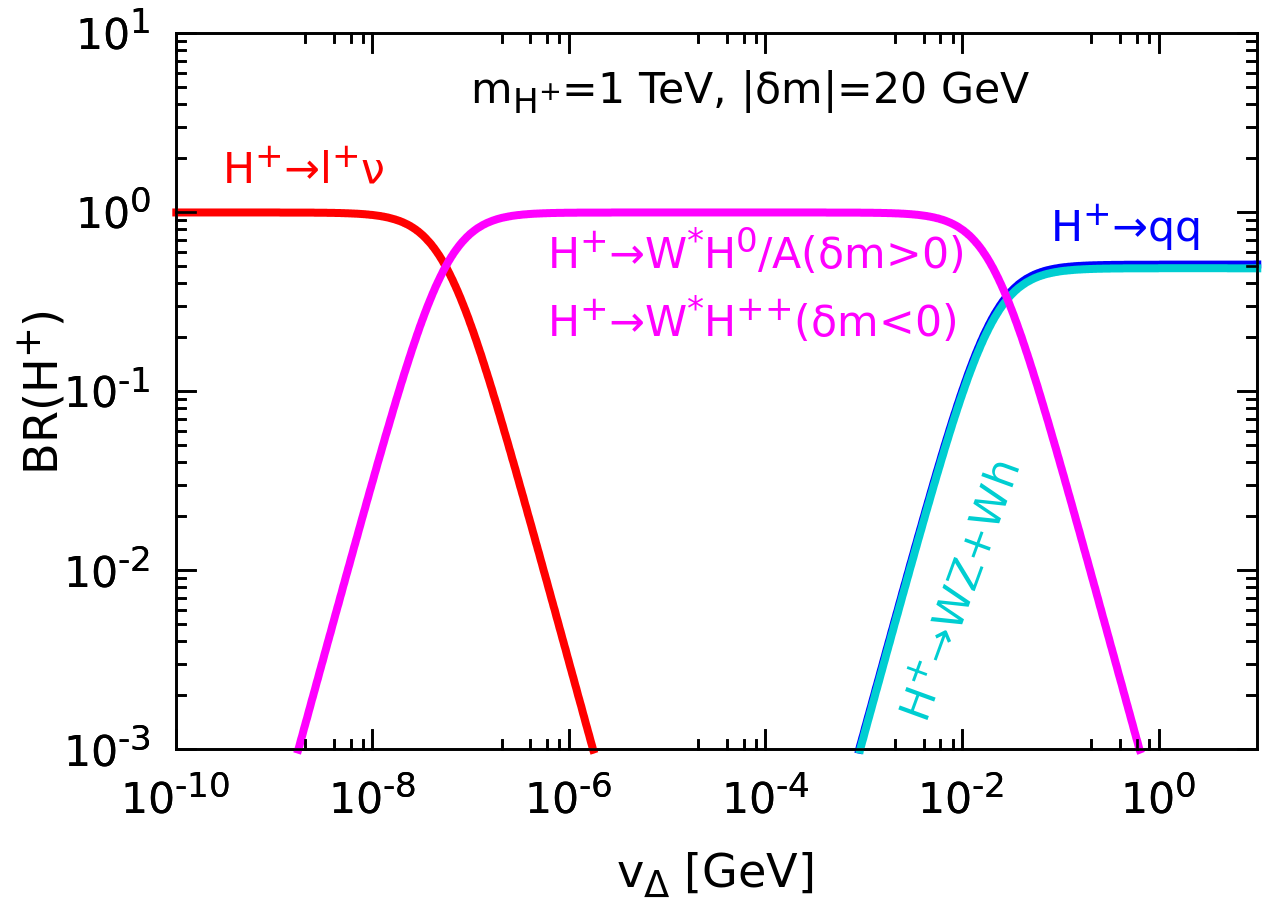}
\end{center}
\caption{Branching ratio of $H^\pm$ for $m_{H^\pm}=250$~GeV~(top panel) and $m_{H^\pm}=1000$~GeV~(bottom panel) as a function of the triplet VEV $v_{\Delta}$.
  The left and right panels are for $|\delta m|=1$~GeV and $|\delta m|=20$~GeV, respectively.}
    \label{fig:BRHp}
\end{figure}
In Fig.~\ref{fig:BRHp}, we have shown the branching ratio of the singly-charged Higgs $H^\pm$ into various channels. 
The singly-charged Higgs boson $H^\pm$ has four decay modes: (i) leptonic decay, i.e. $\ell^\pm\nu$, (ii) hadronic decay, i.e. $q\bar{q'}$, (iii) diboson decay, i.e.
$W^\pm Z$, $W^\pm h$ and (iv) cascade decay, i.e. $H^0/A W^{\pm*}$~($\delta m>0$) or $H^{\pm\pm}W^{\mp*}$~($\delta m<0$). 
Comparing left and right panels in Fig.~\ref{fig:BRHp}, we see that for relatively large mass splitting the cascade decays dominate in the intermediate triplet VEV region. 
One sees that, as for the doubly-charged scalar, the branching ratio pattern for singly-charged scalar decays is quite sensitive to mass splitting and the triplet VEV. 
In order to probe the singly-charged scalar, one should take into account all the complexity of its decays which we sketch in Sec.~\ref{sec:collider-signature}.\\[-.4cm] 

For singly-charged Higgs bosons $H^{\pm}$ in the type-II seesaw model, the coupling to a pair of quarks is suppressed by $\frac{v_\Delta}{v_\Phi}$. 
Both the CMS and the ATLAS official searches~\cite{CMS:2015lsf,ATLAS:2018gfm,ATLAS:2018ntn} employ the $gg\to tbH^{\pm}$ and $gb\to tH^{\pm}$ production channels for the singly-charged Higgs. 
%
%As a result the limits derived from these LHC searches do not apply in the case of type-II seesaw model.
%
The $H^{\pm}$ has also been looked for in the vector boson fusion production channel~\cite{ATLAS:2018iui,CMS:2017fgp}.
However, these production channels depend on the $H^{\pm}-W^{\pm}-Z$ coupling, which is again suppressed by $v_\Delta$, hence not directly applicable to the type-II seesaw model.
In summary, all the limits derived from the LHC searches discussed above do not apply in the case of type-II seesaw model.
Only the combined trully model-independent LEP limit of around 80 GeV on $m_{H^{\pm}}$ applies~\cite{LEPHiggsWorkingGroupforHiggsbosonsearches:2001ogs}.

%%%%%%%%%%%%%%%%%%%%%%%%%%%%%%%%%%%%%%%%%%

\underline{\bf The neutral scalars}\\[-.4cm]
%%%%%%%%%%%%%%%%%%%%%%%%%%%%%%%%%%%%%%%%%%%%%%%%%%%%%%%%%%%%%%

We now discuss the existing constraints on neutral scalars $H^{0}/A$. In order to do that let us first discuss all the relevant $H^{0}/A$ partial decay widths. 
Assuming $\delta m=0$ we have that the $H^0$ decay widths are given as~\cite{Ashanujjaman:2021txz,FileviezPerez:2008jbu,Aoki:2011pz,Rizzo:1980gz,Keung:1984hn,Djouadi:1997rp}: 
\begin{align}
&\Gamma(H^0\to f\bar{f})=\frac{N_c^fm_f^2m_{H^0}}{8\pi v_\Phi^2}\sin^2\alpha\left[\beta\left(\frac{m_f^2}{m_{H^0}^2}\right)\right]^3,
\\
&\Gamma(H^0 \to \nu \nu)=\frac{\kappa m_{H^0} \cos^2\alpha}{8\pi}\sum_{i,j=1}^3|Y^{ij}_\Delta|^2,
\end{align}
\begin{align}
&\Gamma(H^0\to W^+W^-)=\frac{g^4 m_{H^0}^3}{256\pi m_W^4}\left(v_\Phi\sin\alpha -2v_\Delta \cos\alpha\right)^2\left(1-4\frac{m_W^2}{m_{H^0}^2}+12\frac{m_W^4}{m_{H^0}^4}\right)\beta\left(\frac{m_W^2}{m_{H^0}^2}\right),
\\
&\Gamma(H^0\to ZZ)=\frac{g^4 m_{H^0}^3}{512\pi m_W^4}\left(v_\Phi\sin\alpha -4v_\Delta \cos\alpha\right)^2\left(1-4\frac{m_Z^2}{m_{H^0}^2}+12\frac{m_Z^4}{m_{H^0}^4}\right)\beta\left(\frac{m_Z^2}{m_{H^0}^2}\right)\\
&\Gamma(H^0\to h h)=\frac{\lambda_{H^0h h}^2 v_\Phi^2}{8\pi m_{H^0}}\beta\left(\frac{m_{h}^2}{m_{H^0}^2}\right)
\end{align}
where $\tan(2\alpha)=\frac{2B}{A-C}$, $N_c^f$ is the color factor with $N_c^q=3$ and $N_c^\ell=1$, $\kappa=2(1)$ for $i\neq j(i=j)$. 

For nonzero mass splitting, in this case one must have $\delta m < 0$, and the relevant CP--even decay width is
%%$(2).\,\, \lambda_4<0:\,\,\delta m>0\,( m_{H^{\pm\pm}} > m_{H^\pm} > m_{H^0/A})$ and
%%
\begin{align}
\Gamma(H^0 \to H^\pm W^{\mp *}) =\frac{9g^4m_{H^0}}{512\pi^3}\xi_{H^\pm W^\mp H^0}^2G\left(\frac{m_{H^\pm}^2}{m_{H^0}^2},\frac{m_W^2}{m_{H^0}^2}\right).
\end{align}  

On the other hand the relevant decay widths of the CP--odd neutral scalar $A$ with $\delta m=0$ are given as: 
\begin{align}
&\Gamma(A\to f\bar{f})=\frac{N_c^fm_f^2m_{A}}{8\pi v_\Phi^2}\sin^2\beta_0~ \beta\left(\frac{m_f^2}{m_{A}^2}\right),
\\
&\Gamma(A \to \nu \nu)=\frac{m_{A} \cos^2\beta_0}{8\pi}\sum_{i,j=1}^3|Y^{ij}_\Delta|^2,
\\
&\Gamma(A \to h Z)=\frac{g^2m_{A}^3}{64\pi m_W^2} (\cos\alpha \sin\beta_0 -2\sin\alpha \cos\beta_0)^2 \left[\lambda\left(\frac{m_{h}^2}{m_{A}^2},\frac{m_Z^2}{m_{A}^2}\right)\right]^{3/2},
\end{align} 
where $\tan\beta_0=\frac{2v_\Delta}{v_\Phi}$. 

Similarly for the case of nonzero mass splitting one should also consider the following decay width
\begin{align}
\Gamma(A \to H^\pm W^{\mp *}) =\frac{9g^4m_{A}}{512\pi^3}\xi_{H^\pm W^\mp A}^2G\left(\frac{m_{H^\pm}^2}{m_{A}^2},\frac{m_W^2}{m_{A}^2}\right).
\label{eq:AtoHW}
\end{align}  
Again we refer to the Appendix.~\ref{app:decay-width} for all the relevant functions. 
The qualitative features of the branching ratio plots for neutral scalar bosons $H^0/A$ are very similar to those of the singly-charged Higgs case, and will not be presented. 
Similarly to the singly-charged Higgs boson $H^\pm$, the neutral scalar bosons $H^0/A$ have four types of decay modes:
(i) leptonic decays, i.e. $H^0/A\to \ell^\pm\ell^\mp + \nu\nu$,
(ii) hadronic decays, i.e. $H^0/A\to q\bar{q}$,
(iii) diboson decays, i.e. $H^0\to W^\pm W^\mp + Z Z + hh$ or $A\to hZ$ and
(iv) cascade decays, i.e. $H^0/A \to H^{\pm} W^{\mp *}$~($\delta m < 0$).
In the nearly degenerate scenario the cascade decay is very small and depending on the triplet VEV,
either the leptonic decays~($v_\Delta < 10^{-4}$ GeV) or bosonic decays~($v_\Delta > 10^{-4}$ GeV) dominate. 
On the other hand for relatively large mass splitting, the bosonic decays dominate in the intermediate triplet VEV region.
Therefore, in order to probe neutral type-II seesaw scalars in a robust manner one must take into account these decay channels. 

The official ATLAS and CMS searches for neutral scalars $H^0/A$ focus on gluon fusion production \cite{ATLAS:2017eiz,CMS:2018rmh,ATLAS:2018sbw},
associated with b-jet productions \cite{ATLAS:2017eiz,CMS:2018rmh, CMS:2018hir}, and associated production with a vector boson \cite{ATLAS:2018sbw}.
Most of these searches are relevant for two Higgs doublet models or supersymmetric extensions. 
However, these production channels are again suppressed by $v_\Delta$ in the type-II seesaw, since they involve the $H^0/A-q-q$ coupling, rendering them again irrelevant
for constraining the type-II seesaw scalar bosons. 

In summary, in contrast to the doubly-charged scalar boson discussed above~(at least for $\delta m\approx 0$), 
there has been no dedicated LHC searches for neutral scalar bosons ($H^{\pm},H^0/A$) within the context of the triplet seesaw. 
For example, all the above mentioned singly-charged and neutral Higgs boson production channels at the LHC are suppressed in the type-II seesaw model.
Instead the triplet-like scalar bosons are produced \textit{a la} Drell-Yan, Fig.~\ref{fig:feyn-diagH}.
Indeed, Drell-Yan quark-antiquark annihilation through s-channel $\gamma/Z$ and $W$ exchanges at the LHC
produces triplet scalar bosons and hence a number of signatures which we discuss in detail in the next section. 
We find that all of these Drell-Yan production cross-sections can be sizeable.
In particular, the production of the singly-charged scalars in association with the neutral ones,
often overlooked by both CMS and ATLAS searches, has the largest cross-sections. 
  Hence the need for including these production channels when probing the type-II seesaw parameter space.

\section{Collider signatures from type-II seesaw scalars} 
\label{sec:collider-signature}

Neutrino mass generation, especially when mediated by scalars, as in our simple type-II seesaw scheme, provides interesting signatures at high-energy colliders.
Although this is the first and simplest SM-based seesaw scheme~\cite{Schechter:1980gr,Schechter:1981cv}, these signatures are still far under-studied in a comprehensive manner.
By that we mean one in which the proper connection with neutrino mass generation and the associated theoretical and phenomenological constraints is duly taken into account.
The very interesting results recently obtained in~\cite{Miranda:2022xbi} further motivate us to present here a comprehensive theoretical study.

We first sketch the main channels that can produce the type-II seesaw neutrino mass mediators and the relevant cross sections.
%%%
We start by discussing the triplet scalar signatures at the LHC, and then move to future hadron as well as lepton colliders.
Our study illustrates their potential for probing fundamental neutrino properties within the type-II seesaw setup.
 Throughout this section we will see that, depending on the value of the triplet VEV $v_\Delta$ there are three important regimes.
  Two of them are either far below or far above $10^{-4}$~GeV or so.
  There is however an intermediate region around $10^{-4}$~GeV where additional decay channels appear depending on the sign and magnitude of the mass-splitting $\delta m$
  between the triplet scalars. This region can play an important role and should be taken into account.  \\[-.4cm] 

\underline{\bf {Large Hadron Collider constraints} }\\[-.4cm]

As discussed earlier, one must take into account all possible production channels and decay modes in order to fully constrain the simplest triplet seesaw~\cite{Schechter:1980gr,Schechter:1981cv}.
In this section we focus on possible signatures at $pp$ and $e^+e^-$ colliders to either discover or constrain the parameter space. 

\begin{figure}[t]
\begin{center}
\includegraphics[height=3.5cm,width=0.5\textwidth]{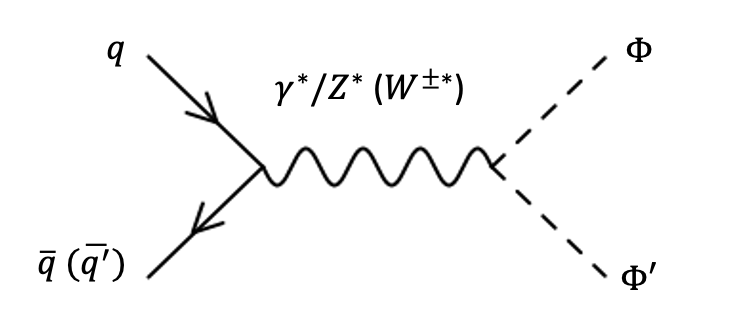}
\end{center}
\caption{Feynman diagrams for Drell-Yan production of the triplet Higgs bosons $\Phi,\Phi'\in\{H^0,A,H^\pm,H^{\pm\pm}\}$ in type-II seesaw.}
    \label{fig:feyn-diagH}
\end{figure}

The relevant pair and associated triplet scalar production $\Phi\Phi'$~(with $\Phi,\Phi'\in\{H^0,A,H^\pm,H^{\pm\pm}\}$) proceeds via the neutral- or charged-current
Drell-Yan mechanism, respectively, involving s-channel $\gamma/Z$ and $W^\pm$ exchange, Fig.~\ref{fig:feyn-diagH}: 
\begin{align}
& q\bar{q'}\to W^* \to H^{\pm\pm} H^\mp, H^\pm H^0, H^\pm A,\\
& q\bar{q}\to \gamma^*/Z^* \to H^{\pm\pm}H^{\mp\mp}, H^\pm H^\mp, H^0 A.
\end{align}

\begin{figure}[b]
\begin{center}
\includegraphics[width=0.49\textwidth]{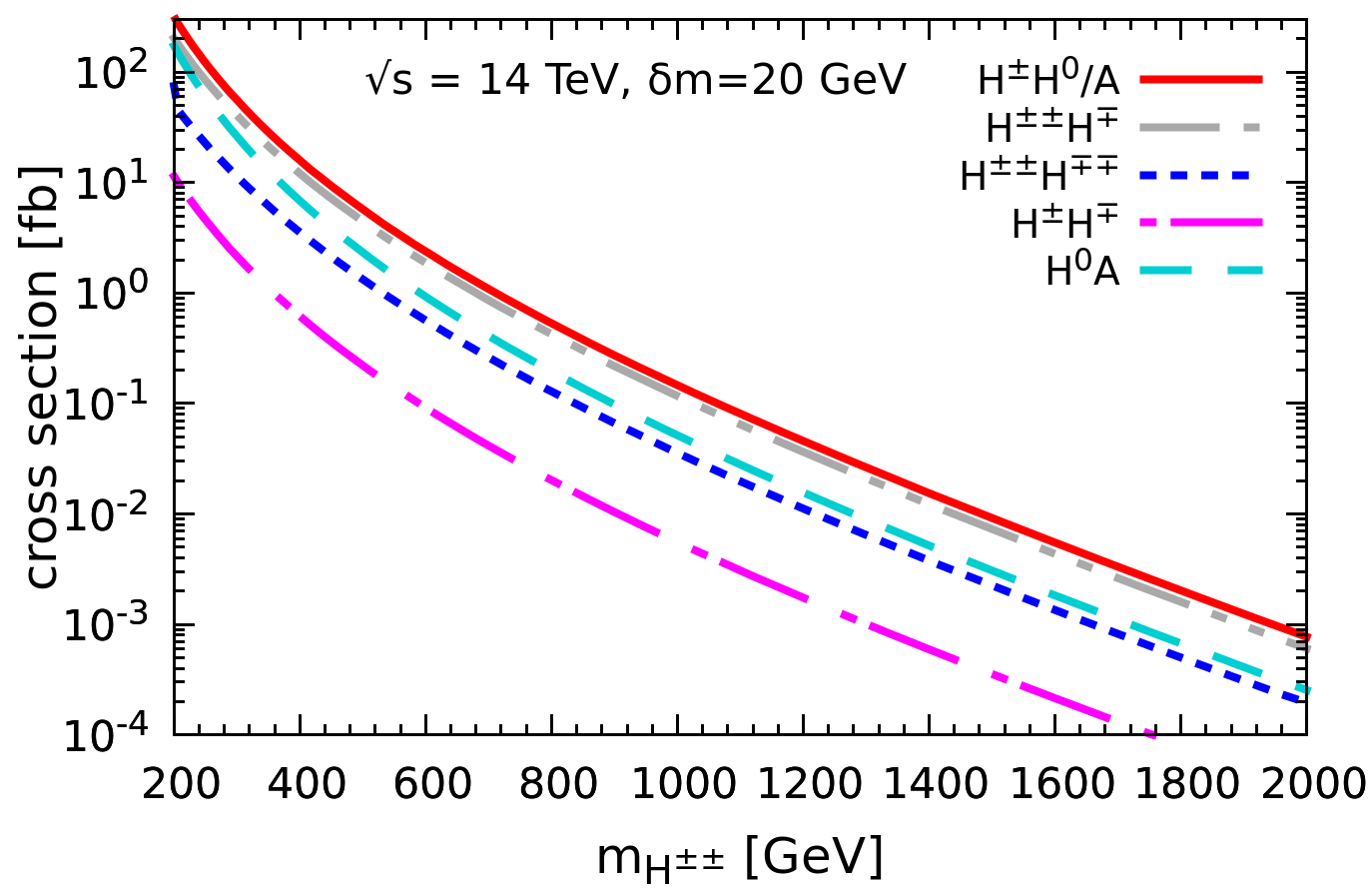}
\includegraphics[width=0.49\textwidth]{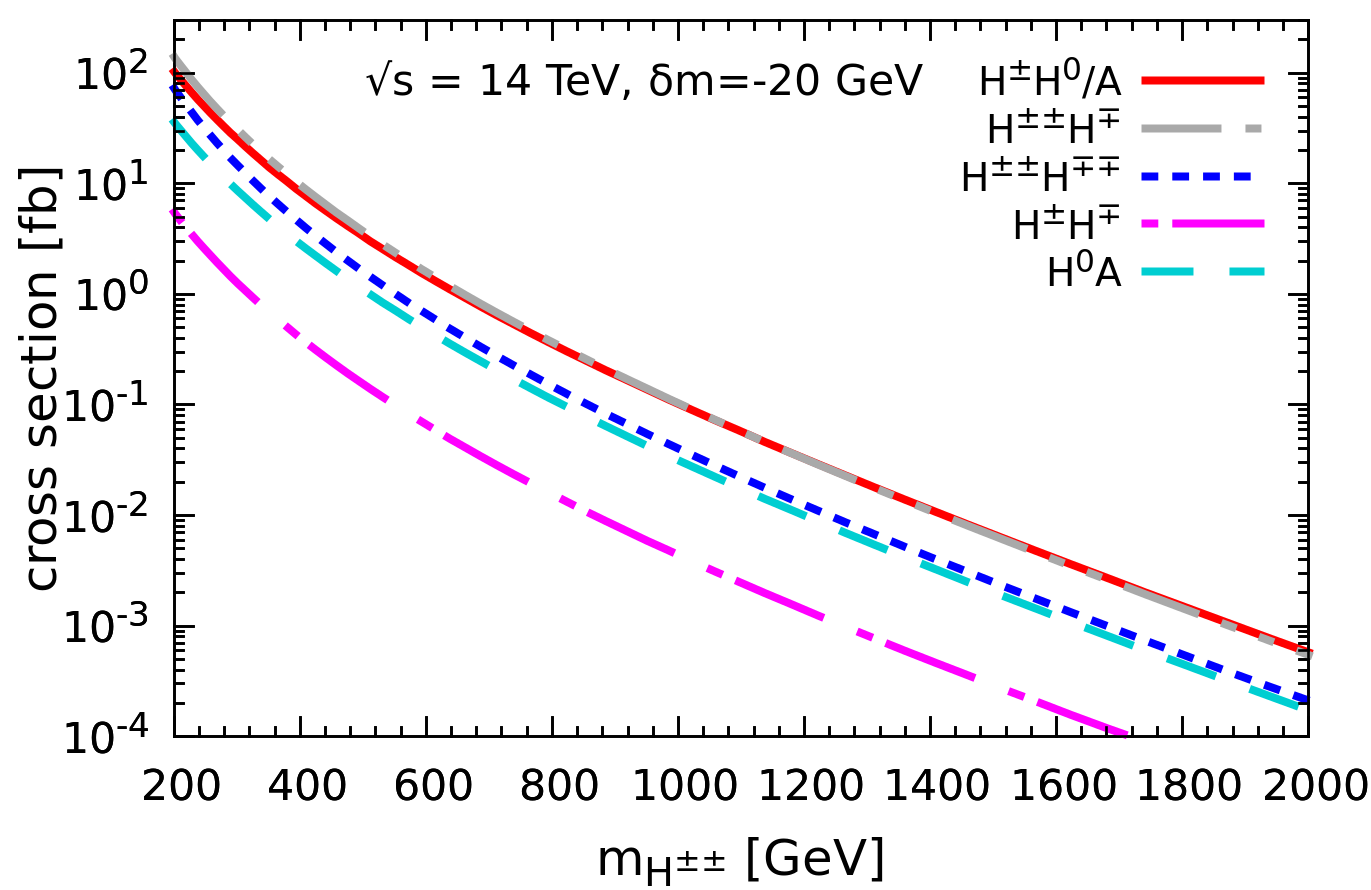}
\end{center}
\caption{Cross section for pair and associated production of triplet seesaw scalars vs the doubly-charged scalar mass $m_{H^{\pm\pm}}$ at a $pp$ collider.
  The center of mass energy is $\sqrt{s}=14$~TeV. The left panel is for $\delta m>0$~($m_{H^{\pm\pm}}>m_{H^\pm}>m_{H^0/A}$) while the right panel is for
  $\delta m<0$~($m_{H^{\pm\pm}}<m_{H^\pm}<m_{H^0/A}$).}
    \label{fig:cross-section14TeV}
\end{figure}

In Fig.~\ref{fig:cross-section14TeV} we display these cross sections versus the mass of $H^{\pm\pm}$, for $\sqrt{s}=14$~TeV.   
Depending on the value and sign of this mass splitting $\delta m$, the cross section also varies.
The left and right panels of Fig.~\ref{fig:cross-section14TeV} correspond to cross sections for positive ($\delta m=20$~GeV) and negative ($\delta m=-20$~GeV) mass splitting, respectively.
We see that most of these production channels have sizable cross sections, suggesting that significant event numbers can be produced. 
This also stresses the need for including all of the possible channels in both the CMS and ATLAS analyses.  \black
For example, we see from Fig.~\ref{fig:cross-section14TeV} that the production cross section for $H^\pm H^0/H^\pm A$, neglected in current ATLAS and CMS analyses,
are substantial, even larger than $H^{\pm\pm}H^{\mp}$ and $H^{\pm\pm}H^{\mp\mp}$, especially in the $\delta m>0$ region. \\[-.4cm]

\underline{\bf Future hadron colliders} \\[-.4cm]

\begin{figure}[!h]
\begin{center}
\includegraphics[width=0.49\textwidth]{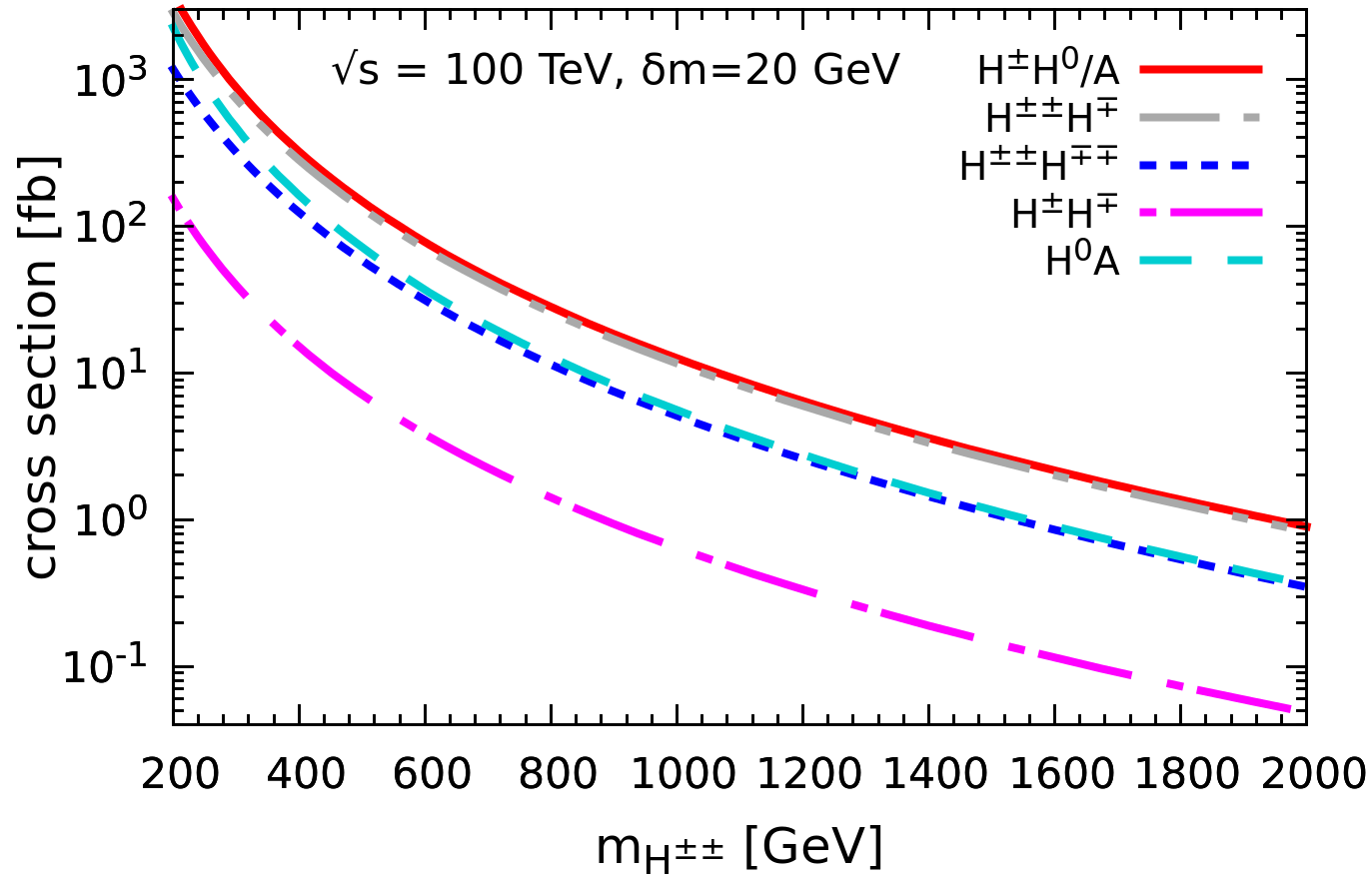}
\includegraphics[width=0.49\textwidth]{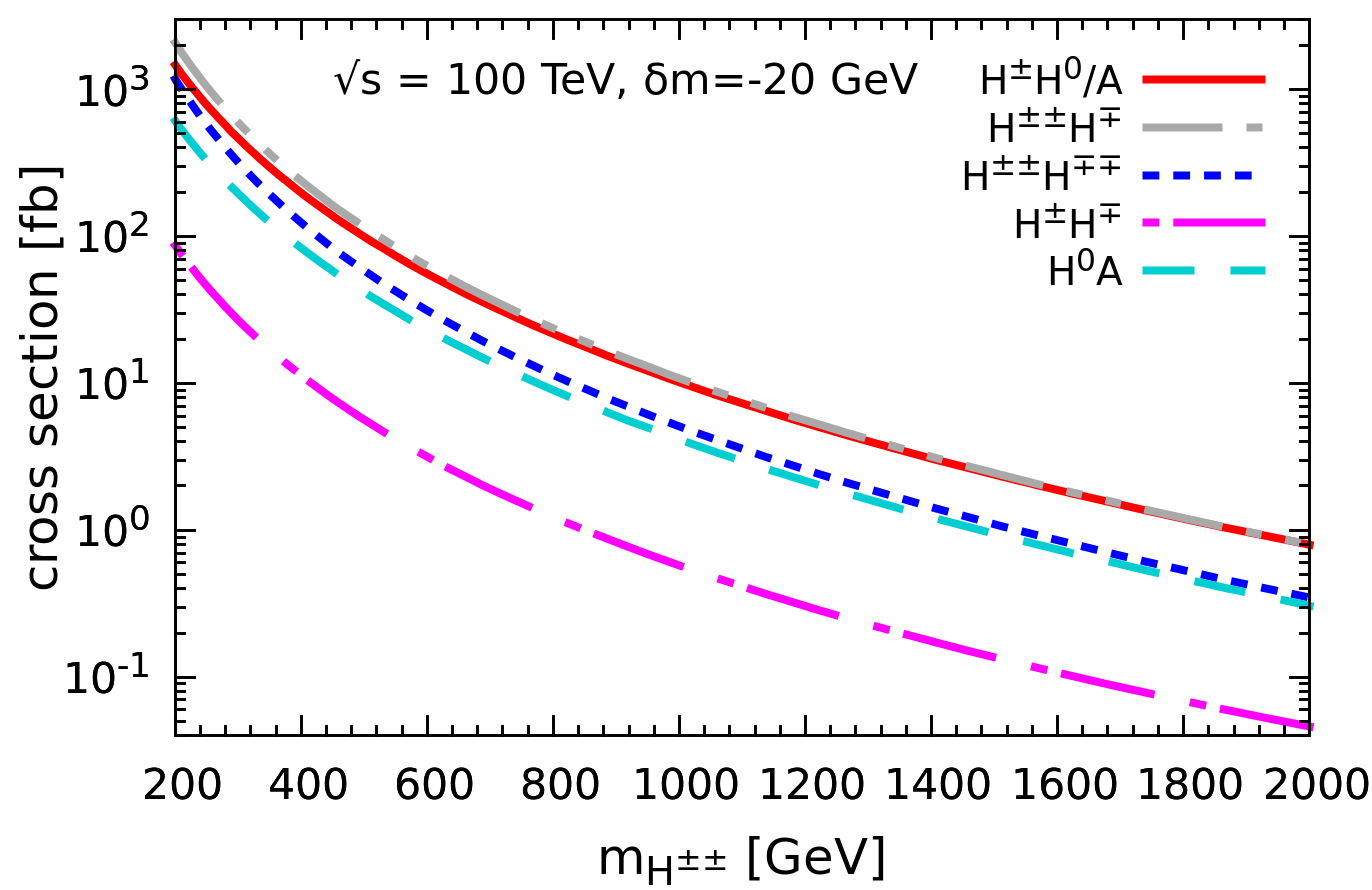}
\end{center}
\caption{Cross section for pair and associated production of triplet seesaw scalars vs the doubly-charged scalar mass at the FCC-hh with $\sqrt{s}=100$~TeV.
  Same conventions as Fig.~\ref{fig:cross-section14TeV}.}
    \label{fig:cross-section100TeV}
\end{figure}

In Fig.~\ref{fig:cross-section100TeV}, we also show the production cross section for a future $pp$ collider with center-of-mass energy $\sqrt{s}=100$~TeV, such as FCC-hh.
As shown in the Fig.~\ref{fig:cross-section100TeV}, the production cross section at such center of mass energy is large enough that multi-TeV Higgs masses can be explored.

The LHC signatures depend also on the decay channels of each scalar boson being produced. Apart from fully invisible processes, all relevant branching ratios are given in Eq.~\eqref{eq:Hpptoll} to Eq.~\eqref{eq:AtoHW}.  
  We have already shown the branching ratio of $H^{\pm\pm}$ and $H^{\pm}$ in Figs.~\ref{fig:BRHpp} and \ref{fig:BRHp}, respectively.
  The behaviour of the $H^0/A$ branching ratios is similar to the singly-charged Higgs.
  The doubly-charged Higgs decays to dilepton, diboson or into cascade mode depending on the triplet VEV $v_\Delta$ and mass splitting $\delta m$.
  The singly-charged Higgs boson $H^\pm$ or neutral boson $H^0/A$ has four decay modes: leptonic decay, hadronic decay, diboson decay and cascade decay,
  i.e. $H^0/A W^*$~($\delta m>0$) or $H^{\pm\pm}W^{\mp*}$~($\delta m<0$) depending on the triplet VEV and mass splitting $\delta m$. \\[-.4cm]

  In what follows we discuss possible signatures at colliders depending on what we assume for the mass splitting
  $\delta m$~\cite{Primulando:2019evb,Ashanujjaman:2021txz,Antusch:2018svb,Mitra:2016wpr}.
There are three regimes to consider:
(A) $\lambda_4=0:\,\,\delta m\approx 0\,( m_{H^{\pm\pm}}\simeq m_{H^\pm}\simeq m_{H^0/A}) $,
(B) $\lambda_4>0:\,\,\delta m<0\,( m_{H^{\pm\pm}} < m_{H^\pm} < m_{H^0/A})$ and
(C) $\lambda_4<0:\,\,\delta m>0\,( m_{H^{\pm\pm}} > m_{H^\pm} > m_{H^0/A})$.
As the official ATLAS and CMS searches already considered the production channel $H^{\pm\pm}H^{\mp\mp}$, we do not repeat here. 

\subsection{$\lambda_4=0:\,\,\delta m\approx 0\,( m_{H^{\pm\pm}}\simeq m_{H^\pm}\simeq m_{H^0/A}) $} 
In this case, all the pair- and associated pair-production, except for $H^\pm H^{\mp}$, have a sizeable cross section. 
For small triplet VEV $v_{\Delta}<10^{-4}$~GeV, $H^{\pm\pm}$, $H^{\pm}$ and $H^0/A$ predominantly decay to $\ell^\pm\ell^\pm$, $\ell^\pm\nu$ and $\nu\nu$,
respectively. 
Hence, to probe this $|\delta m|\approx 0$ and $v_\Delta<10^{-4}$~GeV region, we must include the following three possible multilepton final states:  
\begin{align}
  pp\to H^{\pm\pm}H^\mp \to \ell^\pm\ell^\pm\ell^\mp\nu, \,\,\,\, pp \to H^{\pm}(H^0/A)\to \ell^\pm\nu\nu\nu,\,\,\,\,\text{and}\,\,\,\,\,\,
  pp\to H^0 A\to \nu\nu\nu\nu.
\label{eq:pp-dmzero-set0}
\end{align}
From the above, we see that despite having sizeable cross-sections, $H^\pm H^0/A$ and $H^0A$ can not compete with the three-lepton final state coming from $H^{\pm\pm}H^\mp$
due to their invisible decays.

On the other hand, for larger triplet VEV $v_\Delta>10^{-4}$~GeV, $H^{\pm\pm}$, $H^{\pm}$, $H^0$ and $A$ decay to $W^\pm W^\pm$, $W^\pm Z/h$, $ZZ/WW/hh$ and $hZ$, respectively. 
Therefore, to probe the region $|\delta m|\approx 0$ and $v_\Delta>10^{-4}$~GeV we must use the following multi-lepton final states 
\begin{align}
& pp \to H^{\pm\pm}H^\mp \to W^\pm W^\pm W^\mp (Z/h) \to \ell^{\pm}\ell^\pm\ell^\mp\ell^+\ell^-\nu\nu\nu, \label{eq:pp-dmzero-set1}\\
& pp \to H^{\pm}H^0 \to (W^\pm Z/h) (ZZ/hh/W^\pm W^\mp) \to (\ell^\pm\ell^+\ell^-\nu) (\ell^+\ell^-\ell^+\ell^-/\ell^\pm\ell^\mp\nu\nu), \label{eq:pp-dmzero-set2}\\
& pp \to H^0 A \to (ZZ/hh/W^\pm W^\mp) Zh \to (\ell^+\ell^-\ell^+\ell^-/\ell^\pm\ell^\mp\nu\nu) (\ell^+\ell^-\ell^+\ell^-). \label{eq:pp-dmzero-set3}
\end{align}
\subsection{$\lambda_4>0:\,\,\delta m<0\,( m_{H^{\pm\pm}} < m_{H^\pm} < m_{H^0/A})$}
\label{subsec:dm-negative}
For small mass splitting such as $\delta m\lesssim \mathcal{O}(1\,\text{GeV})$, the signatures resemble the degenerate scenario $\delta m\approx 0$.
On the other hand, for triplet VEV in between $10^{-5}\,\text{GeV}<v_\Delta < 10^{-3}\,\text{GeV}$ and relatively large mass splitting $\delta m$, the cascade decay of $H^\pm$ into $H^{\pm\pm}W^{\mp *}$ is dominant.
This effectively enhances the production cross section for $H^{\pm\pm}$. In the same triplet VEV region, $H^0/A\to H^\pm W^{\mp *}$ decay is also significantly large.
Moreover, for $v_\Delta > 10^{-4}$~GeV, $H^{\pm\pm}\to W^\pm W^\pm$ has large branching ratio.
Therefore the following processes are important in order to probe the region $\delta m<0$ and moderate $v_\Delta\sim 10^{-4}$~GeV: 
\begin{align}
& pp\to H^\pm H^0/A \to H^{\pm\pm} W^{\mp *} H^\pm W^{\mp *} \to H^{\pm\pm} W^{\mp *} H^{\pm\pm} W^{\mp *} W^{\mp *} \to 4 W^\pm + X, \label{eq:pp-dmnegative-set1}\\
& pp \to H^0 A \to H^\pm W^{\mp *} H^\pm W^{\mp *} \to H^{\pm\pm} W^{\mp *} W^{\mp *} H^{\pm\pm} W^{\mp *} W^{\mp *} \to 4 W^\pm + Y, \label{eq:pp-dmnegative-set2} \\ 
& pp \to H^{\pm\pm} H^\mp \to H^{\pm\pm} H^{\mp\mp} W^{\pm *} \to 2W^\pm + 2 W^\mp + Z. \label{eq:pp-dmnegative-set3}
\end{align}
Taking into account the subsequent leptonic decay of the produced on-shell $W^\pm$, these processes will lead to same-sign and opposite-sign tetra-lepton signatures~\cite{Chun:2019hce}.
Again for small $v_\Delta$, $H^{\pm\pm},H^{\pm},H^0/A$ dominantly decay leptonically giving rise to multilepton final states.
  However these can not compete with multilepton final states coming from pair production of $H^{\pm\pm}H^{\mp\mp}$, hence we have not discussed them.

\subsection{$\lambda_4<0:\,\,\delta m>0\,( m_{H^{\pm\pm}} > m_{H^\pm} > m_{H^0/A})$}
\label{subsec:dm-positive}
For small $\delta m$, this scenario will be again the same as the case of $\delta m\approx 0$. 
If the mass splitting $\delta m$ is large, above $1$~GeV, and the triplet VEV is moderate, then $H^{\pm\pm}\to H^\pm W^{\pm *}$ and
$H^\pm\to H^0/A W^{\pm *}$ decay modes dominate over others.
This effectively enhances the $H^0$ and $A$ production cross section. 
Further for moderate triplet VEV $v_\Delta\sim \mathcal{O}(10^{-4}\,\text{GeV})$, $H^0$ and $A$ dominantly decay to $ZZ/W^\pm W^\mp /hh$ and $hZ$, respectively.
Hence, the following are the possible signatures to probe the region of moderate $v_\Delta$ and large enough $\delta m$:
\begin{align}
& pp \to H^\pm (H^0/A) \to (H^0/A) W^{\pm *} (H^0/A) \to (ZZ/W^\pm W^\mp/hh/hZ)W^{\pm *} (ZZ/W^\pm W^\mp/hh/hZ) \nonumber  \\
& \to (\ell^+\ell^-\ell^+\ell^-/\ell^\pm\nu\ell^\mp\nu) (\ell^+\ell^-\ell^+\ell^-/\ell^\pm\nu\ell^\mp\nu) + X , \label{eq:pp-dmpositive-set1} \\
& pp \to H^0 A \to (ZZ/W^\pm W^\mp/hh) hZ \to \ell^+\ell^-\ell^+\ell^-\ell^\pm\nu\ell^\mp\nu, \label{eq:pp-dmpositive-set2} \\
& pp \to H^{\pm\pm} H^\mp \to W^\pm \,\,W^\pm H^0/A W^{\mp *} \to W^\pm W^\pm (ZZ/W^\pm W^\mp/hh/hZ) W^{\mp *} \nonumber \\
& \to \ell^\pm\nu\ell^\pm\nu (\ell^+\ell^-\ell^+\ell^-/\ell^\pm\nu\ell^\mp\nu) + X. \label{eq:pp-dmpositive-set3}
\end{align}

From the discussion in \ref{subsec:dm-negative} and \ref{subsec:dm-positive}, we conclude that the cascade decays~($H^\pm\to H^{\pm\pm} W^{\mp *}, H^0/A\to H^\pm W^{\mp *}$
for $\delta m<0$ and $H^{\pm\pm}\to H^{\pm} W^{\pm *}$, $H^\pm\to H^0/A W^{\pm *}$ for $\delta m>0$) should play a significant role in restricting the model parameter space in
the non-degenerate situation~($|\delta m|\neq 0$) for intermediate region $v_\Delta\sim 10^{-4}$ GeV.

\subsection{Displaced vertices}
\label{sec:displaced-vertices}

In certain parameter regions, the doubly-charged Higgs $H^{\pm\pm}$ can be long-lived, leading to the possibility of displaced vertices
that could be searched at HL-LHC~($\sqrt{s}=14$~TeV) or FCC-hh~($\sqrt{s}=100$~TeV). 
We already saw that for large triplet VEV $v_\Delta$, $H^{\pm\pm}$ dominantly decay to on-shell $W^\pm W^\pm$, but is kinematically forbidden if $m_{H^{\pm\pm}}\lesssim 160$~GeV. 
In that case, $H^{\pm\pm}$ decays mainly via $H^{\pm\pm}\to W^\pm W^{\pm *}\to W^{\pm}f \bar{f'}$ and this rate is proportional to triplet VEV $v_\Delta$.
Depending upon the parameter choices, the decay width can be small, i.e. the decay length can be macroscopic.
For example, with $v_\Delta\sim 10^{-4}$~GeV and $m_{H^{\pm\pm}}\sim 150$~GeV, proper decay length can be order of 1 mm. 
In such parameter region one has very good prospects for displaced vertex searches. For a detailed discussion, see Ref.~\cite{Antusch:2018svb}.\\[-.4cm]

\underline{\bf Future lepton colliders} \\[-.4cm]

As seen in Figs.~\ref{fig:cross-section14TeV} and \ref{fig:cross-section100TeV}, the pair production~($H^{\pm\pm}H^{\mp\mp}$) cross section at hadron colliders becomes small for large
doubly-charged Higgs masses. 
In addition to this, the existence of multiple backgrounds reduces their physics reach for doubly-charged Higgs boson discovery.
A lepton $e^+e^-$ collider, with a considerably cleaner environment, is more suitable for searching $H^{\pm\pm}$.  
In this section, we discuss the possible collider signatures associated to the doubly-charged Higgs boson triplet of the type-II seesaw mechanism.
We examine its production cross section at $e^+ e^-$ colliders, such as CLIC~\cite{CLICdp:2018cto}, FCC-ee~\cite{FCC:2018evy}, ILC~\cite{Barklow:2015tja} and CEPC~\cite{CEPCStudyGroup:2018ghi}.
First note that the doubly-charged Higgs boson $H^{\pm\pm}$ can be produced at $e^+e^-$ colliders through $\gamma/Z$-mediated processes.

\begin{figure}[t]
\begin{center}
\includegraphics[width=0.49\textwidth]{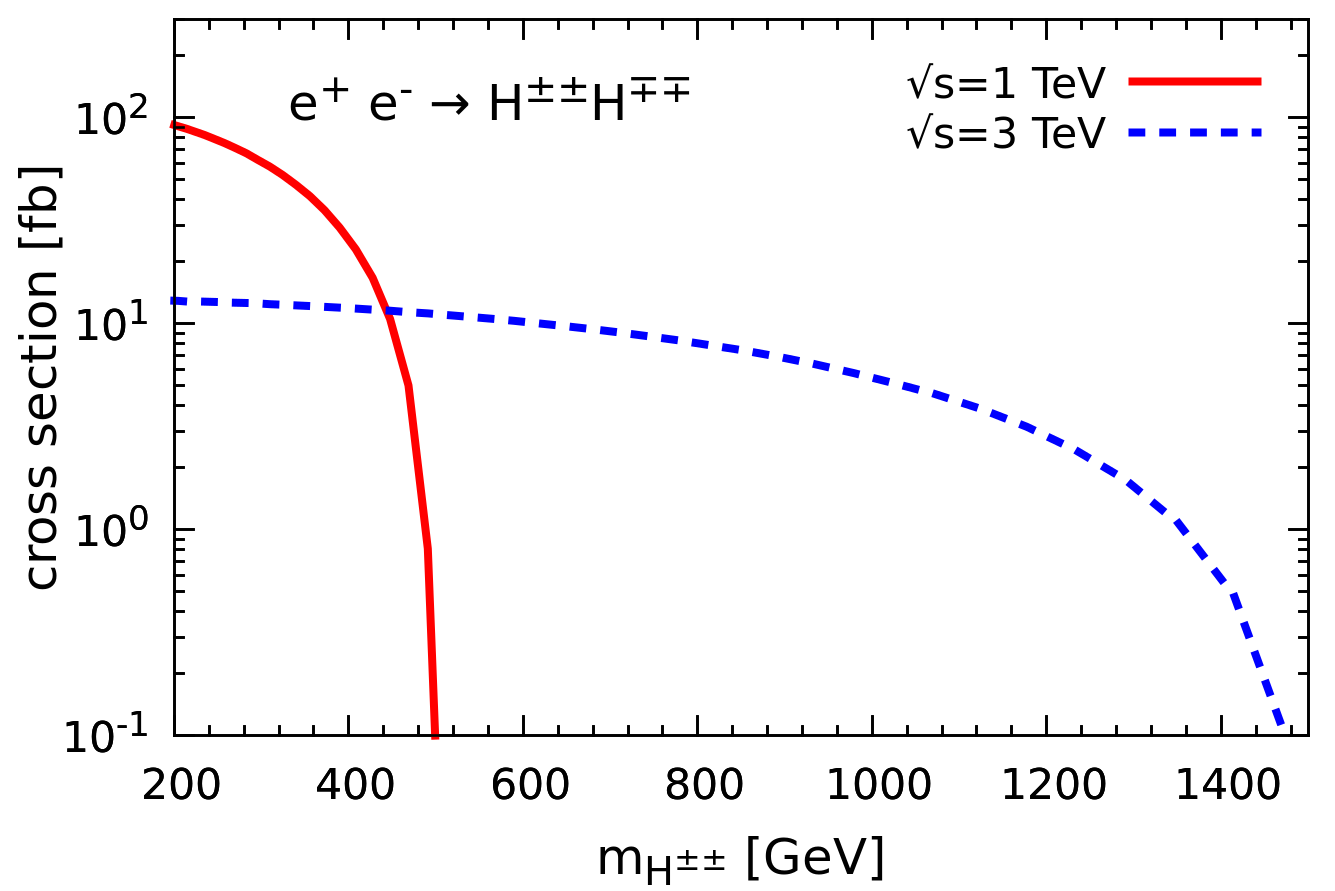}
\end{center}
\caption{The production cross section for $H^{\pm\pm}H^{\mp\mp}$ at $e^+e^-$ collider. The center of mass energies are $\sqrt{s}=1$~TeV~(red-solid line) and 3 TeV~(blue-dashed line).}
    \label{fig:cross-section-ILC}
\end{figure}

In Fig.~\ref{fig:cross-section-ILC}, we show the $H^{\pm\pm}H^{\mp\mp}$ production cross section at an $e^+e^-$ collider for center of mass energies $\sqrt{s}=1$~TeV and 3 TeV, respectively.
For small triplet VEV $v_\Delta < 10^{-4}$~GeV, the $H^{\pm\pm}\to \ell^\pm \ell^\pm$ decay mode dominates and one can have tetra-lepton signatures.  
%%%%%%%%%%%%%%%%%%%%%%%%%%%%%%%%%%%
\begin{figure}[t]
\begin{center}
\includegraphics[width=0.49\textwidth]{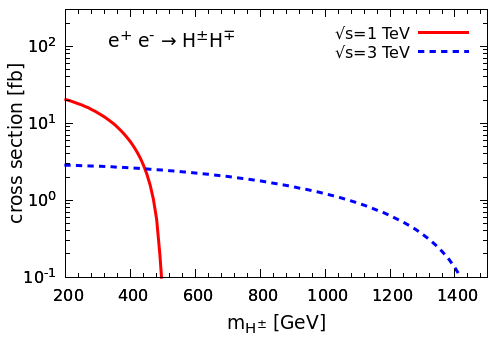}
\includegraphics[width=0.49\textwidth]{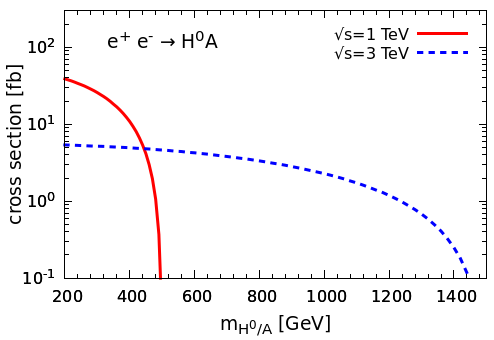}
\end{center}
\caption{The production cross section for $H^{\pm}H^{\mp}$~(left panel) and $H^0/A$~(right panel) at $e^+e^-$ collider. The center of mass energies are $\sqrt{s}=1$~TeV~(red-solid line) and 3 TeV~(blue-dashed line).}
    \label{fig:cross-section-ILC-2}
\end{figure}
For large triplet VEV $v_\Delta > 10^{-4}$~GeV, $H^{\pm\pm}$ will decay dominantly into $W^\pm W^\pm$ gauge bosons. One can consider either the leptonic or hadronic decay modes of $W^\pm$.
Hence to probe the large triplet VEV $v_\Delta > 10^{-4}$~GeV region, one can look for the following multilepton or multijet final states: 
\begin{align}
e^+ e^- \to H^{\pm\pm} H^{\mp\mp} \to W^\pm W^\pm W^\mp W^\mp \to \ell^\pm \ell^\pm \ell^\mp \ell^\mp \nu\nu\nu\nu/ 8 j.
\label{eq:ee-to-lepton-or-jet}
\end{align}
If the $H^{\pm\pm}$ mass $m_{H^{\pm\pm}}\sim\mathcal{O}(1\,\text{TeV})$, the gauge bosons produced from $H^{\pm\pm}$ decay will be highly boosted, leading to a fat-jet~\cite{Agrawal:2018pci}:
\begin{align}
e^{+} e^{-} \to H^{\pm\pm} H^{\mp\mp}\to W^\pm W^\pm W^\mp W^\mp \to 4\,\text{fat-jets}.
\label{eq:ee-to-fatjet}
\end{align}
Note that for such a heavy Higgs boson~($m_{H^{\pm\pm}}\sim 1$~TeV), the produced $W^\pm$ or fat-jet will have large transverse momenta~($p_T\sim 1$~TeV),
which can be used effectively to reduce the SM backgrounds. 

For completeness, in Fig.~\ref{fig:cross-section-ILC-2}, we show the $H^{\pm}H^{\mp}$~(left panel) and $H^0/A$~(right panel) production cross section at an $e^+e^-$ collider
  for center of mass energies $\sqrt{s}=1$~TeV and 3 TeV, respectively. Although smaller than $H^{\pm\pm}H^{\mp\mp}$, the production cross sections are still sizeable.
%%%%%%%%%%%%%%%%%%%%%%%%%%%%%%%%%%%%%%%%%%%%%%%%%%%%%%%%%%%%%%%%%%%%%
\section{Deconstructing the type-II seesaw at colliders}
\label{sec:compl-betw-clfv}

\begin{figure}[!h]
\begin{center}
\includegraphics[width=0.49\textwidth]{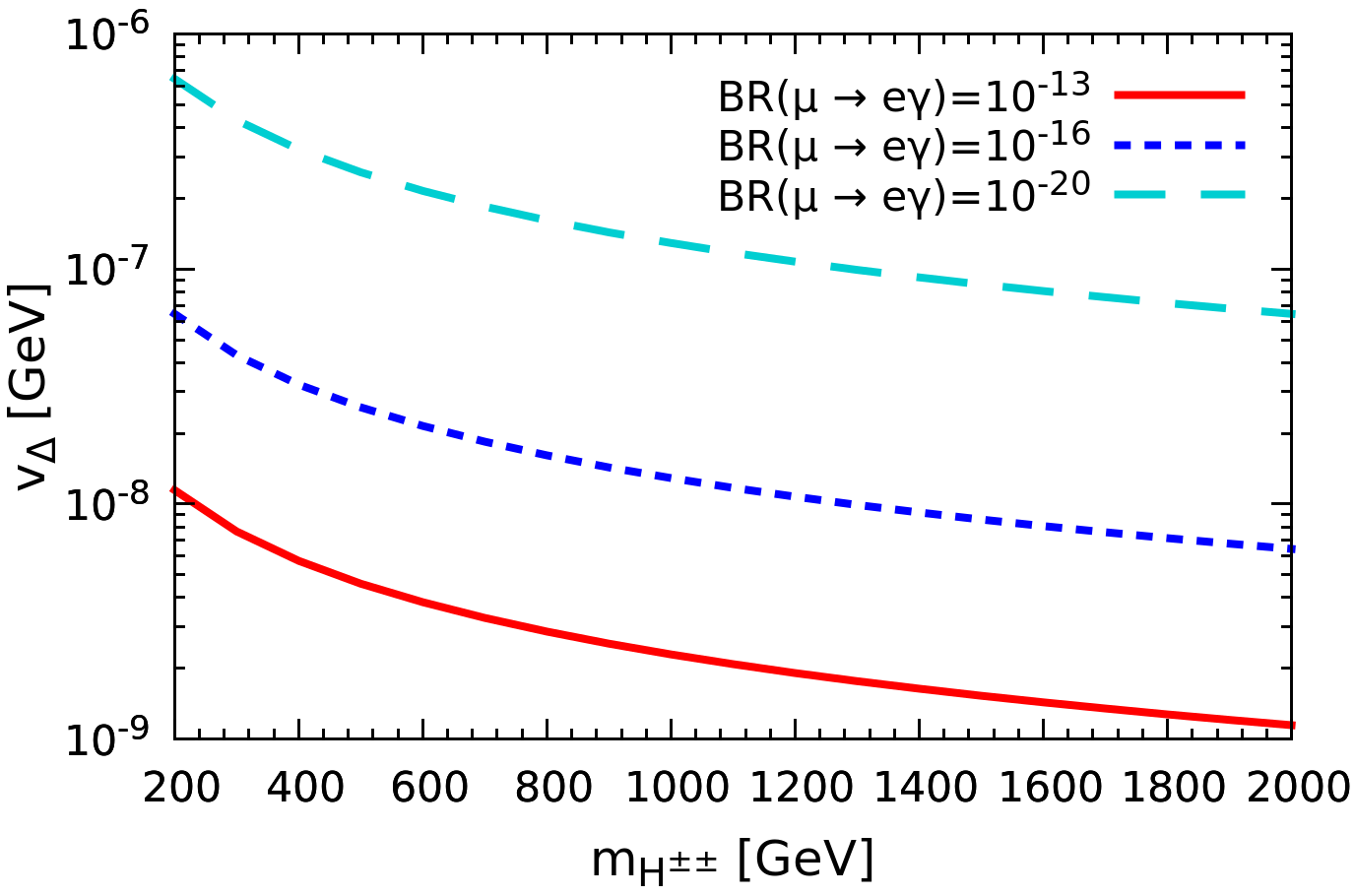}
\end{center}
\caption{Contour of $\text{BR}(\mu\to e\gamma)$ in the $m_{H^{\pm\pm}}-v_\Delta$ plane. The oscillation parameters are fixed to their best fit values.}
    \label{fig:densityLFV}
\end{figure}

As already mentioned, charged lepton flavor violaton is mediated by the same triplet scalar responsible for neutrino mass generation.
A single universal symmetrical Yukawa coupling $Y_\Delta$ is involved, see Eqs.~(\ref{eq:s-18}) and (\ref{eq:BRmutoegamma}), (\ref{eq:BRmutoeee}).
The relevant Yukawa coupling matrix can be parameterised in terms of neutrino oscillation parameters as,
\begin{align}
Y_\Delta=\frac{\sqrt{2}}{v_\Delta} U\textup{diag}\left\{ m_1, m_2, m_3\right\} U^T.
\label{eq:Yukawa}
\end{align}
The structure of the restrictions on $m_\nu = Y_\Delta v_\Delta$ that follow from oscillations and \znbb decay have been given in Fig.~\ref{fig:dbd-bounds}. 
One sees from Eqs.~(\ref{eq:BRmutoegamma}) and (\ref{eq:BRmutoeee}) that the cLFV branching fractions are detemined by the same matrix $Y_\Delta$.
In Fig.~\ref{fig:densityLFV}, we show contours of $\text{BR}(\mu\to e\gamma)$ in the $m_{H^{\pm\pm}}-v_\Delta$ plane, obtained for normal
ordered light neutrino spectrum and best-fit values for the neutrino oscillation parameters~\cite{deSalas:2020pgw,10.5281/zenodo.4726908}. 
One sees how $\text{BR}(\mu\to e\gamma)$ may easily lie within reach of experiment, such as the current measurements of MEG~\cite{MEG:2013oxv}
that lead to the limit $\text{BR}(\mu\to e\gamma)< 4.2\times 10^{-13}$. 
One sees that the $\mu\to e\gamma$ branching ratio can easily exceed current senstivities for small triplet VEV $v_\Delta$.
The same happens for other cLFV processes.

Hence, given the neutrino mass ordering, the magnitude of the Yukawa coupling is completely determined by $v_\Delta$. 
We checked that for the triplet VEV range $10^{-10}\,\text{GeV}\leq v_\Delta\leq 10^{-3}\,\text{GeV}$ relevant for Fig.~\ref{fig:densityLFV},
the Yukawa coupling always remains within the perturbative regime.

\underline{\bf Distinguising the neutrino mass ordering at colliders} \\[-.4cm]

We have already discussed that the decays of the doubly-charged Higgs boson to leptonic final states are determined by the Yukawa coupling $Y_\Delta$. The decay width is given by
\begin{equation}
\Gamma (H^{\pm \pm} \to l^{\pm}_i l^{\pm} _j)=\Gamma_{l_i l _j}=\frac{m_H^{\pm \pm} } {(1+\delta_{ij}) 16 \pi}  |Y_{\Delta}^{ij}|^2.
\end{equation}
where the Yukawa coupling is determined by Eq.~(\ref{eq:Yukawa}). 
Current measurements of neutrino oscillation parameters~\cite{deSalas:2020pgw,10.5281/zenodo.4726908} restrict the pattern of diagonal and off-diagonal entries of Yukawa coupling $Y_\Delta$.
Note that for $v_\Delta < 10^{-4}$~GeV, $H^{++}$ decays predominantly to leptons and as $\Gamma_{\ell_i\ell_j}\propto |Y_{\Delta}^{ij}|^2$, the patterns of various leptonic channels will exactly
follow the pattern of $Y_\Delta^{ij}$, partially determined by oscillation experiments. 

\begin{figure}[t]
\begin{center}
\includegraphics[width=0.49\textwidth]{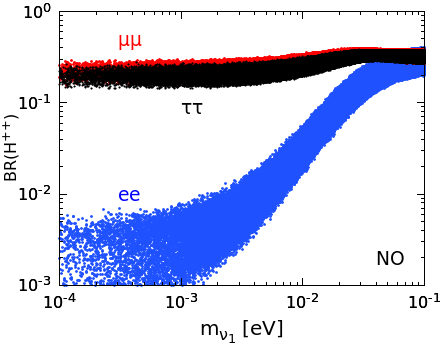}
\includegraphics[width=0.49\textwidth]{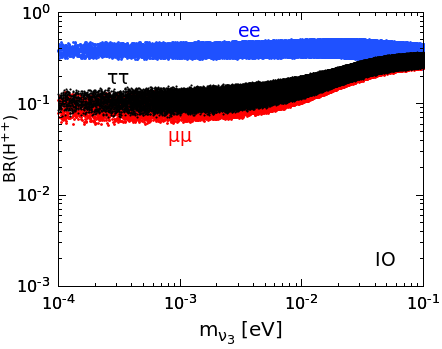}
\end{center}
\caption{$H^{++}$ branching ratio to same-flavour like-sign dilepton final states versus the lightest neutrino mass for {\bf NO} (left panel), and {\bf IO} (right panel).
  Oscillation parameters are varied within their $3\sigma$ ranges. We have fixed triplet VEV, $v_\Delta < 10^{-4}$~GeV and $m_{H^{\pm\pm}}=1$~TeV. Note that for {\bf NO}, di-muon and di-tau final states are always larger than di-electron final state, whereas for {\bf IO}, di-electron final states are always larger than di-muon and di-tau final sates. }
    \label{fig:BRHppdiag}
\end{figure}

In Fig.~\ref{fig:BRHppdiag}, we show the branching fractions for the decays into same-flavour leptonic final states.
To obtain this we randomly varied the mixing angles within their $3\sigma$ ranges of allowed values~\cite{deSalas:2020pgw,10.5281/zenodo.4726908},
varying $\delta_{\text{CP}}$ in the range $[-\pi:\pi]$, and choosing the triplet VEV as $v_\Delta<10^{-4}$~GeV.
Since there is no infomation on Majorana phases we set them to zero. 
\begin{figure}[t]
\begin{center}
\includegraphics[width=0.49\textwidth]{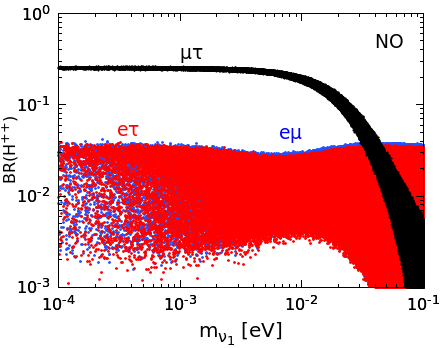}
\includegraphics[width=0.49\textwidth]{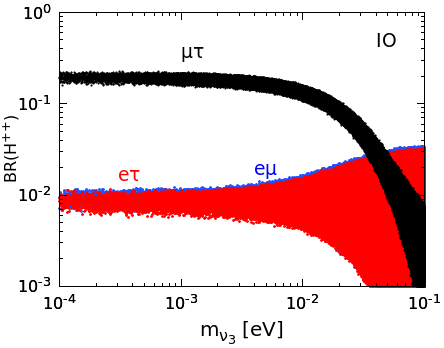}
\end{center}
\caption{Same as Fig.~\ref{fig:BRHppdiag} but now for $H^{\pm\pm}$ decay to different flavour like-sign dilepton final states.
     Except for ``large'' values of the lightest neutrino mass above few$\times 10^{-2}$eV or so, in both {\bf NO} and {\bf IO} cases, the $\mu\tau$ like-sign final state is dominant.}
    \label{fig:BRHppoffdiag}
\end{figure}
%%%%%%%%%%%%%%%%%%%%%%%%%%%%%
Depending on the ordering of the light-neutrino mass spectrum we obtain the following decay branching ratio patterns~\cite{Cai:2017mow}:
\begin{align}
&\text{BR}(H^{++}\to \mu\mu), \text{BR}(H^{++}\to \tau\tau)\gg \text{BR}(H^{++}\to ee)\,\,\,\,\text{for {\bf NO}},\\
& \text{BR}(H^{++}\to ee)\gg \text{BR}(H^{++}\to \mu\mu), \text{BR}(H^{++}\to \tau\tau)\,\,\,\,\text{for {\bf IO}}.
\end{align}
The above two relations suggest that, depending on the ordering of the light neutrino masses, $H^{++}$ can decay to either $\mu\mu$, $\tau\tau$~(for {\bf NO}) or $ee$~({\bf IO})
with relatively large strength. Hence, it might be possible to probe the ordering~({\bf NO} or {\bf IO}) by looking into the decay patterns of $H^{++}$ to same-flavour leptonic final states.\\[-.4cm]

In Fig.~\ref{fig:BRHppoffdiag}, we show the $H^{++}$ decay branching fraction into different-flavour like-sign dilepton final states.
  One finds that for both {\bf NO} and {\bf IO} cases the $H^{++}$ charged-lepton-flavour-violating branching ratios % to the $e\mu$, and $e\tau$ final states
  are of the same order of magnitude as the lepton-flavour-conserving ones, and especially the $\mu\tau$ channel can be quite large for both mass spectra.
  %%%%%%%%%%%%%%%%  
  \begin{figure}[t]
\begin{center}
\includegraphics[width=0.45\textwidth]{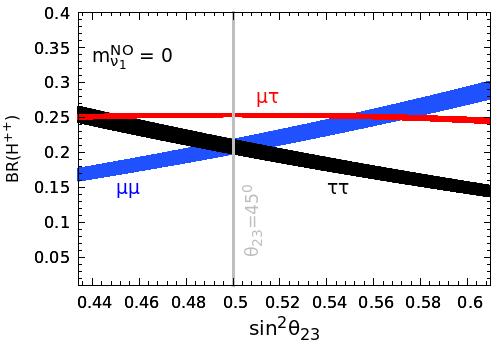}
\includegraphics[width=0.45\textwidth]{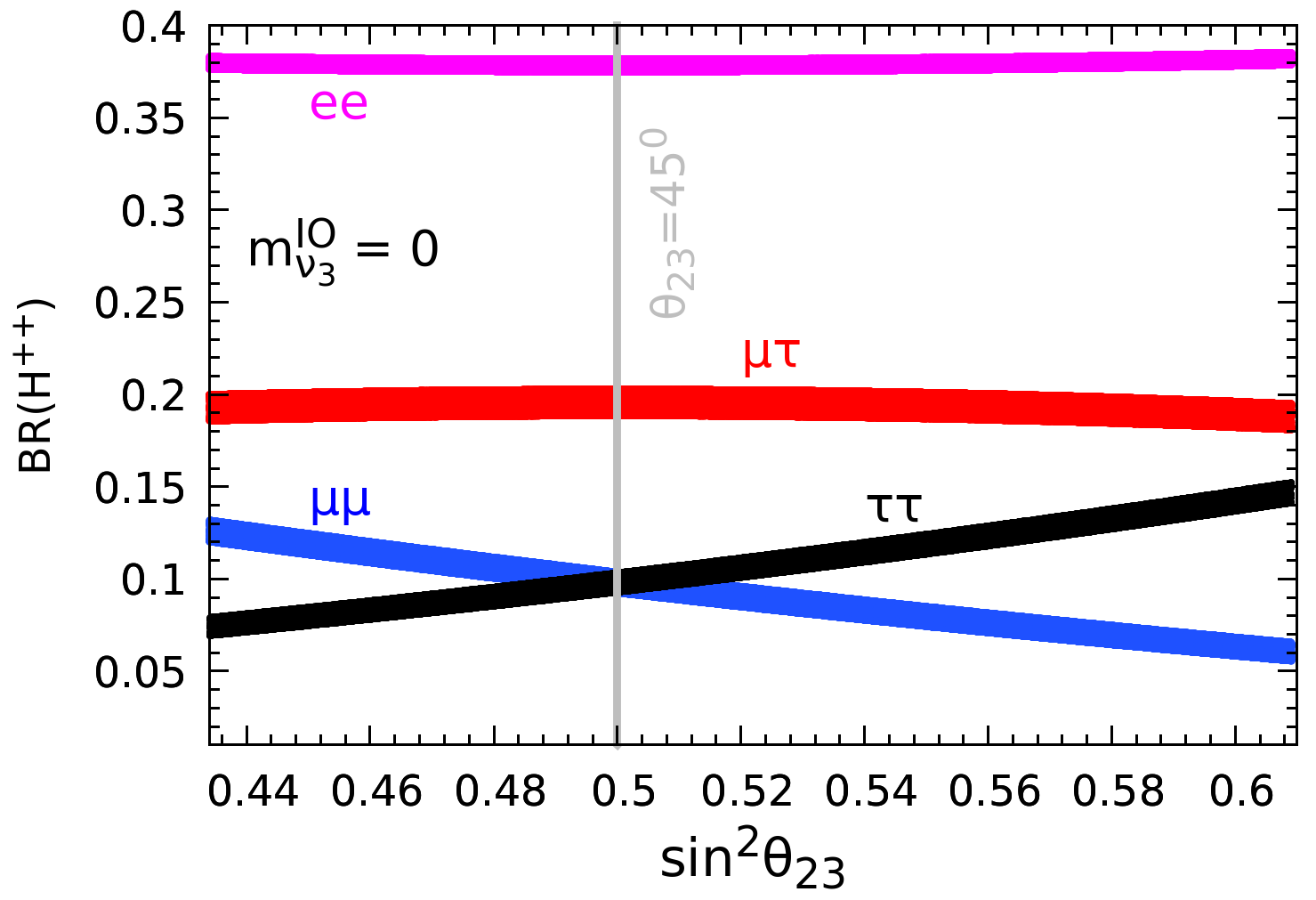}
\end{center}
\caption{ 
  $H^{\pm\pm}$ decay branching ratios versus the atmospheric angle within its allowed $3\sigma$ range, for both {\bf NO}~(left panel) and {\bf IO}~(right panel), repectively.
  The lightest neutrino mass is fixed to zero, with triplet VEV $v_\Delta<10^{-4}$~GeV, so that the dilepton decay channel is dominant.
  The Dirac phase $\delta_{\text{CP}}$ is varied in the range $[-\pi:\pi]$ and other oscillation parameters are fixed to their best fit values~\cite{deSalas:2020pgw,10.5281/zenodo.4726908}.
One sees that like-sign di-muon and di-tau branching fractions correlate with the octant in different ways depending on the mass ordering. }
    \label{fig:Atmos}
  \end{figure}

  \vskip .2cm
  Likewise, in Fig.~\ref{fig:Atmos}, we display the leptonic branching ratios~(only those which are appreciable) in terms of the allowed $3\sigma$ range of the ``atmospheric'' mixing angle
    $\theta_{23}$, both for {\bf NO}~(left panel) and {\bf IO}~(right panel). 
 As before, the triplet VEV is chosen as $v_\Delta<10^{-4}$~GeV, $\delta_{\text{CP}}$ is varied in the range $[-\pi:\pi]$, and other oscillation parameters are fixed to their best fit
 values~\cite{deSalas:2020pgw,10.5281/zenodo.4726908}.
 The vertical line in each panel denotes $\theta_{23}=45^\circ$. 
 One sees that for the {\bf NO} case, electron final-states are penalized with respect to those into muons and taus 
 and one can have sizeable branching ratios for $\mu\mu$ and $\mu\tau$ final states. On the other hand, for the case of {\bf IO}, the $ee$ final state is dominant.
 One sees that like-sign di-muon and di-tau branching fractions correlate with the atmospheric octant in different ways depending on the neutrino mass ordering.

Note that, although $H^{\pm\pm}$ pair production at $pp$ or $e^+e^-$ colliders is gauge-mediated, its decays to like-sign dileptons are determined
by the Yukawa coupling $Y_\Delta$. 
This coupling matrix is also responsible for inducing \lfv processes at low energies.
As a result one expects that collider observables such as the cross sections $\sigma(pp/e^+e^-\to H^{\pm\pm}H^{\mp\mp}\to \ell_i^{\pm}\ell_j^{\pm}\ell_k^\mp\ell_m^\mp)$,
where at least one of the final-state leptons differs from the others,
may strongly correlate with low-energy flavour violating phenomena such as neutrino oscillations or rare $\ell_i\to \ell_j\gamma$ decays.
\begin{figure}
\begin{center}
\includegraphics[width=0.40\textwidth]{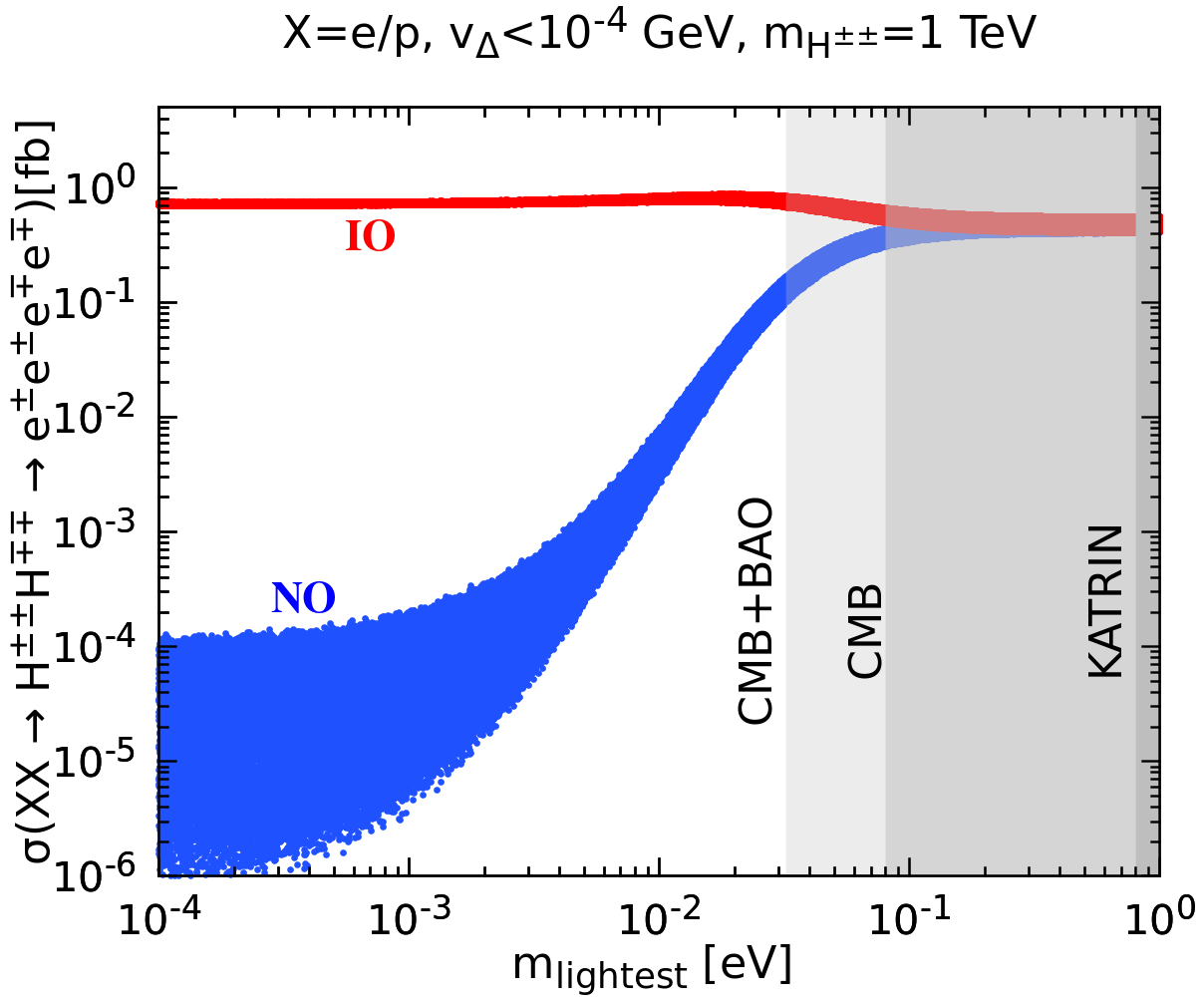}
\includegraphics[width=0.40\textwidth]{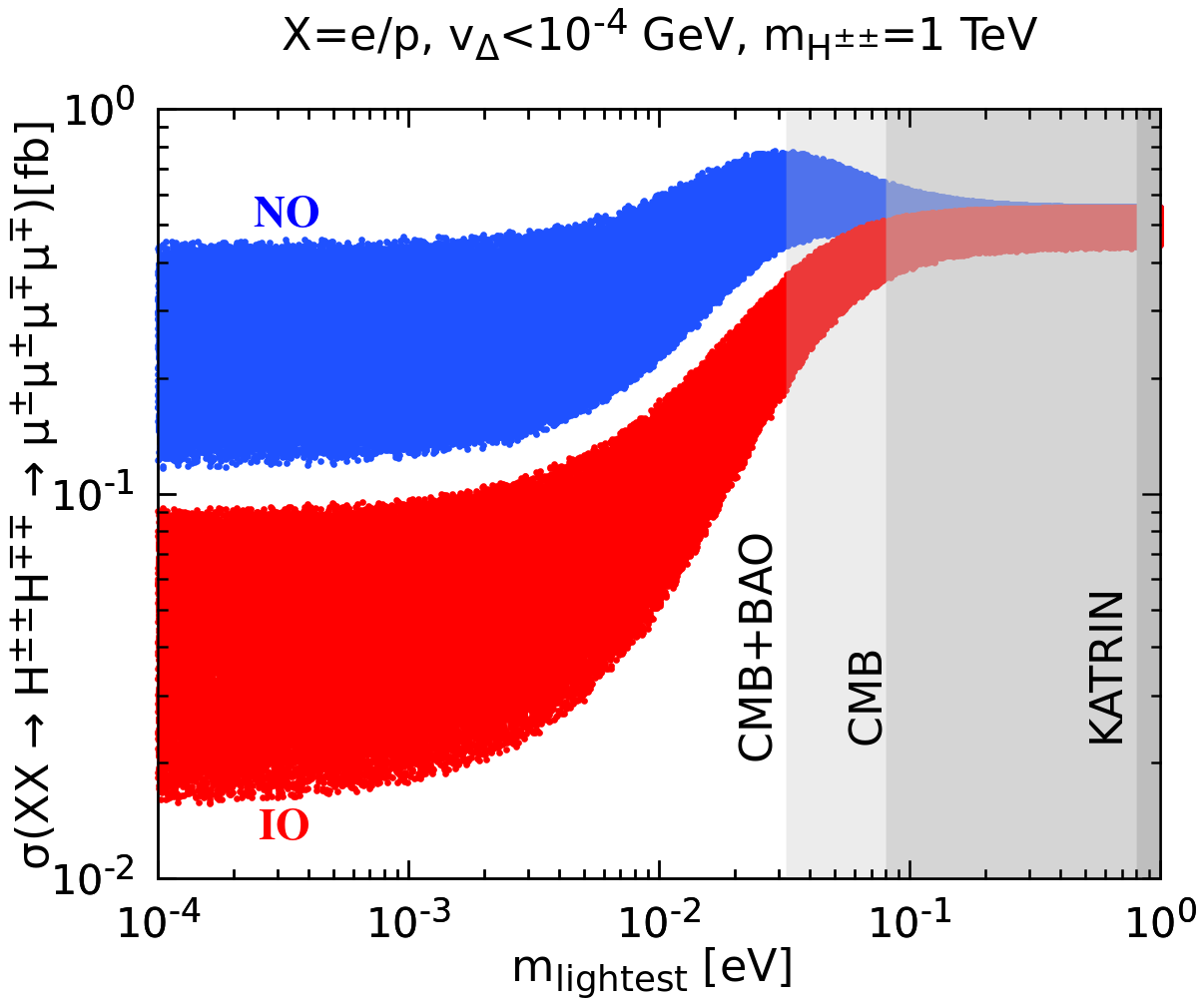}
\includegraphics[width=0.40\textwidth]{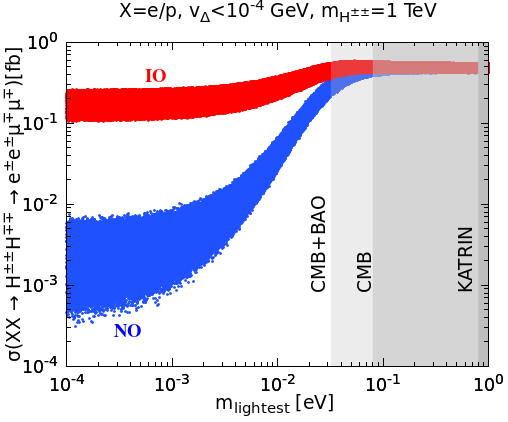}
\includegraphics[width=0.40\textwidth]{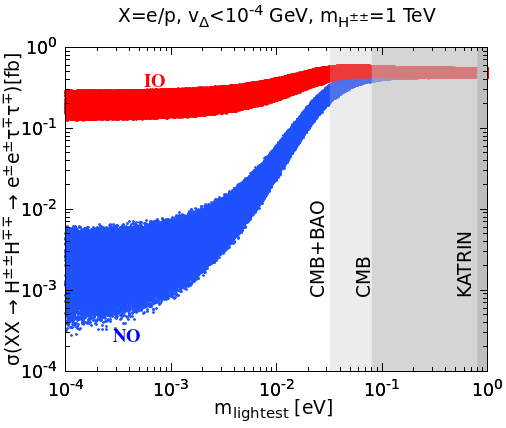}
\includegraphics[width=0.40\textwidth]{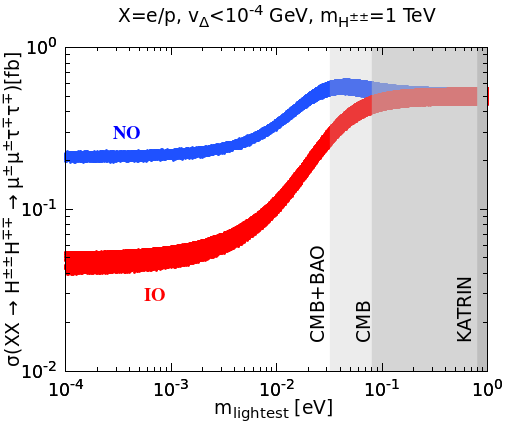}
\end{center}
\caption{
  Probing neutrino mass and mass ordering in neutrino oscillations through the 4-lepton cross sections such as $e^{\pm}e^{\pm}e^{\mp}e^{\mp}$, $\mu^{\pm}\mu^{\pm}\mu^{\mp}\mu^{\mp}$, $e^{\pm}e^{\pm}\mu^{\mp}\mu^{\mp}$, $e^{\pm}e^{\pm}\tau^{\mp}\tau^{\mp}$ and $\mu^{\pm}\mu^{\pm}\tau^{\mp}\tau^{\mp}$. 
  Red points correspond to inverted neutrino mass-ordering {\bf IO}, while blue points correspond to {\bf NO}. 
  Here we chose the doubly-charged Higgs mass as $m_{H^{\pm\pm}}=1$~TeV, suitable for a generic X-collider that can be either FCC-hh~\cite{FCC:2018vvp} or a lepton
  collider~(CLIC~\cite{CLICdp:2018cto}/FCC-ee~\cite{FCC:2018evy}/ILC~\cite{Barklow:2015tja}, CEPC~\cite{CEPCStudyGroup:2018ghi}).
  Oscillation parameters are varied within their $3\sigma$ ranges~\cite{deSalas:2020pgw,10.5281/zenodo.4726908}. % and the triplet VEV is taken as $v_\Delta \leq 10^{-4}\,\text{GeV}$.
  Vertical shaded bands indicate the KATRIN \cite{Aker:2021gma}, CMB \cite{Aghanim:2018eyx}, and CMB+BAO \cite{eBOSS:2020yzd} limits on the lightest neutrino mass. We see from these results how one can probe the mass ordering through these multilepton final sates at future colliders. 
}  
    \label{fig:BRLFVandCross}
  \end{figure}

  In Fig.~\ref{fig:BRLFVandCross} we display $\sigma(pp,e^+ e^-\to H^{\pm\pm}H^{\mp\mp}\to eeee/\mu\mu\mu\mu/ee\mu\mu/ee\tau\tau/\mu\mu\tau\tau)$ versus the lightest neutrino mass,
  both for {\bf NO} and {\bf IO} mass spectra. We fix the doubly-charged Higgs mass at $m_{H^{\pm\pm}} = 1$ TeV, so that both the 3 TeV $e^+e^-$ as well as 100 TeV $pp$ colliders
  have similar cross section (5 fb), as seen in Figs.~\ref{fig:cross-section100TeV} and \ref{fig:cross-section-ILC}. We denote by ``X'' such future ``generic'' collider.
  As before, rather than fixing reference values for the oscillation parameters, we varied them within their allowed $3\sigma$ ranges~\cite{deSalas:2020pgw,10.5281/zenodo.4726908},
  keeping the triplet VEV small, $v_\Delta< 10^{-4}$~GeV.
In this case the Yukawa coupling $Y_\Delta$ is large, so $H^{\pm\pm}$ decays predominantly to like-sign dileptons,
hence the four-lepton cross sections can be rather large. 
 Indeed, the cross sections into final states such as $e^{\pm}e^{\pm}e^{\mp}e^{\mp}$, $e^{\pm}e^{\pm}\mu^{\mp}\mu^{\mp}$ and
$e^{\pm}e^{\pm}\tau^{\mp}\tau^{\mp}$~($\mu^{\pm}\mu^{\pm}\mu^{\mp}\mu^{\mp}$ and $\mu^{\pm}\mu^{\pm}\tau^{\mp}\tau^{\mp}$) are always larger for {\bf IO}~({\bf NO})
and also that the difference increases with a smaller mass of the lightest neutrino.
Moreover, comparing blue and red points in all of the panels, one notes that the magnitudes of this multilepton signal differences for normal 
and inverted neutrino mass ordering strongly correlate with the magnitude of the lightest neutrino mass.
This provides a way to probe the neutrino mass ordering through these four-lepton final states in a robust manner, throughout the parameter region allowed by neutrino oscillations.

%\newpage
\underline{\bf Lepton flavour violation first at colliders?}\\[-.4cm]

It was suggested in the LEP days that charged lepton flavour~\cite{Bernabeu:1987gr} and leptonic CP symmetry~\cite{Rius:1989gk} non-conservation could be first seen at high-energies.
  %  %
  This proposal was scrutinized in the context of the LEP-1 experiments and subsequently revived in~\cite{AguilarSaavedra:2012fu,Das:2012ii,Deppisch:2013cya}
  for the case of the LHC~\cite{ATLAS:2014vur} (see also~\cite{Cottin:2021lrq,Beltran:2021hpq,Kriewald:2021qej,FCC:2018evy,FCC:2018byv} for recent references).
  It has recently been noted that the type-II seesaw mechanism provides a neater seesaw realization of the same idea~\cite{Miranda:2022xbi}.
We now provide a complete discussion including also the relevant theoretical aspects.
First of all, comparing Figs.~\ref{fig:BRHppdiag}, \ref{fig:BRHppoffdiag} and Fig.~\ref{fig:Atmos} one sees that same-flavour ($\mu\mu$) and
different-flavour ($\mu\tau$) doubly-charged Higgs decay branching fractions have very similar values.
Hence one can conclude that within the type-II seesaw mechanism, lepton flavour violating 4-lepton cross sections need not be suppressed when compared to those for same-flavour leptonic final states.

%%%%%%%%%%%%%%%%%%%%%%%%%%%%%%%%%%%%%%%%%%%%%%%%%%
\begin{figure}[!h]
\begin{center}
\includegraphics[width=0.40\textwidth]{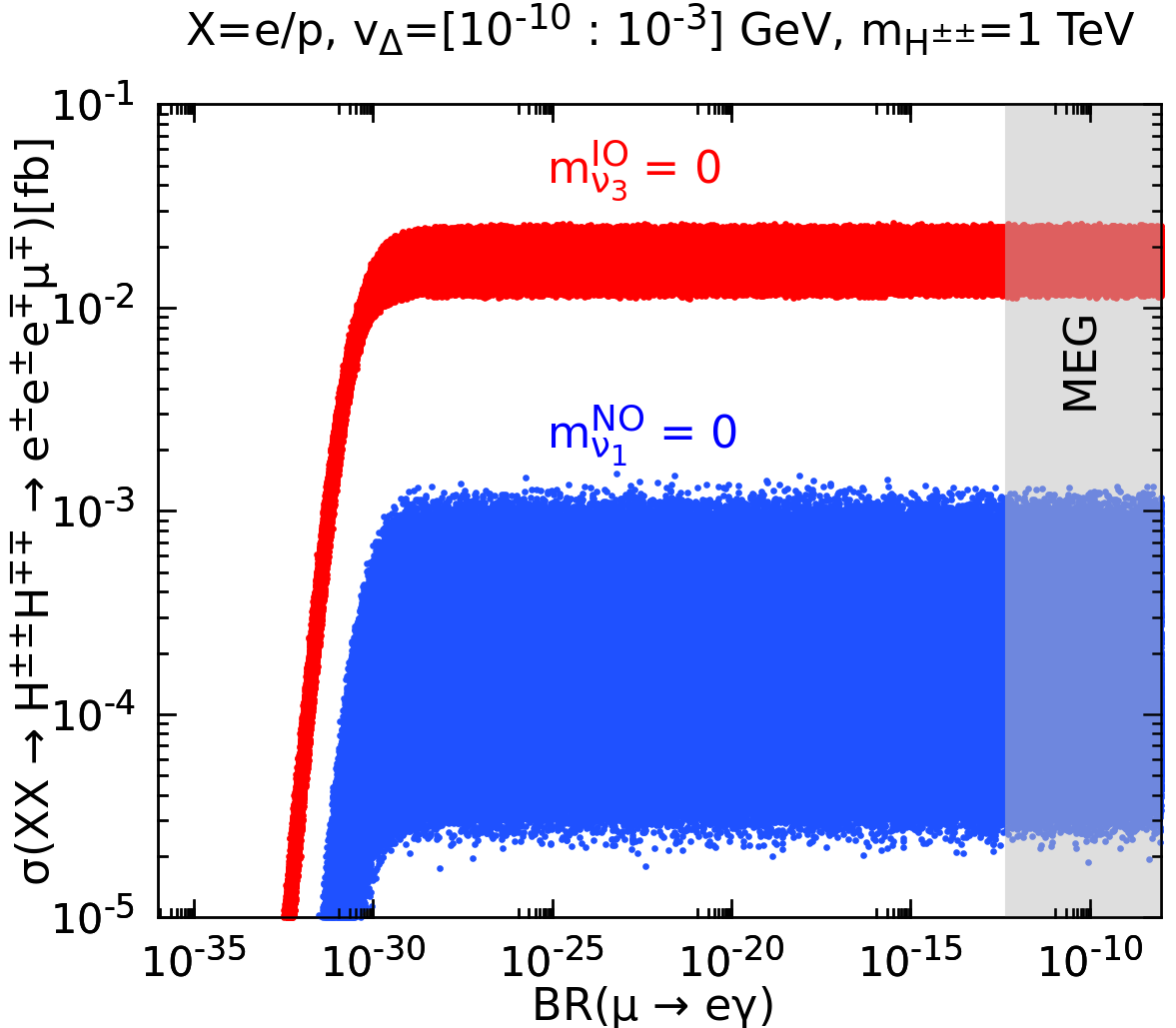}
\includegraphics[width=0.40\textwidth]{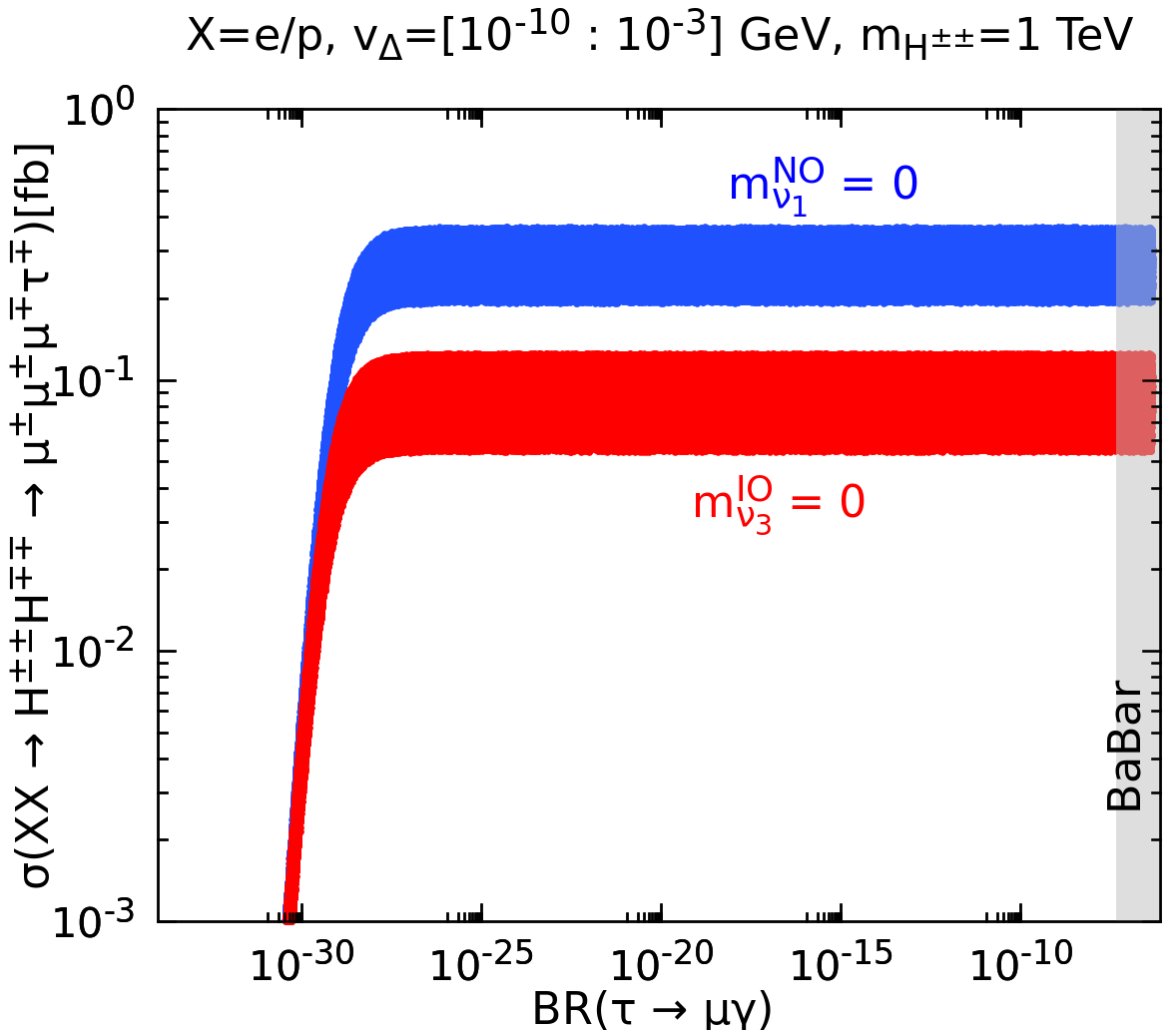}
\includegraphics[width=0.40\textwidth]{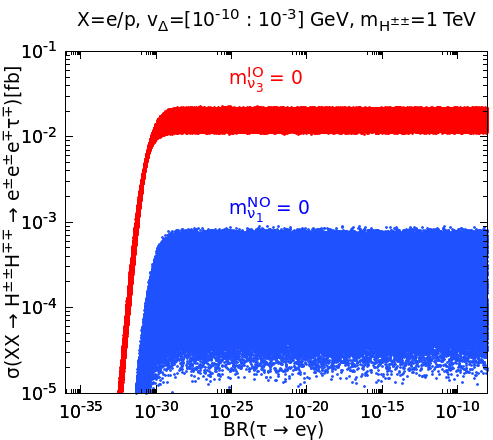}
\includegraphics[width=0.40\textwidth]{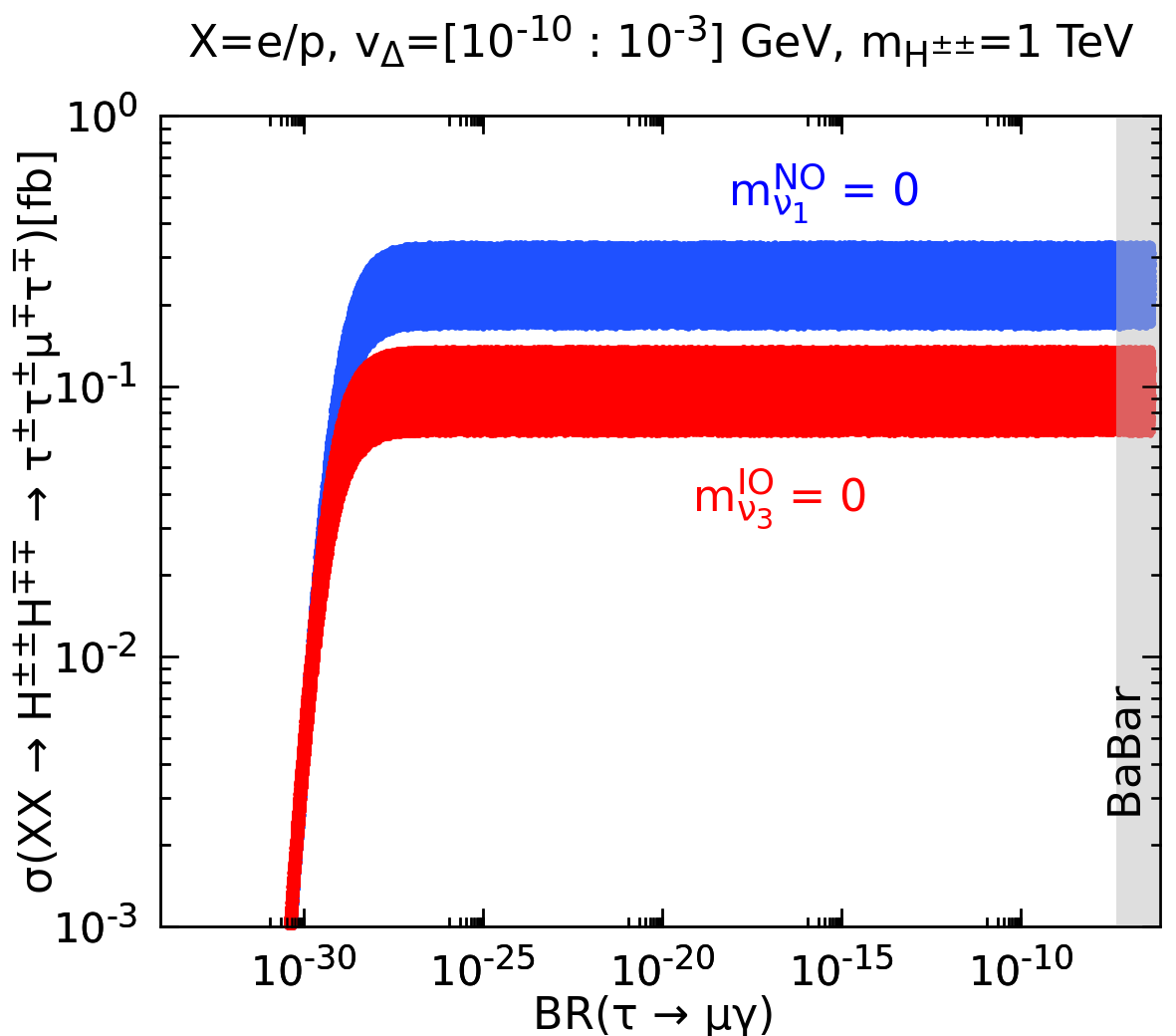}
\includegraphics[width=0.40\textwidth]{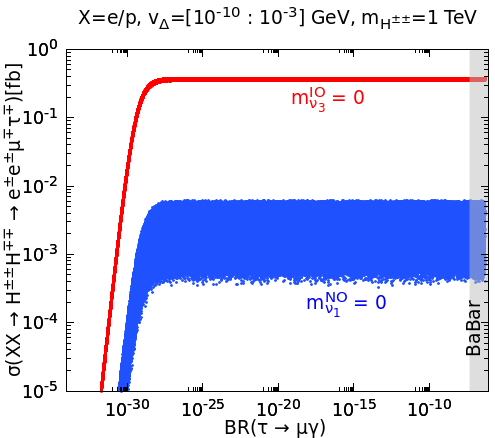}
\end{center}
\caption{\Clfvg multilepton cross sections versus $\text{BR}(\mu\to e\gamma)$, $\text{BR}(\tau\to e\gamma)$ or $\text{BR}(\tau\to \mu\gamma)$. 
  Red points correspond to {\bf IO}, while blue points correspond to {\bf NO}. 
  We have varied the oscillation parameters within their $3\sigma$ ranges~\cite{deSalas:2020pgw,10.5281/zenodo.4726908}. 
  The gray band is excluded by the MEG limit~\cite{MEG:2013oxv} or the BaBar limit~\cite{BaBar:2009hkt}. We see specially that for final states such as $e^{\pm}e^{\pm}e^{\mp}\mu^{\mp}$, $e^{\pm}e^{\pm}e^{\mp}\tau^{\mp}$ and $e^{\pm}e^{\pm}\mu^{\mp}\tau^{\mp}$ one can clearly distinguish {\bf NO} from {\bf IO} for all ``interesting'' values of $\text{BR}(\mu\to e\gamma)$, $\text{BR}(\tau\to e\gamma)$ and $\text{BR}(\tau\to \mu\gamma)$, respectively.}
    \label{fig:BRLFVandCross0}
  \end{figure}
\begin{figure}
\begin{center}
\includegraphics[width=0.49\textwidth]{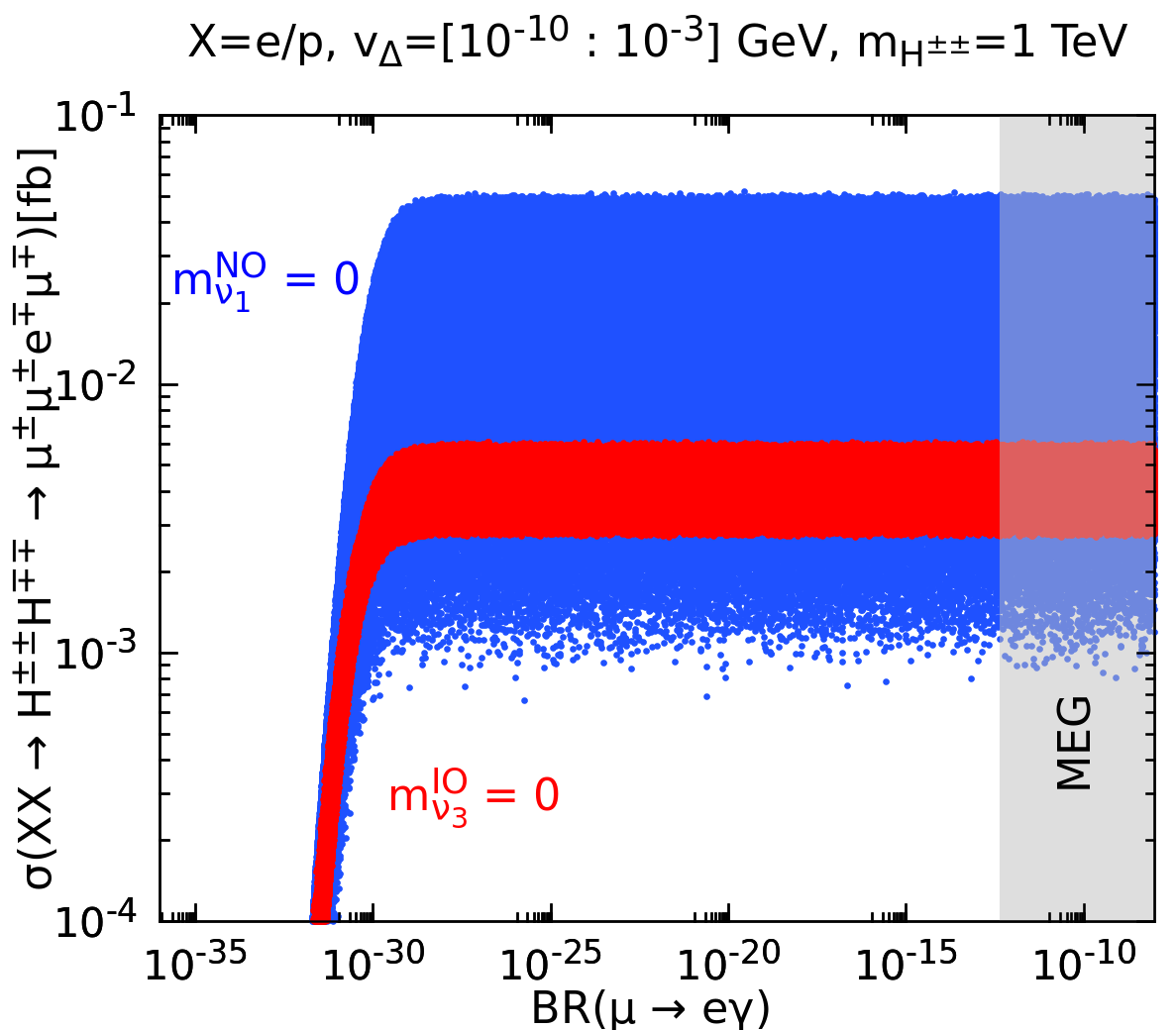}
\includegraphics[width=0.49\textwidth]{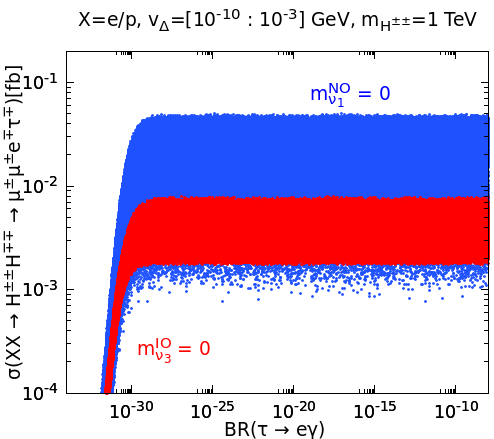}
\end{center}
\caption{
  \Clfvg multilepton cross sections versus $\text{BR}(\mu\to e\gamma)$ or $\text{BR}(\tau\to e\gamma)$. Although these \Clfvg cross sections are sizeable even for very tiny $\text{BR}(\mu\to e\gamma)$ or $\text{BR}(\tau\to e\gamma)$ values, here it is difficult to distinguish the mass orderings, as they overlap. }
    \label{fig:BRLFVandCross1}
  \end{figure}

  The fact that, as mentioned, \clfv may be first observed as a high-energy phenomenon is seen neatly from Fig.~\ref{fig:BRLFVandCross0}, where we show the correlation between
    $\sigma(e^+ e^-\to H^{\pm\pm}H^{\mp\mp}\to eee\mu/\mu\mu \mu\tau/\tau\tau\mu\tau/eee\tau/ee\mu\tau)$ and $\text{BR}(\mu\to e\gamma)$, $\text{BR}(\tau\to \mu\gamma)$ or $\text{BR}(\tau\to e\gamma)$
    both for {\bf NO} and {\bf IO}. For simplicity we took the lightest neutrino mass $m_{\nu_{1(3)}}=0$.
    Indeed, one sees that the various flavour-violating multi-lepton cross-sections can be sizeable,
      even when the low-energy $\text{BR}(\mu\to e\gamma)$, $\text{BR}(\tau\to \mu\gamma)$ or $\text{BR}(\tau\to e\gamma)$ decay branching ratios lie well below detectability.

    In Fig.~\ref{fig:BRLFVandCross0}, comparing red and blue points, we see that the cross section into $e^{\pm}e^{\pm}e^{\mp}\mu^{\mp}$, $e^{\pm}e^{\pm}e^{\mp}\tau^{\mp}$ and $
      e^{\pm}e^{\pm}\mu^{\mp}\tau^{\mp}$ final states (see left panels and lower one) is always larger in the case of {\bf IO} compared to {\bf NO},
      whereas for final state such as $\mu^{\pm}\mu^{\pm}\mu^{\mp}\tau^{\mp}$ and $\tau^{\pm}\tau^{\pm}\mu^{\mp}\tau^{\mp}$ the behaviour is the opposite.
      Moreover, although the cross sections can be large they are somewhat less sensitive to mass ordering. 

  Similarly, in Fig.~\ref{fig:BRLFVandCross1} we show how the 4-lepton cross sections $\sigma(e^+ e^-\to H^{\pm\pm}H^{\mp\mp}\to \mu^\pm\mu^\pm e^{\mp} \mu^\mp)$ (left panel) and
  $\sigma(e^+ e^-\to H^{\pm\pm}H^{\mp\mp}\to \mu^{\pm}\mu^{\pm}e^{\mp}\tau^{\mp})$ (right panel) correlate with $\text{BR}(\mu\to e\gamma)$ and $\text{BR}(\tau\to e\gamma)$, respectively.
  One sees that although the cross-sections for these flavour-violating four-lepton final states can be sizeable even for tiny values of
  $\text{BR}(\mu\to e\gamma)$ or $\text{BR}(\tau\to e\gamma)$, these can not probe the inverted mass ordering, since its prediction for the
  cross section can also occur for the case of normal ordering. However, predicted lepton flavour violating multi-lepton cross section values
  within the blue region and outside the red one would be a signature for normal mass ordering.
  This can be understood from the doubly-charged Higgs decay branching ratios into flavor conserving and flavor violating channels, depicted in
  Figs.~\ref{fig:BRHppdiag} and \ref{fig:BRHppoffdiag}.
We also note that there are other multilepton final states, such as $\tau^{\pm}\tau^{\pm}e^{\mp}\mu^{\mp}$ and $\tau^{\pm}\tau^{\pm}e^{\mp}\tau^{\mp}$, whose cross sections
exhibit a similar behaviour as the one discussed above. 

 In conclusion, we have shown that multilepton signatures can be used to probe the neutrino mass ordering.
  Even when not useful as a probe of the neutrino mass ordering, our four-lepton cross sections to different-flavour
  leptonic final states can probe the fundamental flavour structure of the neutrino sector, and perhaps provide
  a way of reconstructing the neutrino mixing matrix at collider experiments~\footnote{For example, this happens in supersymmetric models with bilinear violation of R-parity~\cite{Romao:1991ex,Hirsch:2000ef,Diaz:2003as,Hirsch:2004he}, where the atmospheric mixing angle can be independently probed at collider experiments~\cite{deCampos:2007bn,DeCampos:2010yu,deCampos:2012pf}.}. 

\section{non-standard neutrino interactions }
\label{sec:nsi-from-type}

A feature of low-scale neutrino mass generation models is the existence of potentially sizeable non-standard neutrino interactions (NSI).
These can be induced either at the tree level or through radiative corrections~\cite{Boucenna:2014zba},
and can manisfest phenomenologically in a variety of ways~\cite{Wolfenstein:1977ue,GonzalezGarcia:1998hj,Fornengo:1999zp,Guzzo:2001mi,Huber:2001zw,Huber:2001de,Huber:2002bi,Barranco:2005ps,
  Barranco:2007ej,Deniz:2010mp,Lindner:2016wff,Akimov:2017ade,Coloma:2017ncl,Liao:2017uzy,Kosmas:2017tsq,Denton:2018xmq,Altmannshofer:2018xyo,Miranda:2019skf,Ohlsson:2012kf,Miranda:2015dra,Farzan:2017xzy,Proceedings:2019qno}.
 Neutral-current NSI (denoted as NC-NSI) can be expressed as follows: 
\begin{equation}
{\cal L}_{{\rm NC-NSI}}=-2\sqrt{2}G_{F}\overline{\nu}_{\alpha}\gamma^{\mu}{\cal P}_{L}\nu_{\beta}\overline{f}\gamma_{\mu}{\cal P}_{X}\epsilon_{\alpha\beta}^{f X}f,\label{eq:NSI_def_NC}
\end{equation}
factorizing $G_{F}$, for convenience.  They lead to flavour changing (FC) neutrino-matter interactions, as well as to non-universal (NU) interactions.
Note that in the above equation ${\cal P}_{L}=\left(1-\gamma^{5}\right)/2$, ${\cal P}_{R}=\left(1+\gamma^{5}\right)/2$, $X=L$ or $R$, $\nu_{\alpha}$ are neutrinos of flavor $\alpha$
and $\beta$ ($\alpha,\ \beta=e$, $\mu$, $\tau$) respectively. Here $f$ denote charged SM fermions such as electrons ($e^{-}$) and quarks ($u$, $d$).  
The NC-NSIs  in Eq.~(\ref{eq:NSI_def_NC}) are particularly responsible for the scattering processes $\nu_{\alpha}+f\to\nu_{\beta}+f$.
Therefore they induce matter effects on neutrino oscillations, modifying their propagation properties~\cite{Wolfenstein:1977ue},
as well as the elastic neutrino-electron scattering \cite{Barranco:2007ej,Deniz:2010mp}\footnote{Neutrino-nucleus scattering, e.g. CEvNS~\cite{Akimov:2017ade,Lindner:2016wff,Coloma:2017ncl,Liao:2017uzy,Kosmas:2017tsq,Denton:2018xmq,Altmannshofer:2018xyo,Miranda:2019skf} is not affected in our case,
  as the triplet couplings are purely leptonic.}.

\begin{figure}[!h] \centering
\includegraphics[width=0.7\textwidth]{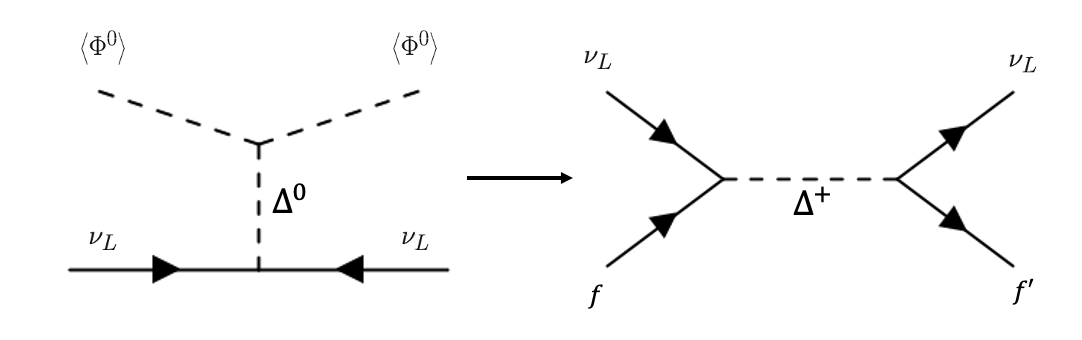}
\caption{\label{fig:nuNSI} Effective non-standard neutrino interactions arising in generic neutrino mass generation schemes. }
 \end{figure}

Within the simplest type-II seesaw, the same triplet that mediates neutrino mass generation also induces NSI as illustrated in Fig. \ref{fig:nuNSI}. 
% %
To see this, notice that the singly-charged Higgs of the triplet $\Delta$ has Yukawa interactions with neutrinos and charged leptons described by the Lagrangian shown previously in
Eq.~(\ref{app:x-3}). More specifically  
\begin{equation}
{\cal L}_{\Delta^{\pm}}=\sqrt{2}Y_{\Delta\alpha\beta}\nu_{L\alpha}^{T}C\ell_{L\beta}\Delta^{+}+{\rm h.c.},\label{eq:-7}
\end{equation}

 Integrating out the physical field $H^{\pm} \approx \Delta^{\pm}$, and applying Fierz transformations\footnote{In the Dirac notation, we have the Fierz transformation
   $\overline{f}_{1}{\cal P}_{L}f_{2}\overline{f}_{3}{\cal P}_{R}f_{4}=-\frac{1}{2}\overline{f}_{1}\gamma_{\mu}{\cal P}_{R}f_{4}\overline{f}_{3}\gamma^{\mu}{\cal P}_{L}f_{2}$.}
 to this charged current interaction, we get the effective operator
\footnote{There is a rotation between doublet and triplet Higgs scalars, but the rotation angle is neutrino-mass suppressed.}:
\begin{equation}
{\cal L}\supset-\frac{\textcolor{red}{i}}{m_{H^{\pm}}^{2}}Y_{\Delta\alpha'\beta'}^{*}Y_{\Delta\alpha\beta}\overline{\nu_{L\alpha'}}\gamma^{\mu}\nu_{L\alpha}\overline{\ell_{L\beta'}}\gamma_{\mu}\ell_{L\beta}.
\label{eq:LagTypeIa}
\end{equation}
Taking $\beta=\beta'$ in Eq.~(\ref{eq:LagTypeIa}) we can identify the NC-NSI of neutrinos with charged leptons as given in Ref.~\cite{Malinsky:2008qn}.
Comparing Eq. (\ref{eq:LagTypeIa}) and Eq.~(\ref{eq:NSI_def_NC}), and again assuming $m_{H^{\pm\pm}} \approx m_{H^{\pm}} \approx M_{\Delta}^{2}$, one can extract the corresponding NC-NSI parameters: 
\begin{equation}
    \epsilon_{\alpha\beta}^{\ell L}=-\frac{Y_{\Delta\alpha \ell}^{*}Y_{\Delta\beta \ell}}{2\sqrt{2}G_F M_{\Delta}^{2}}.
\label{eq:-19}
\end{equation}

%\newpage
 
\underline{\bf Constraining NSI from cLFV}\\[-.4cm]

\begin{figure}[t] 
\begin{center}
\includegraphics[width=0.475\textwidth]{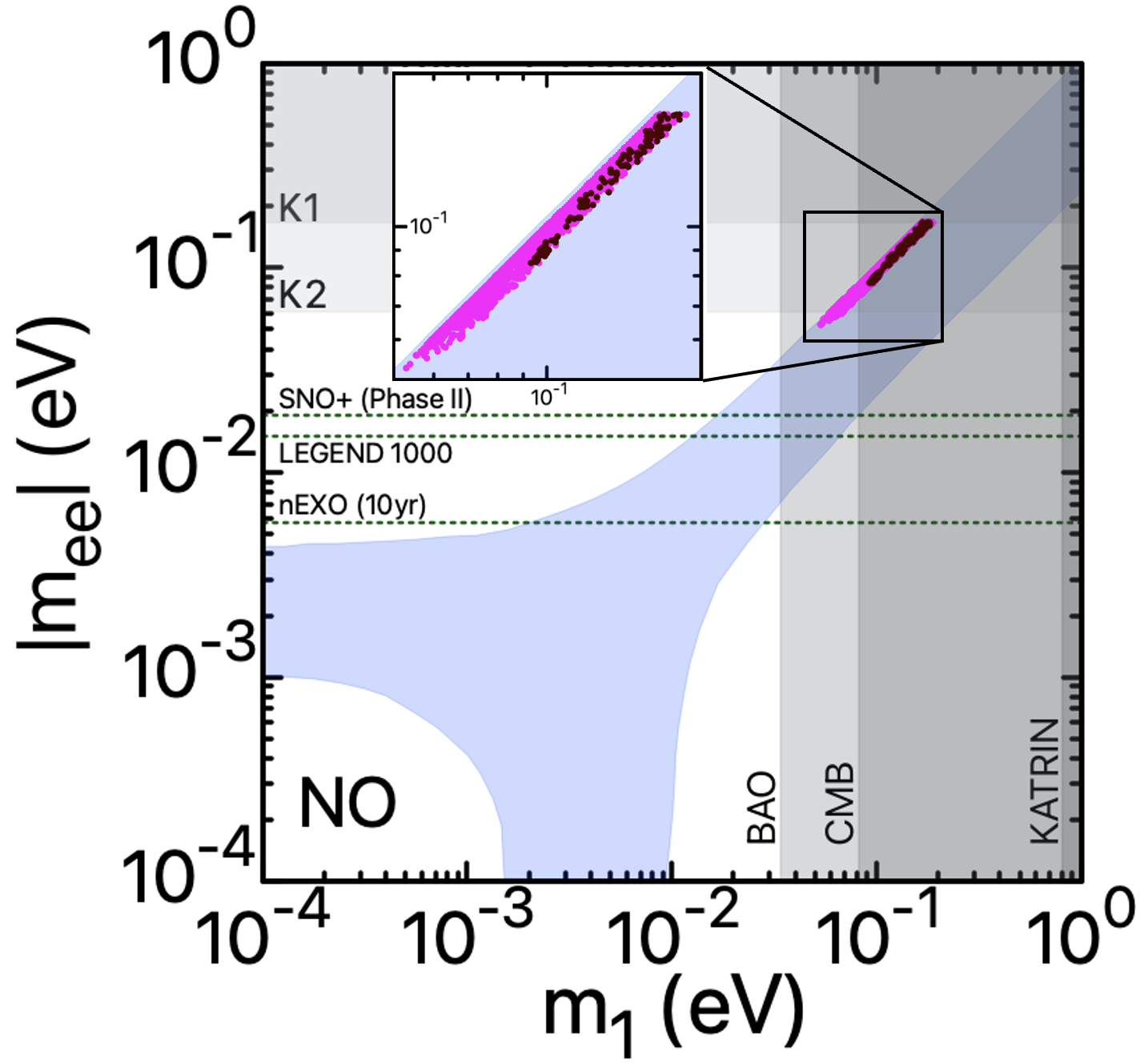}
\includegraphics[width=0.49\textwidth]{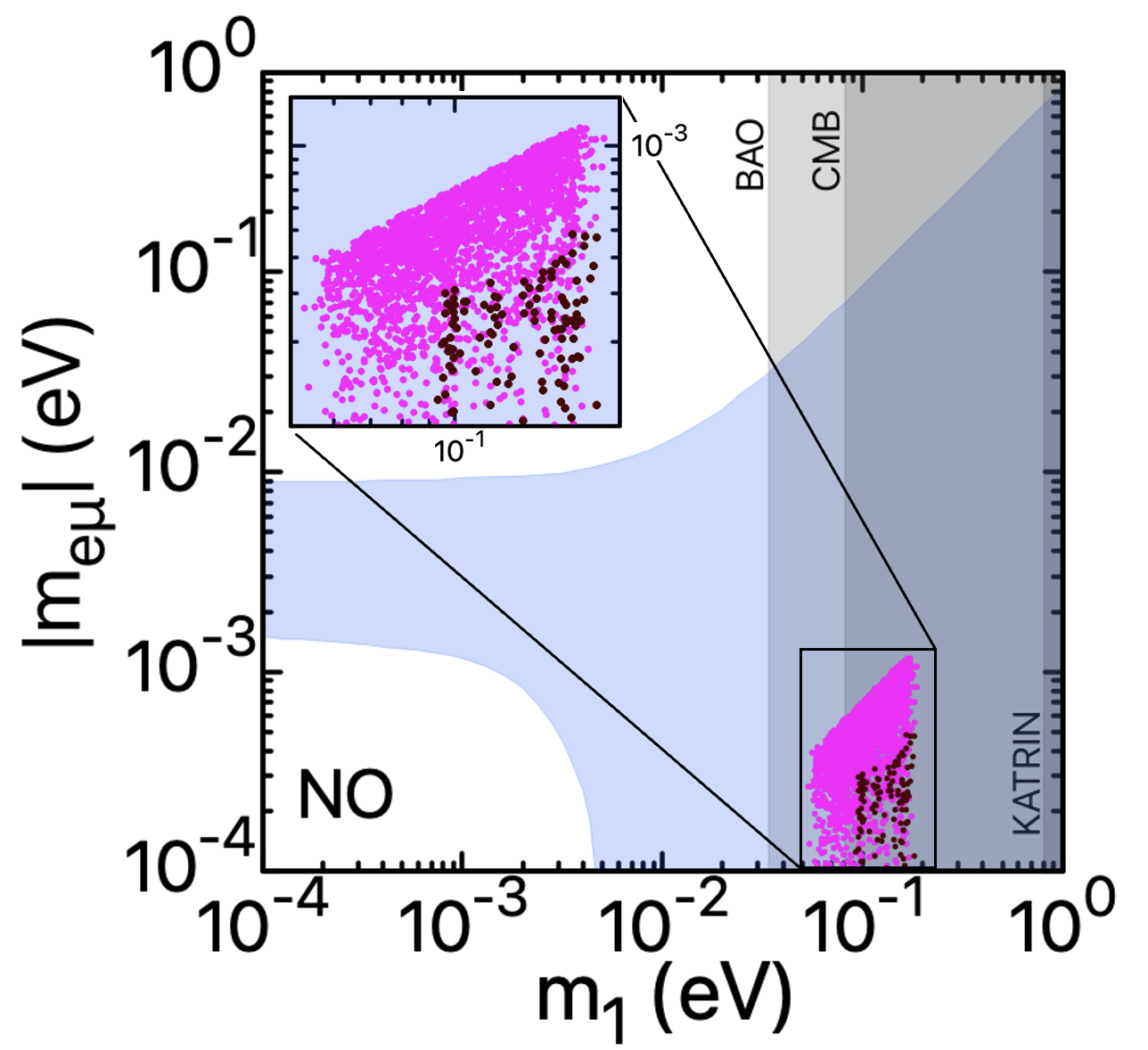}
\end{center}
\caption{
  The blue regions are the allowed neutrino mass matrix element magnitudes $|m_{ee}|$ (left) and $|m_{e\mu}|$ (right) in the flavor basis,
  as a function of $m_{1}$ for normal neutrino mass ordering ({\bf NO}).
  The mixing angles, squared mass differences, and $\delta$ are taken within their 3$\sigma$ limits as reported in \cite{deSalas:2020pgw}.
  The vertical shaded regions illustrate KATRIN's limit for $m_{1}$ as well as the cosmological limits.
  The left panel coincides with the effective mass from neutrinoless double beta decay $|m_{ee}|$ and the shaded horizontal bands are excluded by KAMLAND-Zen \cite{KamLAND-Zen:2016pfg}.
  Magenta dots represent the regions where the corresponding matrix elements satisfy the cLFV limits with the NSI parameter $|\varepsilon_{ee}^{eL} | > 1\times 10^{-4}$.
  Brown dots represent the analogue for $|\varepsilon_{e\tau}^{eL}|$.}
    \label{Flavor-Mass-scan}
\end{figure}

As in Fig.~\ref{fig:dbd-bounds} we can now determine the allowed regions for the magnitudes of the neutrino mass matrix elements obeying the constraints discussed in the sections above.
In order to determine the allowed modulus for each of the elements of the mass matrix in the flavor basis we have performed a scan over the parameter space. 
For this analysis we consider the range $m_1 < 0.8$~eV given by KATRIN's upper limit for $m_{\beta}$~\cite{Aker:2021gma}.
We generate random values for the neutrino mixing $U$ parameters within their 3$\sigma$ limits, taking into account the most recent global neutrino oscillation analysis~\cite{deSalas:2020pgw}.
We vary the Majorana phases randomly in the range between 0 and 2$\pi$.  
The results obtained by this scan for $|m_{ee}|$ and $|m_{e\mu}|$ are shown as blue regions in Fig.~\ref{Flavor-Mass-scan}. % 
They are given in the flavor basis as a function of $m_{1}$, for the case normal neutrino mass ordering. 
Notice that the left panel in the figure coincides wih the absolute value $|m_{ee}|$ of the effective Majorana neutrino mass in neutrinoless double beta decay.
Notice that for the determination of the allowed NC-NSI strength we have taken into account consistency with the triplet VEV and neutrino mass matrix restrictions,
Eqs. (\ref{V-triplet-approx}), (\ref{eq:s-18}), and (\ref{eq:-19}).
Moreover, we have also included consistency with LFV constraints in Table~\ref{table:lfv}, where we show the explicit Yukawa coupling combinations relevant for each of the decay modes,
and the corresponding experimental branching ratio limit denoted by $L_{i}$.  

One can determine the conditions that maximize the value of any of the six different NSI parameters, for example the diagonal NSI parameter $|\varepsilon_{ee}^{eL}|$.
From the previous definitions for $\varepsilon_{ee}^{eL}$ (Eq.~\ref{eq:-19}) and mass matrix element ($|m_{ee}|= v_{\Delta}|Y_{\Delta ee}|/ \sqrt{2} $) we have: 
\begin{equation}
    \frac{|m_{ee}|^{2}}{\sqrt{2}G_{F}} = |\varepsilon_{ee}^{eL}| \,  M_{\Delta}^{2}v_{\Delta}^{2}.
\label{M-times-v}
\end{equation}
\begin{table*}[t]
\begin{ruledtabular}
\begin{tabular}{cccc}
 NSI & Explicit Form & Estimated Limit ({\bf NO}) & Estimated Limit ({\bf IO})   \tabularnewline
\hline 
\vvvv $|\epsilon_{ee}^{e L}|$ & $(2\sqrt{2}G_{F})^{-1}M_{\Delta}^{-2} |Y_{\Delta e e}^{*}Y_{\Delta e e}|$ & $< 8.0\times 10^{-4}$ & $< 8.0\times 10^{-4}$\tabularnewline
\vvvv $|\epsilon_{e\mu}^{e L}|$ & $(2\sqrt{2}G_{F})^{-1}M_{\Delta}^{-2} |Y_{\Delta e e}^{*}Y_{\Delta\mu e}|$ & $< 7.0\times 10^{-7}$ & $< 7.0\times 10^{-7}$ \tabularnewline
\vvvv $|\epsilon_{e\tau}^{e L}|$ & $(2\sqrt{2}G_{F})^{-1}M_{\Delta}^{-2} |Y_{\Delta e e}^{*}Y_{\Delta \tau e}|$ & $< 2.0\times 10^{-4}$ & $< 2.1\times 10^{-4}$ \tabularnewline
\vvvv $|\epsilon_{\mu\mu}^{e L}|$ & $(2\sqrt{2}G_{F})^{-1}M_{\Delta}^{-2} |Y_{\Delta\mu e}^{*}Y_{\Delta\mu e}|$ & $< 6.8\times 10^{-6}$ & $< 2.5\times 10^{-6}$ \tabularnewline
\vvvv $|\epsilon_{\mu\tau}^{e L}|$ & $(2\sqrt{2}G_{F})^{-1}M_{\Delta}^{-2} |Y_{\Delta\mu e}^{*}Y_{\Delta\tau e}|$ & $< 4.8\times 10^{-6}$ &$< 2.5\times 10^{-6}$ \tabularnewline
\vvvv $|\epsilon_{\tau\tau}^{e L}|$ & $(2\sqrt{2}G_{F})^{-1}M_{\Delta}^{-2} |Y_{\Delta\tau e}^{*}Y_{\Delta\tau e}|$ & $< 9.5\times 10^{-5}$&  $< 9.9\times 10^{-5}$ \tabularnewline
\end{tabular}
\end{ruledtabular}
\caption{\label{tab:eps_lfv} Constraints on NSI parameters from cLFV processes. See text for details.}
\label{table:nsi}
\end{table*}
Besides NSI parameters, we also express cLFV branching ratios in terms of the neutrino mass matrix elements.
For example, from Table~\ref{tab:lfv} we see that from $\mu\to e\gamma$ we have the restriction: 
\begin{equation}
    \frac{|m^\dagger m|_{e\mu}}{G_F} < L_1 \,   M_{\Delta}^{2}v_{\Delta}^{2},
   \end{equation} 
   where $L_1$ is the corresponding limit quoted in Table~\ref{tab:lfv}. 
Using the previous equations to eliminate the product $M_{\Delta}v_{\Delta}$, we find:
\begin{equation}
  \frac{|m^{\dagger}m|_{\alpha\beta}}{|m_{ee}|^{2}} < \frac{L_{i}}{2\sqrt{2}~|\varepsilon_{ee}^{eL}|} ,
  \hspace{2cm}       i = 1,3,4
\label{Condition-1}
\end{equation} 
\begin{equation}
  \frac{|m^{\dagger}|_{\alpha\beta}|m|_{\sigma\rho}}{|m_{ee}|^{2}} < \frac{L_{i}}{2\sqrt{2}~|\varepsilon_{ee}^{eL}|} ,
  \hspace{1.5cm}       i \neq 1,3,4
\label{Condition-2}
\end{equation}
where $L_{i}$ are the limits found from cLFV in Table~\ref{table:lfv} and $\alpha$, $\beta$, $\sigma$, and $\rho$ represent the relevant flavors for each $L_i$, also given in Table~\ref{table:lfv}. 
In order to ensure an efficient parameter scan, we will use these ratios as a guidance to search for potentially large NSI parameters that satisfy all the cLFV experimental conditions,
as well as those from the neutrino data. We have performed the scan both for {\bf NO} as well as for the {\bf IO} case.
The latter is also viable with current oscillation data~\cite{deSalas:2020pgw}, and is characterized by having the lightest neutrino corresponding to $m_{3}$. \\[-.4cm]

To perform the scan, we start by studying the restrictions on NU-NSI parameters in the {\bf NO} case. 
In order to determine each of the $|m_{\alpha\beta}|$ we generate random values for the neutrino parameters in the mixing matrix $U$, within their 3$\sigma$ ranges from Ref.~\cite{deSalas:2020pgw}. 
Then we give different values of $|\varepsilon_{ee}^{eL}|$ until we find the maximum for which the ten conditions in Eqs. (\ref{Condition-1}) and
(\ref{Condition-2}) are satisfied. 
This gives us the strength of the NSI that is consistent with oscillation data as well as cLFV constraints. 
No combination of parameters was found with an NSI strength of order $10^{-3}$ or larger. 
However, we found several combinations in parameter space that induced an NSI of order $10^{-4}$, which are indicated in Fig. \ref{Flavor-Mass-scan} as magenta dots.
The panels in this figure give an idea of the $|m_{\alpha\beta}|$ elements required to have an NSI of order $10^{-4}$. 
For simplicity we only show $|m_{ee}|$ (left) and $|m_{e\mu}|$ (right), but similar plots can be obtained for each $|m_{\alpha\beta}|$. 
In the left panel we also indicate as shaded horizontal bands, marked as K1 and K2, those excluded by the KAMLAND-Zen collaboration~\cite{KamLAND-Zen:2016pfg}.
Notice that a diagonal NSI of order $10^{-4}$ is reachable for masses just below KAMLAND-Zen limits and the current cosmological limits from the CMB~\cite{Aghanim:2018eyx} (see figure). 
We followed the same procedure to maximize the possible value for each of the other two diagonal NSIs, for which we found upper limits below $10^{-4}$.
The best result for each case is shown in Table~\ref{tab:eps_lfv}.  
\begin{figure}[t]
\begin{center}
\includegraphics[width=0.5\textwidth]{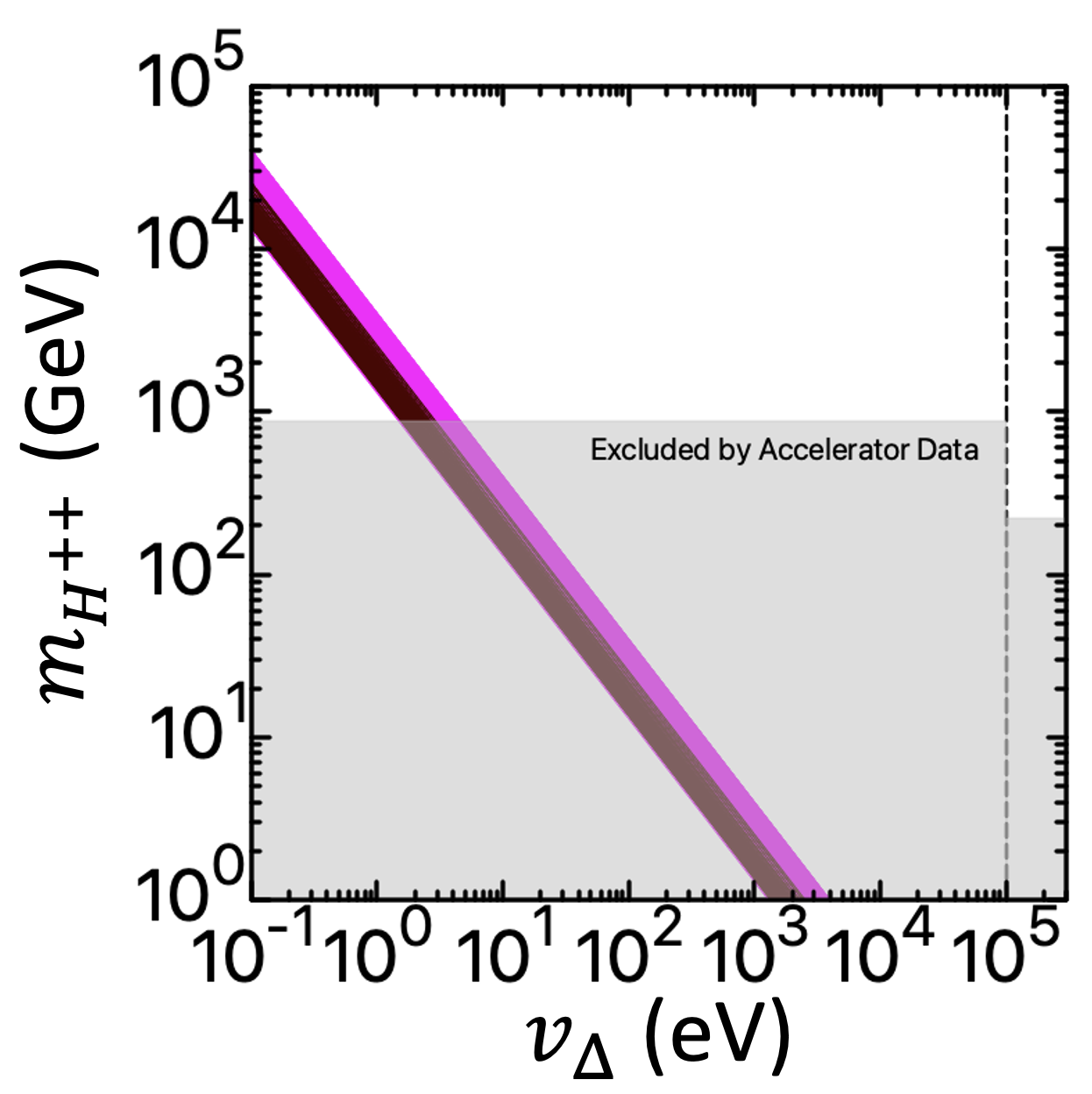}
\end{center}
\caption{
  Combinations of $M_{\Delta}$ and $v_{\Delta}$ where one can have $|\varepsilon_{ee}^{eL}| > 1\times 10^{-4}$ consistent with cLFV limits and normal neutrino mass ordering, {\bf NO}. 
  The gray region is excluded by collider data (see Section \ref{sec:constraints}), and the dashed line indicates $v_{\Delta} = 0.1$ MeV. 
  The brown points correspond to $|\varepsilon_{e\tau}^{eL}| > 1\times 10^{-4}$. For an explanation of the different excluded regions in grey, see text.}
    \label{Mv}
\end{figure}

For the case of non-diagonal NSIs, we follow a similar method.
First, by comparing the definition of the non-diagonal NSIs with the second, third, and forth rows in Table \ref{table:lfv}, one sees that we expect these NSIs to lie below 8$\times 10^{-4}$.
The most severely constrained is $|\varepsilon_{e\mu}^{eL}| < L_2 / (2\sqrt{2}) = 7.1\times 10^{-7}$, because of the stringent limits on $\mu \to 3e$.
Similarly, for the case of $|\varepsilon_{e\tau}^{eL}|$ we again generate random values for $U$ in order to produce the neutrino mass matrix elements $|m_{\alpha\beta}|$. 
Then we look for the maximum value of $|\varepsilon_{e\tau}^{eL}|$ so that Eqs. (\ref{Condition-1}) and (\ref{Condition-2}), with $ |m_{ee}|^{2} \to |m_{e\tau}^{*}||m_{ee}|$, are satisfied. 
Brown dots in Fig.~\ref{Flavor-Mass-scan} allow for $|\varepsilon_{e\tau}^{eL}| > 1\times 10^{-4}$. 
A similar analysis was done for $|\varepsilon_{\mu\tau}^{eL}|$. The maximum limits found with this method for all the NSI are given in Table~\ref{tab:eps_lfv}. 
All in all, there are only three NSIs that can exceed $ \approx 10^{-4}$, still far from the current sensitivity of reactor and solar neutrino experiments.
\begin{figure}[t]
\begin{center}
\includegraphics[width=0.493\textwidth]{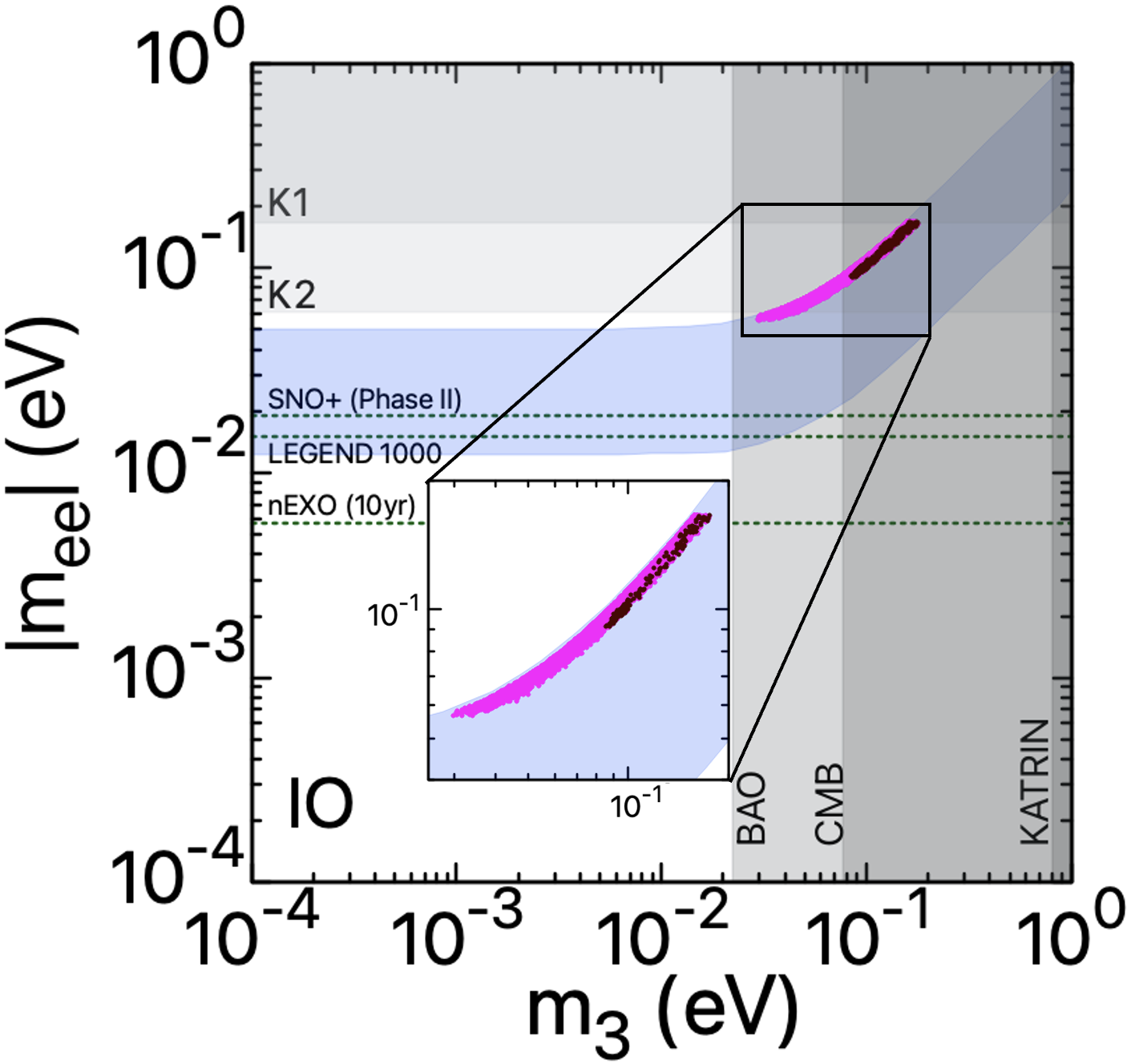}
\includegraphics[width=0.49\textwidth]{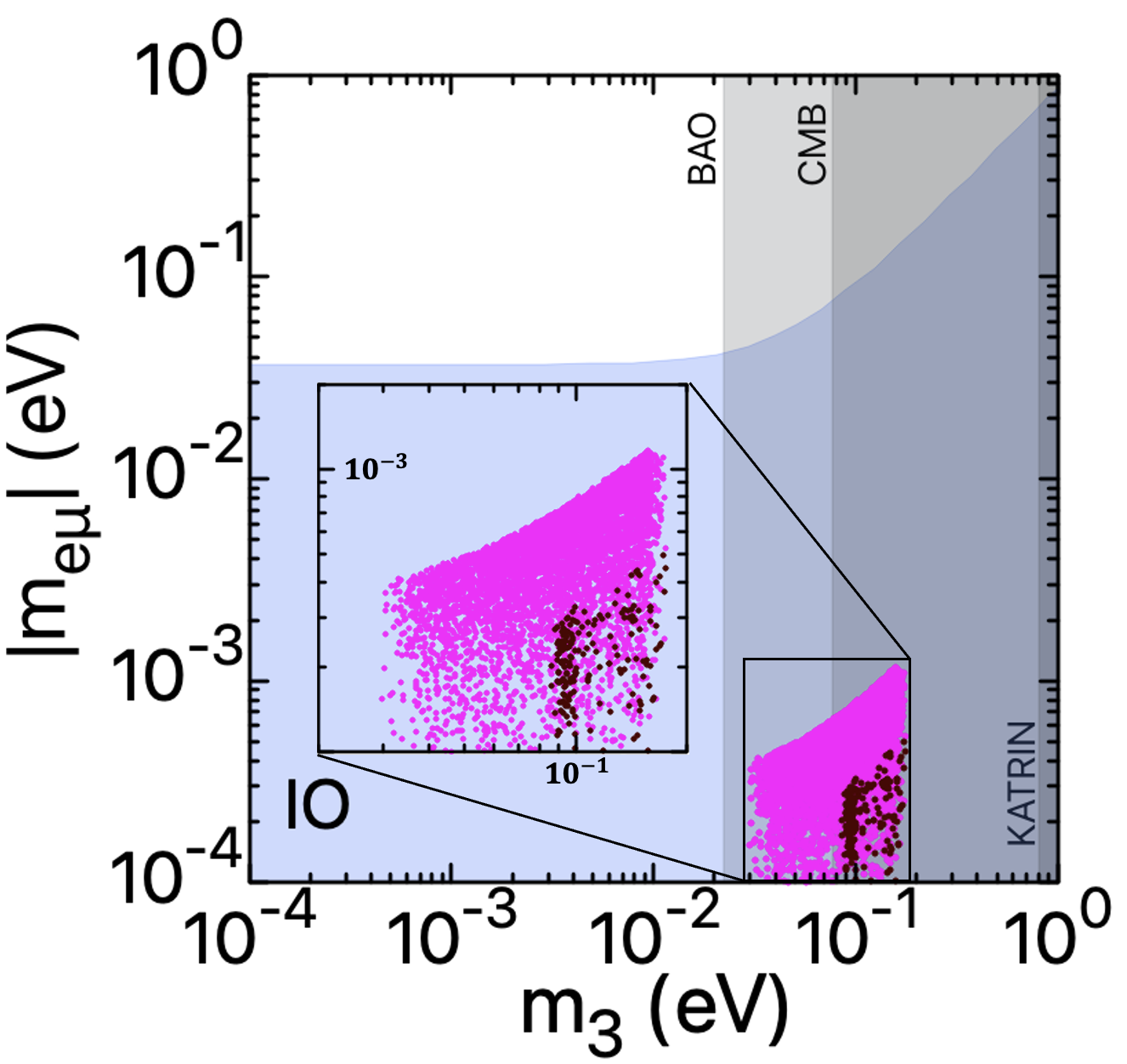}
\end{center}
\caption{
  Same as Fig.~\ref{Flavor-Mass-scan} but for the case of inverted neutrino mass ordering, {\bf IO}.
  Magenta dots denote the regions where the corresponding matrix elements obey cLFV limits with the NSI parameter $|\varepsilon_{ee}^{eL} | > 1\times 10^{-4}$.
  Brown dots correspond to $|\varepsilon_{e\tau}^{eL}|$.}
    \label{Flavor-Mass-IO}
\end{figure}

From the previous analysis, we can also get information about the triplet mass and VEV that could induce an NSI of a desired order. 
From Eq. (\ref{M-times-v}), notice that for a given NSI, there is an inverse relation between $M_{\Delta}$ and $v_{\Delta}$.
This means that there are different combinations of the product of these parameters that reproduce the same NSI of a desired order. 
For instance, the magenta dots in Fig.~\ref{Flavor-Mass-scan} correspond to combinations for which $|\varepsilon_{ee}^{eL}| > 1\times 10^{-4}$.
For each of these dots we can generate a plot $M_{\Delta}$ vs $v_{\Delta}$ with the NSI value fixed. 
Fig.~\ref{Mv} shows the resulting curves, which combined produce the magenta region of allowed values for the mass and the VEV that can induce an NSI of order $10^{-4}$.
The shaded grey regions in this figure correspond to the excluded regions for $m_{H^{++}}$ from collider constraints discussed in sec.~\ref{sec:collider-constraints}.
For lower triplet VEV $v_\Delta< 10^{5}$ eV, the doubly-charged Higgs mass constraint is $m_{H^{++}}> 870$ GeV, while for $v_\Delta > 10^{5}$ eV, we have a weaker bound $m_{H^{++}}> 220$ GeV.\\[-.3cm] 

Our analysis has been focused on the assumption of normal ordering of the neutrino masses.
The alternative case of inverse ordering leads to the blue regions in Fig.~\ref{Flavor-Mass-IO}. 
One sees the $|m_{ee}|$ (left) and $|m_{e\mu}|$ (right) values obtained from oscillation data, when varying the neutrino mixing parameters within their 3$\sigma$ ranges. 
Following a similar analysis we conclude that in the {\bf IO} case we can only have NSI above $10^{-4} $ and consistent with cLFV limits for $|\varepsilon_{ee}^{eL}|$ and $|\varepsilon_{e\tau}^{eL}|$. 
The magenta dots show the regions for $|\varepsilon_{ee}^{eL}|$, and the brown dots those for $|\varepsilon_{e\tau}^{eL}|$.
Table~\ref{tab:eps_lfv} summarizes the largest NSI strengths consistent with current experimental restrictions. 

Finally, we comment on the consistency of our NSI results with current limits on $\mu \to e$ conversion in nuclei~\cite{Kosmas:2000nj,Deppisch:2005zm}.
 Within the type-II seesaw mechanism, both the singly- and doubly-charged scalars contribute to $\mu \to e$ conversion.
 For relatively light nuclei the conversion rate is given by
\begin{equation}
\mathrm{BR}(\mu \to e)\approx \frac{\alpha^{5}}{36\pi^{4}}\frac{m_{\mu}^{5}}{\Gamma_{\mathrm{capt}}}Z_{\rm eff}^{4}ZF^{2}(-m_{\mu}^{2})\left | (Y_{\Delta}^{\dagger}Y_{\Delta})_{e\mu}\left [ \frac{5}{24m_{H^{+}}^{2}}+\frac{1}{m_{H^{++}}^{2}} \right ] + \frac{1}{m_{H^{++}}^{2}}\sum_{\alpha} Y^{\dagger}_{\Delta e\alpha}Y_{\Delta \alpha\mu} f(m_{\alpha},m_{H^{++}})\right |^{2},
\label{CR}
\end{equation}
where $F(q^{2})$ is the nuclear form factor, $m_{\mu}$ is the muon mass, $Z_{\rm eff}$ is an effective atomic charge, and the function $f$ is given by \cite{Raidal:1997hq,Ma:2000xh}:
\begin{equation}
f(m_{\alpha},m_{H^{++}}) = 4\frac{m_{\alpha}^{2}}{m_{\mu}^{2}}+ \mathrm{ln}\left ( \frac{m_{\alpha}^{2}}{m_{H^{++}}^{2}} \right ) + \left ( 1-2\frac{m_{\alpha}^{2}}{m_{\mu}^{2}}  \right )\sqrt{1+4\frac{m_{\alpha}^{2}}{m_{\mu}^{2}}}\mathrm{ln}\left ( \frac{\sqrt{m_{\mu}^{2}m_{H^{++}}^{-2}}+ \sqrt{m_{\mu}^{2}m_{H^{++}}^{-2}+4m_{\alpha}^{2}m_{H^{++}}^{-2}}}{\sqrt{m_{\mu}^{2}m_{H^{++}}^{-2}}- \sqrt{m_{\mu}^{2}m_{H^{++}}^{-2}+4m_{\alpha}^{2}m_{H^{++}}^{-2}}} \right ).
\end{equation}
For the case of Ti, the most stringent experimental limit comes from SINDRUM II, i.e. $\mathrm{BR}(\mu \to e) < 4.2\times 10^{-12}$ \cite{SINDRUMII:1993gxf}.
To test the consistency of this experimental bound with NSIs of order $10^{-4}$, we considered the {\bf NO} oscillation parameters for which
  $|\varepsilon_{ee}^{eL}| = 8\times10^{-4}$ is the largest allowed value from Table II. We found that the predicted branching ratio in
  Eq.~(\ref{CR}), $\mathrm{BR}(\mu \to e) = 1.6\times 10^{-13}$,  is one order of magnitude smaller than the experimental limit for this choice, assuming
  $m_{{H}^{++}}^{2} \approx m_{{H}^{+}}^{2}$ again, which is consistent with the current limit of SINDRUM II. 
Hence the results in Table~\ref{tab:eps_lfv} obtained for the $\mu \to e\gamma$ case seem unaffected by the inclusion of nuclear $\mu \to e$ conversion bounds.
A visible modification of the NSI sensitivity would require an improvement of the current experimental bound on the latter by over an order of magnitude.

\section{Summary and outlook}
\label{sec:outlook}

We have examined the possibility of vindicating and ``deconstructing'' the simplest type-II seesaw mechanism, with explicit violation of lepton number,
in which a Higgs triplet mediates neutrino mass generation in the Standard Model~\cite{Schechter:1980gr,Schechter:1981cv}, see Fig.~\ref{fig:neutrino-mass}.
We have first discussed the existing restrictions from electroweak precision data, i.e. from the $S$, $T$, $U$ parameters,
and electroweak symmetry breaking, including perturbative unitarity and stability of the vacuum, see Figs.~\ref{fig:quartic-from-neutrino-mass}, \ref{fig:contribution-lambda} and \ref{fig:RGE}.
Constraints from neutrino oscillations were examined in Fig.~\ref{fig:dbd-bounds} where one sees also present limits and future sensitivities from cosmology and neutrinoless double beta decay searches.

Charged \lfv resulting from diagrams in Figs.~\ref{fig:Tree-Level1} and \ref{fig:Tree-Level2} lead to stringent restrictions, especially from $\mu \to e\gamma$.
Note that, as seen in Fig.~\ref{fig:BRHpp}, in the limit of small triplet vacuum expectation value, the doubly-charged Higgs boson has dominant lepton number and \clfvg decays.
The scalars mediating neutrino mass generation may also be produced at present and future hadron colliders, see Figs.~\ref{fig:feyn-diagH}, \ref{fig:cross-section14TeV} and \ref{fig:cross-section100TeV}.
Indeed, we discussed current restrictions coming from collider searches. 
In addition, we have discussed some of the novel experimental opportunities associated to the scalar mediator of neutrino mass generation, 
extending the discussion recently given in~\cite{Miranda:2022xbi}.
The decay pattern of the doubly- and singly-charged Higgs bosons is shown in Figs. \ref{fig:BRHpp} and \ref{fig:BRHp}.
Such decays could lead to new impressive experimental signatures at high-energy hadron colliders, 
e.g. those given in Eqs.~(\ref{eq:pp-dmzero-set1}) and (\ref{eq:pp-dmzero-set3}), or in Eqs.~(\ref{eq:pp-dmnegative-set1}) and (\ref{eq:pp-dmnegative-set2}).
Likewise, these scalar mediators would also be produced at lepton colliders, the cross section for scalar production is given in Fig.~\ref{fig:cross-section-ILC} and Fig.~\ref{fig:cross-section-ILC-2}.
Especially interesting signatures in this case are those given in Eqs.~(\ref{eq:ee-to-lepton-or-jet}) and (\ref{eq:ee-to-fatjet}). For comparison, Fig.~\ref{fig:densityLFV} illustrates the cLFV sensitvities in terms of the scalar mass, and the triplet seesaw VEV.

A remarkable feature of the triplet seesaw mechanism is the possibility of full ``deconstruction'' free from the ambiguities that characterize the type-I seesaw variant and which 
are seen neatly in the Casas-Ibarra description~\cite{Casas:2001sr}. 
The latter simply reflect the large number of parameters characterizing the type-I seesaw mixing matrix, given in~\cite{Schechter:1980gr,Schechter:1981cv}.

As recently summarized in~\cite{Miranda:2022xbi}, one can probe neutrino oscillation physics at collider energies through the pattern of the type-II scalar boson mediator decays. 
This is seen, for example, in Figs.~\ref{fig:BRHppdiag}, \ref{fig:BRHppoffdiag} and \ref{fig:Atmos}, where one can see that the doubly-charged Higgs boson has
lepton-flavour-violating decay rates, comparable to the lepton-flavour-conserving ones. 

Indeed, Figs.~\ref{fig:BRLFVandCross0} and ~\ref{fig:BRLFVandCross1} clearly show that \clfv could be observed first as a high-energy phenomenon,
as the corresponding signal cross section can be sizeable even when low-energy
\clfvg processes, such as $\mu \to e\gamma$, have very small rates. High-energy probes clearly complement the efforts towards \clfv searches in other high-intensity facilities such as BaBar.

The simplest seesaw mechanism leads to promising signatures not only at hadron colliders such as LHC and FCC, but also at future $e^+ e^-$ colliders such as ILC, CLIC or CEPC in China.
Indeed, the type-II seesaw illustrates how high-energy signatures can be complementary to neutrino oscillation studies.
For example, Fig.~\ref{fig:BRLFVandCross} shows how the rates for multilepton final-state events coming from pair-production of the doubly-charged Higgs boson % $H^{\pm\pm}H^{\mp\mp}$
may be used as a probe of the neutrino mass ordering, currently not yet robustly determined by oscillation experiments.

The results found here illustrate the complementarity and interplay of the high-energy and high-intensity frontiers in particle physics, providing encouragement for dedicated simulation studies
in order to evaluate the potential of these proposed facilities in probing the neutrino sector.

We have also examined the possibility of having sizeable non-standard neutrino interactions (see Fig.~\ref{fig:nuNSI}) given the current constraints from \lfv searches.
We saw in Figs.~\ref{Flavor-Mass-scan}, \ref{Mv} and \ref{Flavor-Mass-IO} that effects in neutrino propagation associated to non-standard interactions are below the current detectability,
though they could lie above the $10^{-4}$ level, e.g. $|\varepsilon_{ee}^{eL} | > 1\times 10^{-4}$ providing a real challenge for the future.

\begin{acknowledgments}
  Work of J.V. is supported by the Spanish grants PID2020-113775GB-I00 (AEI/10.13039/501100011033) and PROMETEO/2018/165 (Generalitat Valenciana) and by CONACYT-Mexico under grant A1-S-23238. O. G. M. has been supported by SNI (Sistema Nacional de Investigadores). S.M. has been supported by KIAS Individual Grants (PG086001) at Korea Institute for Advanced Study. X.J.X has been supported in part by the National Natural Science Foundation of China under grant No. 12141501.
\end{acknowledgments}

\appendix 
\section{Renormalisation group equations}
\label{app:RGEs}
In our work we have used the package SARAH~\cite{Staub:2015kfa} to perform the renormalization group analysis of the type-II seesaw model.
The $\beta$ function of a given parameter $c$ is given by,
\begin{align*}
 \mu_r\frac{dc}{d\mu_r}\equiv\beta_{c}=\frac{1}{16\pi^{2}}\beta_{c}^{(1)}+\frac{1}{(16\pi^{2})^{2}}\beta_{c}^{(2)} \, .
\end{align*}
where $\mu_r$ is the running scale and $\beta_{c}^{(1)}$, $\beta_{c}^{(2)}$ are the one-loop and two-loop renormalization group terms.
Below we list only one-loop RGEs for various quartic couplings and Yukawa couplings $Y_\Delta$, $y_t$. \\

\underline{\bf Quartic couplings: }\\[-.4cm]
\begin{align}
\beta_{\lambda}^{(1)} & =\frac{27}{50}g_{1}^{4}+\frac{9}{5}g_{1}^{2}g_{2}^{2}+\frac{9}{2}g_{2}^{4}-\left(\frac{18}{5}g_{1}^{2}+18g_{2}^{2}\right)\lambda+6\lambda^{2}+12\lambda_{1}^{2}+12\lambda_{1}\lambda_{4}+5{\lambda_4}^{2}\nonumber \\
 & +12\lambda y_{t}^{2}-24y_{t}^{4}\,,\\
\beta_{\lambda_1}^{(1)} & =\frac{27}{25}g_{1}^{4}-\frac{18}{5}g_{1}^{2}g_{2}^{2}+6g_{2}^{4}-\left(\frac{9}{2}g_{1}^{2}+\frac{33}{2}g_{2}^{2}\right)\lambda_{1}+3\lambda \lambda_{1}+\lambda \lambda_4+4\lambda_{1}^{2}\nonumber \\
 & +16\lambda_{2}\lambda_{1}+12\lambda_{3}\lambda_{1}+{\lambda_{4}}^{2}+6\lambda_{2}\lambda_{4}+2\lambda_{3}\lambda_{4}+6\lambda_{1}y_{t}^{2} + 4 \lambda_1 \text{Tr}\Big({Y_\Delta^{\dagger}  Y_\Delta}\Big)\,,\\
\beta_{\lambda_4}^{(1)} & =\frac{36}{5}g_{1}^{2}g_{2}^{2}-\left(\frac{9}{2}g_{1}^{2}+\frac{33}{2}g_{2}^{2}\right)\lambda_{4}+\lambda \lambda_{4}+8\lambda_{1}\lambda_{4}+4{\lambda_{4}}^{2}+4\lambda_{2}\lambda_{4}\nonumber \\
 & +8\lambda_{3}\lambda_{4}+6\lambda_{4}y_{t}^{2} +4 \lambda_{4} \mbox{Tr}\Big({Y_\Delta^{\dagger}  Y_\Delta}\Big) \,,\\
\beta_{\lambda_2}^{(1)} & =\frac{108}{50}g_{1}^{4}-\frac{72}{10}g_{1}^{2}g_{2}^{2}+15g_{2}^{4}-\left(\frac{36}{5}g_{1}^{2}+24g_{2}^{2}\right)\lambda_{2}+2\lambda_{1}^{2}+2\lambda_{1}\lambda_{4}\nonumber \\
 & +28\lambda_{2}^{2}+24\lambda_{2}\lambda_{3}+6{\lambda_{3}}^{2} -4 \mbox{Tr}\Big({Y_\Delta^{\dagger}  Y_\Delta  Y_\Delta^{\dagger}  Y_\Delta}\Big)  + 8 \lambda_2 \mbox{Tr}\Big({Y_\Delta^{\dagger}  Y_\Delta}\Big) \,,\\
\beta_{\lambda_3}^{(1)} & =\frac{144}{10}g_{1}^{2}g_{2}^{2}-6g_{2}^{4}+{\lambda_{4}}^{2}-\left(\frac{36}{5}g_{1}^{2}+24g_{2}^{2}\right)\lambda_{3}+24\lambda_{2}\lambda_{3}+18{\lambda_{3}}^{2}\\
&+8 \lambda_3 \mbox{Tr}\Big({Y_\Delta^{\dagger} Y_\Delta}\Big) +4 \mbox{Tr}\Big({Y_\Delta^{\dagger}  Y_\Delta  Y_\Delta^{\dagger}  Y_\Delta}\Big)\,.
\end{align}
%\newpage
\underline{\bf Yukawa couplings: }\\[-.4cm]
\begin{align}
\beta_{Y_\Delta}^{(1)} & =  
\frac{1}{10} \Big(60 \,{Y_\Delta  Y_\Delta^{\dagger}  Y_\Delta}  + Y_\Delta \Big(20 \mbox{Tr}\Big({Y_\Delta^{\dagger}  Y_\Delta}\Big)  -9 \Big(5 g_{2}^{2}  + g_{1}^{2}\Big)\Big)\Big),\\
\beta_{y_t}^{(1)}&  =  
\frac{3}{2} y_t^3 + y_t \Big(3 y_t^2 -8 g_{3}^{2}  -\frac{17}{20} g_{1}^{2}  -\frac{9}{4} g_{2}^{2} \Big),
\end{align}
\begin{align}
\beta_{\kappa}^{(1)}  =  
-\frac{27}{10} g_{1}^{2} \kappa -\frac{21}{2} g_{2}^{2} \kappa +\kappa \lambda +4 \kappa \lambda_1 +6 \kappa \lambda_{4}  +6 \kappa y_t^2 +2 \kappa \mbox{Tr}\Big({Y_\Delta^{\dagger}  Y_\Delta}\Big).
\end{align}
\section{Calculation of the oblique parameters $S$, $T$ and $U$ parameters}
\label{app:STU}
In this appendix, we present the explicit expressions used to evaluate the oblique parameters ($S$, $T$, $U$) for type-II seesaw.
Similar expressions have already been presented in Ref.~\cite{Chun:2012jw}.
Since there was a typo in the $U$ parameter in Ref.~\cite{Chun:2012jw}, we re-drive the expressions using the result in Ref.~\cite{Lavoura:1993nq}
which computed the oblique parameters for generic scalar multiplets assuming their VEVs are negligibly small. Applying their results to the Higgs triplet, we obtain:
\begin{align}
S & =-\frac{1}{3\pi}\sum_{k=-1}^{1}\left[k\ln m_{k}^{2}+6(k-s_{W}^{2}Q_{k})^{2}\xi\left(\frac{m_{k}^{2}}{m_{Z}^{2}},\frac{m_{k}^{2}}{m_{Z}^{2}}\right)\right],\label{eq:m-11}\\
T & =\frac{1}{16\pi s_{W}^{2}}\sum_{k=-1}^{1}\left[(2+k-k^{2})\eta\left(\frac{m_{k}^{2}}{m_{W}^{2}},\frac{m_{k-1}^{2}}{m_{W}^{2}}\right)\right],\label{eq:m-12}\\
U & =\frac{1}{6\pi}\ln\left(\frac{m_{H^{\pm}}^{4}}{m_{H^{\pm\pm}}^{2}m_{H^{0}}^{2}}\right)\nonumber \\
 & +\frac{1}{\pi}\sum_{k=-1}^{1}\left[(-2-k+k^{2})\xi\left(\frac{m_{k}^{2}}{m_{W}^{2}},\frac{m_{k-1}^{2}}{m_{W}^{2}}\right)+2(k-s_{W}^{2}Q_{k})^{2}\xi\left(\frac{m_{k}^{2}}{m_{Z}^{2}},\frac{m_{k}^{2}}{m_{Z}^{2}}\right)\right],\label{eq:m-13}
\end{align}
where 
\begin{equation}
m_{k}=\{m_{H^{0}},\ m_{H^{\pm}},\ m_{H^{\pm\pm}}\},\ Q_{k}=\{0,\ 1,\ 2\}\ \ \text{for}\ k=\{-1,\ 0,\ 1\}.\label{eq:m-15}
\end{equation}
Note that in Eqs.~\eqref{eq:m-12} and \eqref{eq:m-13} when $k=-1$, although $m_{k-1}$ is undefined, terms containing $m_{k-1}$ always vanish. The $\eta$ function is defined as
\begin{equation}
\eta(x,y)=x+y-\frac{2xy}{x-y}\log\left(\frac{x}{y}\right).\label{eq:m-16}
\end{equation}
For small mass splitting, it approximately reduces to 
\begin{equation}
\eta(x,y)\approx\frac{(y-x)^{2}}{3x},\ \ {\rm for}\ \frac{|y-x|}{x}\ll1\thinspace.\label{eq:m-17}
\end{equation}
The full form of the $\xi$ function can be found in Refs.~\cite{Chun:2012jw} and~\cite{Lavoura:1993nq}.
Since we are only interested in cases where the new particles are heavy and the mass splitting is small, we take the following approximate form: 
\begin{equation}
\xi(x,y)\approx\frac{1}{15}\frac{1}{4x-1}-\left(\frac{2(y-x)}{15}+\frac{1}{21}\right)\left(\frac{1}{4x-1}\right)^{2},\ \ {\rm for}\ \frac{|y-x|}{x}\ll1\ll4x-1\thinspace.\label{eq:m-14}
\end{equation}
Using the above approximate forms  and taking only the dominant contributions, we obtain the results in Eqs.~\eqref{eq:m-S}-\eqref{eq:m-U}.
%%%%%%%%%%%%%%%%%%
\section{Relevant definitions for decay-width calculations}
\label{app:decay-width}
Here we present the analytical forms of the quantities introduced when computing the triplet scalar decay branching ratios in Sec.~\ref{sec:collider-constraints}.
The relevant formulae are given below.
\begin{align}
&\xi_{H^\pm W^\mp \hat{\varphi}}=\cos\alpha\sin\beta_\pm-\sqrt{2}\sin\alpha\cos\beta_\pm,~ \sin\alpha\sin\beta_\pm+\sqrt{2}\cos\alpha\cos\beta_\pm,~ \sin\beta_0\sin\beta_\pm+\sqrt{2}\cos\beta_0\cos\beta_\pm,\nonumber\\
& \text{ for } \hat{\varphi}= h^0,H^0 \text{ and }A
\end{align} 

\begin{align}
&\lambda_{H^0 hh}=\frac{1}{4v_\Phi^3}\Big[2v_\Delta\left\{-2M_\Delta^2+v_\Phi^2(\lambda_1+\lambda_4)\right\} \cos^3\alpha+v_\Phi^3\left\{-3\lambda+4(\lambda_1+\lambda_4)\right\} \cos^2\alpha\sin\alpha\nonumber \\
& +4v_\Delta\big\{2M_\Delta^2+v_\Phi^2(3\lambda_2+3\lambda_3-\lambda_1-\lambda_4)\big\}\cos\alpha\sin^2\alpha-2v_\Phi^3(\lambda_1+\lambda_4)\sin^3\alpha\Big].
\end{align}

\begin{align}
&\lambda(x,y)=(1-x-y)^2-4xy~,
\\
&\beta(x)=\sqrt{\lambda(x,x)}=\sqrt{1-4x}~,
\\
&F(x)=\frac{3(1-8x+20x^2)}{\sqrt{4x-1}}\cos^{-1}\left(\frac{3x-1}{2x^{3/2}}\right) - \frac{(1-x)(2-13x+47x^2)}{2x}-\frac{3}{2}(1-6x+4x^2)\log x~, 
\\
&G(x,y)=\frac{1}{12y}\Bigg[2\left(-1+x\right)^3-9\left(-1+x^2\right)y+6\left(-1+x\right)y^2 -6\left(1+x-y\right)y\sqrt{-\lambda(x,y)}\Bigg\{\tan^{-1}\left(\frac{1-x+y}{\sqrt{-\lambda(x,y)}}\right) \notag
\\
&+\tan^{-1}\left(\frac{1-x-y}{\sqrt{-\lambda(x,y)}}\right)\Bigg\}-3\left(1+\left(x-y\right)^2-2y\right)y\log x\Bigg]~,
\\
&H(x,y)=\frac{\tan^{-1}\left(\frac{1-x+y}{\sqrt{-\lambda(x,y)}}\right) +\tan^{-1}\left(\frac{1-x-y}{\sqrt{-\lambda(x,y)}}\right)}{4x \sqrt{-\lambda(x,y)}} \Big\{-3x^3+(9y+7)x^2-5(1-y)^2x+(1-y)^3\Big\} \notag
\\
&+\frac{1}{24xy}\Big\{(-1+x)(2+2x^2+6y^2-4x-9y+39xy)-3y(1-3x^2+y^2-4x-2y+6xy)\log x\Big\}~.
\end{align}

\bibliographystyle{utphys}
\bibliography{bibliography}

\end{document}